# TOWARDS A EUROPEAN QUANTUM ACT

A TWO-PILLAR FRAMEWORK FOR REGULATION AND INNOVATION

Mauritz Kop[1]

## ABSTRACT

Quantum technologies, encompassing computing, sensing, networking, and AI hybrids, promise transformative advancements but also pose significant dual-use risks. Realizing their economic potential while mitigating inherent risks and upholding fundamental values necessitates a robust, anticipatory, and harmonized regulatory framework at the EU level, grounded in the precautionary principle. Looking beyond the familiar, such a framework must be *sui generis*, as foundational quantum mechanical phenomena—including superposition, entanglement, and tunneling—defy the intuitive classical assumptions of factual certainty, locality, and causality that underpin existing legal paradigms. The unprecedented capabilities and risks stemming from these non-classical properties demand a tailored legal structure and bespoke regulatory architecture, contributing to the emerging *lex specialis* for quantum information technologies.

This contribution outlines the rationale for a dedicated European Quantum Act (EU QA), responding to strategic imperatives such as the Quantum Europe Strategy. It draws valuable lessons from legislative strategies in the semiconductor (EU/US Chips Acts) and AI (EU AI Act, UK pro-innovation approach, US AI Action Plan, and China's AI+ Plan) domains, as well as governance models from the nuclear sector (IAEA/NPT). A central aspect of this analysis involves a strategic comparison with U.S. and Chinese technology policy, which can be viewed through the lens of an emerging 'American Digital Silk Road'—an effort to build a techno-economic sphere of influence by setting the global rules of the road. Understanding this analogy provides the EU with valuable insights into the importance of adopting a "full-stack" industrial policy.

In response, our analysis concludes that the EU Quantum Act should be a two-pillar instrument, combining New Legislative Framework (NLF)-style regulation with an ambitious Chips Act-style industrial and security policy. This method simultaneously fosters innovation through strategic investment while imposing clear, risk-based tiers and regulatory sandboxes to avoid overregulation. The framework is designed to be modular and adaptive across the technology's lifecycle. For its industrial policy, the Act can draw from the EU and US Chips Acts to model funding, secure supply chains, and accelerate the "lab-to-market" pipeline, potentially enhanced by a DARPA-style agency to incentivize high-risk, high-reward research. A comparative analysis of global innovation systems (US, EU, China) informs the need for the EU QA to strategically support both fundamental research and commercialization, particularly for dual-use technologies, which is presented as a responsible act of deterrence necessary for national security and technological sovereignty, thereby bolstering the EU's long-term economic competitiveness. This strategic investment is a prerequisite for the EU to become an indispensable partner in a transatlantic tech alliance, ensuring democratic values inform the next technological era.

---

[1] Mauritz Kop is the Founder of the Stanford Center for Responsible Quantum Technology and a Senior Fellow at the Centre for International Governance Innovation (CIGI). He also serves as Expert at the von Neumann Commission. Contact: mkop@stanford.edu.

Key recommendations for the Act include establishing a dedicated EU Office of Quantum Technology Assessment (OQTA) for expert oversight; adopting a "standards-first" philosophy to embed fundamental rights directly into the technology's architecture through certifiable quality management systems (QT-QMS) that can lead to a CE mark, a model learned from successful precedent in medical device regulation; designing novel privacy-enhancing techniques (PETs) that protect against identity-theft; creating a 10-year intellectual property right for foundational quantum algorithms and software under mandatory FRAND license terms; ensuring strategic preparedness for quantum cybersecurity to mitigate the 'Q-Day' threat through a new legal duty of 'Anticipatory Data Stewardship'; utilizing a data-driven methodology for supply chain security, for instance through a Quantum Criticality Index (QCI); and implementing harmonized, use-case specific export controls.

Our contribution articulates the concept of a 'quantum event horizon' to underscore the deep uncertainty and risk of technological lock-in, arguing for a proactive strategy that combines heavy investment in responsible innovation with adaptive governance. To address the risk of technology outpacing classical guardrails, the article explores forward-looking governance paradigms, including novel 'algorithmic regulation' mechanisms and the imposition of a legally enforceable fiduciary duty upon advanced AI systems to act as 'quantum-agentic stewards'. The paper further argues that these transformative technologies may necessitate -at a future stage- new legal doctrines such as the 'Quantum Lex Machina', and new philosophical foundations, which it then proceeds to examine.

On the international stage, the EU should proactively engage in discussions towards a global non-proliferation framework for quantum and AI weapons of mass destruction, inspired by the IAEA/NPT model and overseen by an 'International Quantum Agency (IQA)'. This vision culminates in the 'Qubits for Peace' initiative, a global governance structure designed to ensure quantum technologies are developed safely, ethically, and for the benefit of all humanity. The article concludes by consolidating the analysis into a detailed legislative blueprint for a prospective EU Quantum Act.

TABLE OF CONTENTS

## 1. INTRODUCTION: THE NEED FOR A QUANTUM REGULATORY FRAMEWORK

The advent of quantum technologies—spanning computing, sensing, networking, and AI hybrids—heralds a period of profound societal and economic change. This section introduces the significant opportunities these technologies present across various sectors, while simultaneously highlighting the inherent risks, particularly their dual-use character, that necessitate a dedicated and proactive European regulatory framework. The rationale for establishing a specific EU Quantum Act is presented here, setting the context for the detailed analysis in the subsequent sections.

## 1.1. The Transformative Potential of Quantum Technologies

Quantum mechanics is poised to revolutionize numerous fields through technologies that harness the principles of superposition, entanglement, and tunneling.[2] Quantum computing promises unprecedented breakthroughs in complex domains such as drug discovery, materials science, and financial modeling by exploring vast computational spaces simultaneously.[3] In parallel, quantum sensors offer levels of precision that could transform medical diagnostics and autonomous navigation by detecting minute variations previously thought impossible.[4] Quantum networking aims to establish unconditionally secure communication channels based on the principle that observing a quantum state inevitably alters it, and enable distributed quantum computingquantum networking aims to establish unconditionally secure communication channels and enable distributed quantum computing.[5] The convergence of quantum and artificial intelligence is expected to yield further advances in data analysis and complex problem-solving. Consequently, the anticipated impacts are vast, spanning finance (trading optimization), MedTech (diagnostics, drug development), energy (storage, grid management), logistics (supply chain optimization), space (navigation, communication), and defense (sensing, secure communication, computation).[6]

Many of these technologies possess a dual-use character, applicable for both beneficial civilian and potentially harmful military purposes.[7] A sensor with the precision to aid medical diagnostics could also be used for missile guidance; a computer powerful enough to accelerate drug discovery could also be used to break classical encryption. This latter risk, often termed "Q-Day," represents a tangible threat, as adversaries can harvest encrypted data today for decryption with a future quantum computer.[8] This inherent duality, alongside the prospect of significant societal disruption, necessitates careful ethical and security consideration. Proactive governance is therefore essential, drawing lessons from the regulatory experiences with other transformative technologies like AI, genetics, nanotechnology, semiconductors, and nuclear energy.[9]

---

[2] *See* Kaye, P., Laflamme R., & Mosca, M., An Introduction to Quantum Computing, Oxford University Press. (2007); and Michael A. Nielsen & Isaac L. Chuang, Quantum Computation and Quantum Information (2010) Cambridge University Press, https://doi.org/10.1017/CBO9780511976667

[3] *See* John Preskill, *Quantum Computing in the NISQ Era and Beyond*, 2 Quantum 79 (2018), https://quantum-journal.org/papers/q-2018-08-06-79/; Bova, F. and Goldfarb, A., and Melko, R., Quantum Economic Advantage, Management Science, vol 69(2), pages 1116-1126, (2023), https://www.nber.org/papers/w29724; National Academies of Sciences, Engineering, and Medicine. Quantum Computing: Progress and Prospects. Edited by Mark Horowitz and Emily Grumbling, Washington, DC: The National Academies Press, pp 158-159, (2019) https://doi.org/10.17226/25196; and De Jong, E., Own the Unknown: An Anticipatory Approach to Prepare Society for the Quantum Age, Digital Society, Quantum-ELSPI TC, 1, Springer Nature, (2022), https://link.springer.com/article/10.1007/s44206-022-00020-4].

[4] *See* C. L. Degen, F. Reinhard & P. Cappellaro, *Quantum Sensing*, 89 Rev. Mod. Physics 035002 (2017), https://journals.aps.org/rmp/abstract/10.1103/RevModPhys.89.035002

[5] *See* Stephanie Wehner, David Elkouss & Ronald Hanson, *Quantum Internet: A Vision for the Road Ahead*, 362 Science 303 (2018), https://www.science.org/doi/10.1126/science.aam9288; and DeNardis, L., Quantum Internet Protocols (2022), http://dx.doi.org/10.2139/ssrn.4182865.

[6] *See* Krelina, M. Quantum technology for military applications. EPJ Quantum Technol. 8, 24 (2021). https://doi.org/10.1140/epjqt/s40507-021-00113-y; and TUM Think Tank, An artful approach to quantum applications in medicine, 19 Nov 2024, https://tumthinktank.de/en/output/medical-quantum-wonderland/.

[7] *See* e.g., Pau Alvarez-Aragones, *The New Arms Race in Dual-Use Technologies*, IE INSIGHTS (Sept. 2, 2024), https://www.ie.edu/insights/articles/the-new-arms-race-in-dual-use-technologies/; Zhou Q., The subatomic arms race: Mutually assured development, Harvard International Review, (2021), https://hir.harvard.edu/the-subatomic-arms-race-mutually-assured-development/; Giles, M., The US and China are in a quantum arms race that will transform warfare, MIT Technology Review (2019), https://www.technologyreview.com/2019/01/03/137969/us-china-quantum-arms-race/; and Inglesant P., et al, Responsible Innovation in QT applied to Defence and National Security, NQIT, (2018).

### 1.2. Rationale for a Dedicated EU Quantum Act

The combination of transformative promise and inherent risk underscores the urgent need for a dedicated, harmonized EU regulatory framework. The European Commission has affirmed this imperative by adopting the "Quantum Europe Strategy" in July 2025, which explicitly announces the intention to present a legislative proposal for a European Quantum Act in 2026.[10] Such an Act would provide clear and consistent rules, fostering innovation while proactively addressing risks before they materialize and ensuring alignment with European values.[11] Navigating the quantum event horizon and the associated Collingridge dilemma requires proactive engagement now, during the technology's formative stages. The design of this EU Quantum Act must be firmly rooted within the specific European legal and political context, acknowledging how differing traditions shape regulatory outcomes, particularly concerning fundamental rights and the application of principles like precaution, as comparative studies by scholars such as Mökander and Floridi illustrate.[12] In the words of Perrier, this need for a technologically specific legal instrument is underscored by the argument that effective governance cannot be 'technology neutral'; it must address the unique affordances and operational realities of the technology in question to provide legal and strategic certainty.[13]

---

[8] *See* e.g., Yaakov Weinstein & Brandon Rodenburg, Quantum Computing: Quantifying the Current State of the Art to Assess Cybersecurity Threats, MITRE (2025), https://www.mitre.org/news-insights/publication/intelligence-after-next-quantum-computing-quantifying-current-state-art; Wimmer, M., Moraes, T., Quantum Computing, Digital Constitutionalism, and the Right to Encryption: Perspectives from Brazil. DISO 1, 12 (2022). https://doi.org/10.1007/s44206-022-00012-4; Rand, L., and Rand, T., The "Prime Factors" of Quantum Cryptography Regulation, 3 Notre Dame J. on Emerging Tech. 37 (2022); and Rogers, M., and Minerbi, N. The Quantum Computing Arms Race is not Just About Breaking Encryption Keys, NextGov, (2022), https://www.nextgov.com/ideas/2022/06/quantum-computing-arms-race-not-just-about-breaking-encryption-keys/368834/.
[9] *See* e.g., Johnson, W. G. (2018). Governance tools for the second quantum revolution. Jurimetrics, 59, 487, https://papers.ssrn.com/sol3/papers.cfm?abstract_id=3350830.
[10] European Commission, Communication from the Commission, Quantum Europe Strategy: Quantum Europe in a Changing World, COM(2025) 363 final (July 2, 2025), https://digital-strategy.ec.europa.eu/en/library/quantum-europe-strategy. *See* also European Declaration on Quantum Technologies (Dec. 6, 2023); Eur. Quantum Flagship, Strategic Research and Industry Agenda (SRIA) 2030 (2024); and Eur. Quantum Indus. Consortium (QuIC), Strategic Industry Roadmap (SIR) (2025).
[11] *See* Kai Zenner et al., The 'European Way' - A Blueprint for Reclaiming Our Digital Future (2025). https://papers.ssrn.com/sol3/papers.cfm?abstract_id=5251254.
[12] Mökander, Jakob and Juneja, Prathm and Watson, David and Floridi, Luciano, The US Algorithmic Accountability Act of 2022 vs. The EU Artificial Intelligence Act: What can they learn from each other? (August 18, 2022), Minds & Machines (2022). https://doi.org/10.1007/s11023-022-09612-y.

This is profoundly true for quantum technologies, whose operational reality is rooted in non-classical physics. This break from classical intuition began with Max Planck, widely considered the father of quantum theory, whose 1900 work on black-body radiation first introduced the revolutionary concept that energy is emitted and absorbed in discrete "quanta".[14] This foundational insight unleashed a torrent of theoretical development that culminated in the Copenhagen interpretation of quantum mechanics in the late 1920s, a framework developed principally by Niels Bohr and Werner Heisenberg. This interpretation rests on principles that directly challenge classical determinism. Heisenberg's uncertainty principle, for instance, established that it is impossible to simultaneously know certain pairs of a particle's properties—such as its exact position and momentum—with arbitrary precision.[15] Complementing this, Bohr's principle of complementarity asserted that quantum phenomena possess contradictory yet essential properties (e.g., behaving as both a wave and a particle) that cannot be observed at the same time but are all required for a complete description.[16] The mathematical underpinnings for this strange new reality were rigorously formalized by John von Neumann, whose 1932 work established the Hilbert space formalism that remains the standard mathematical language of quantum mechanics.[17] Later, the work of Richard Phillips Feynman provided a powerful new perspective with his path integral formulation and Feynman diagrams, which became indispensable tools for both conceptualizing and calculating quantum interactions, particularly in quantum electrodynamics.[18]

Our existing legal frameworks are implicitly built upon classical assumptions of cause and effect. The very nature of superposition (the ability to exist in multiple states at once), entanglement (non-local correlations),[19]—and tunneling (the capacity to breach classical barriers) creates unprecedented capabilities and, in turn, risks that demand a *sui generis* legal approach. The qualitative and quantitative exceptionality of these quantum effects—intrinsic to any technology built upon quantum mechanics—creates profound challenges for factual certainty, liability, and explainability across a range of sectors. To illustrate this inadequacy, it is useful to consider a range of current and prospective quantum applications, scenarios, and use cases; these examples, categorized by escalating risk in alignment with the EU AI Act's pyramid of criticality, demonstrate why existing laws fail when confronted with quantum phenomena.[20]

---

[13] Perrier, E., Quantum Information Technologies and Public International Law - A Strategic Perspective, 3 Stanford Center for RQT Research Series (2024). https://purl.stanford.edu/vs906nm4136.
[14] Planck, M. (1901). Ueber das Gesetz der Energieverteilung im Normalspectrum. *Annalen der Physik, 309*(3), 553–563, https://doi.org/10.1002/andp.19013090310
[15] Heisenberg, W. (1927). Über den anschaulichen Inhalt der quantentheoretischen Kinematik und Mechanik. *Zeitschrift für Physik, 43*(3-4), 172–198, https://doi.org/10.1007/BF01397280.
[16] Bohr, N. (1928). The Quantum Postulate and the Recent Development of Atomic Theory. *Nature, 121*(3050), 580–590, https://doi.org/10.1038/121580a0.
[17] von Neumann, J. (1932). *Mathematische Grundlagen der Quantenmechanik*. Springer, https://link.springer.com/book/10.1007/978-3-642-61409-5.
[18] Feynman, R. P. (1948). Space-Time Approach to Non-Relativistic Quantum Mechanics. *Reviews of Modern Physics, 20*(2), 367–387, https://doi.org/10.1103/RevModPhys.20.367.
[19] Einstein, A., Podolsky, B., & Rosen, N. (1935). Can Quantum-Mechanical Description of Physical Reality Be Considered Complete?. *Physical Review, 47*(10), 777–780, https://doi.org/10.1103/PhysRev.47.777.

This need for a bespoke framework that contributes to the emerging *lex specialis* for quantum information technologies is rooted in a fundamental conflict between the physics of the quantum realm and the classical, Newtonian worldview of a deterministic and local reality, that underpins our legal systems. Its principles lead to an erosion of factual certainty through superposition, the end of locality through entanglement, and a profound challenge to causality through tunneling. For centuries, the law has been built on foundational assumptions that seem self-evident: that an object is in one place at one time (locality), that it has a definite state even if unobserved (realism), and that information requires a physical medium to travel between two points (causality). In other words, our legal systems are designed for a world of predictable billiard balls, where facts are certain and cause and effect are linear. Quantum mechanics fails this classical test. It fundamentally challenges these foundational assumptions. Superposition, for example, allows for a massive parallelism that powers quantum computing, but by enabling a system to exist in multiple states at once, it also creates profound challenges for verification, liability, and explainability, eroding the legal concept of factual certainty. Similarly, entanglement creates instantaneous correlations between particles regardless of distance—a phenomenon Albert Einstein famously decried as "spukhafte Fernwirkung" or "spooky action at a distance"—that defies classical notions of locality. This could have significant implications for secure communications, distributed computing, and even new forms of surveillance.

---

[20] Illustrative examples include: **1. Healthcare Liability (High Risk – Challenging Factual Certainty):** A quantum sensor, leveraging superposition to map probabilistic neurological decay, indicates a 70% likelihood of a patient developing a severe disease. Under consumer protection laws demanding diagnosis based on demonstrable facts, is a physician liable for prescribing a high-risk preventative treatment, or for *not* prescribing it? Medical liability law is built on a binary state of injury, not a quantum probability function. **2. Financial Regulation (High Risk – Challenging Causality):** A quantum computer, using tunneling algorithms to bypass market barriers and superposition to optimize a portfolio across millions of possibilities, triggers a flash crash. The machine's decision path is a computational "black box" with no linear, auditable trail of cause and effect, making it impossible to assign liability under regulations that require proving a direct causal link between an action and market harm. **3. Product Safety (High Risk – Challenging Factual Certainty):** A quantum simulation designs a new alloy, certifying it as 99.999% safe but flagging a 0.001% probability of catastrophic failure under conditions physically impossible to replicate in classical testing. If a failure occurs, is the manufacturer liable for a risk that was, for all practical purposes, purely theoretical and non-falsifiable pre-incident? This pits the probabilistic certainty of quantum simulation against the classical demand for empirical proof in product liability law. **4. Digital Forensics (High Risk – Challenging Evidence):** A malicious actor uses a quantum computer to "tunnel" through the encryption of a secure state database, altering information without breaking its cryptographic shell. Unlike a classical hack that leaves a trail, this quantum breach leaves no discernible evidence. A crime becomes impossible to prove under rules of evidence that rely on a chain of causality and material proof, threatening the integrity of all digital systems. **5. International Law (Unacceptable Risk – Challenging Locality):** A state actor uses an entanglement-based network to execute a defensive cyber-operation. An action on a qubit in its own territory instantaneously disables critical infrastructure in an aggressor state without any signal traversing the space between them. Which jurisdiction's laws apply to an act that is both here and there at once? This creates a non-local *casus belli* that fundamentally breaks the framework of international law, which is predicated on territoriality. **6. Autonomous Systems (Unacceptable Risk – Challenging All Tenets):** A hybrid quantum-classical AI agent, tasked with national security, uses its quantum core to analyze potential future threats in superposition. It identifies a high-probability attack and, using an entangled key, non-locally and pre-emptively neutralizes it. The system acts on a probabilistic future (eroding factual certainty), its decision is opaque (breaking causality), and its effect is instantaneous and borderless (violating locality). Such a "quantum-agentic" system, operating beyond classical cause and effect, represents a total paradigm breakdown, rendering concepts of meaningful human control, accountability, and the laws of armed conflict obsolete.

Specifically, properties like superposition, entanglement, and tunneling defy traditional legal concepts of factual certainty, locality, and causality, demanding a bespoke regulatory architecture. This mismatch creates novel and complex challenges, particularly for private international law and civil liability. For instance, if an action initiated in one jurisdiction causes an instantaneous effect in another via an entangled particle, which state's laws apply? How is causality established for civil liability purposes when the connection is non-local? These fundamental questions highlight that traditional legal frameworks are ill-equipped to resolve cross-border disputes involving quantum effects. Addressing these gaps may require legal innovations, such as developing doctrines of 'probabilistic causation' for liability, establishing international conventions on jurisdictional primacy in cases of entangled effects, or adapting 'market share liability' principles for scenarios where quantum effects obscure direct causal links.

Therefore, a *sui generis* framework in both public and private law is not merely preferable but necessary, as simply adapting existing laws—designed for a classical world—is inadequate to address the qualitative and quantitative exceptionality of these quantum effects.[21] This bespoke legal framework should not be viewed as a regulatory burden, but as a strategic enabler of technological sovereignty and a competitive advantage, providing the legal certainty and predictability necessary to attract long-term investment and build public trust.

Given that the precautionary principle is a cornerstone of EU consumer protection and environmental and health regulation, enshrined in Article 191 of the Treaty on the Functioning of the European Union (TFEU),[22] its application is particularly salient for addressing the profound, uncertain, and potentially irreversible societal risks that come with the suite of quantum technologies. These are not merely incremental risks but qualitatively different ones, stemming directly from the non-intuitive nature of quantum mechanics itself. The precautionary principle mandates a proactive stance, which must in turn be carefully balanced by the principles of proportionality and subsidiarity to incentivize innovation and progress.

---

[21] Kop, M. *et al. Towards Responsible Quantum Technology* (Harvard Berkman Klein Center for Internet & Society, 2023), https://cyber.harvard.edu/publication/2023/towards-responsible-quantum-technology.

[22] Consolidated Version of the Treaty on the Functioning of the European Union art. 191, May 9, 2008, 2008 O.J. (C 115) 47, https://eur-lex.europa.eu/LexUriServ/LexUriServ.do?uri=CELEX:12008E191:EN:HTML

In addition to provisions tailored to quantum's unique functionality, the EU Quantum Act should strategically synthesize elements from existing technology best practices, regulations, and innovation systems. An effective approach requires synthesizing the lessons of past technological revolutions: it must combine the *anticipatory foresight* and international coordination of the nuclear model for managing systemic risks; the *ethical integration* and public engagement learned from biotechnology and AI's struggles to ensure social legitimacy; and the *adaptive flexibility* and multi-stakeholder approach that allowed the internet to innovate and scale. Finally, the historical challenges of nanotechnology regulation—including overregulation, overpromising, and a lack of public awareness—underscore the critical need for proactive dialogue and the avoidance of overly burdensome rules that could stifle innovation.

Furthermore, the EU Quantum Act must learn from the strategic missteps observed in the discourse surrounding Artificial Intelligence. The intense focus of Silicon Valley on achieving artificial general intelligence (AGI), coupled with conflicting and often alarming narratives ranging from utopian promises to existential threats, has created a 'widening schism' between the tech elite and a general public that remains skeptical of the hype.[23] To avoid this pitfall, the EU's quantum strategy must be grounded not in speculative, far-off goals, but in transparent, tangible benefits. By prioritizing public trust and demonstrating real-world value from the outset, the Act can build the societal consensus necessary for long-term success, a stark contrast to a model that risks alienating the very public it purports to serve.

The EU Quantum Act could emulate the EU and US Chips Acts to establish models for funding, "lab-to-market" innovation, and supply chain security.[24] From the EU AI Act, and the principles of the proposed AI Liability Directive which were partially integrated into the main Act, it could draw upon the risk-based product safety regime, reinforcement mechanisms, and clear liability rules.[25] The principles-based, flexible approach of the UK's Pro-Innovation AI Regulation offers a model for adaptable governance that can evolve with the technology.[26]

---

[23] Eric Schmidt & Selina Xu, *Silicon Valley Needs to Stop Obsessing Over Superhuman A.I.*, N.Y. Times, Aug. 19, 2025, https://www.nytimes.com/2025/08/19/opinion/artificial-general-intelligence-superintelligence.html.

[24] *See* European Commission, European Chips Act, https://commission.europa.eu/strategy-and-policy/priorities-2019-2024/europe-fit-digital-age/european-chips-act_en; CHIPS and Science Act of 2022, Pub. L. No. 117-167, 136 Stat. 1366; Fact Sheet: CHIPS and Science Act Will Lower Costs, Create Jobs, Strengthen Supply Chains, and Counter China, The White House, August 9, 2022, available at https://www.whitehouse.gov/briefing-room/statements-releases/2022/08/09/fact-sheet-chips-and-science-act-will-lower-costs-create-jobs-strengthen-supply-chains-and-counter-china/; European Commission, European Chips Act - Questions and Answers (Sept. 21, 2023), https://ec.europa.eu/commission/presscorner/detail/en/qanda_23_4519; PubAffairs Bruxelles, European Chips Act enters into force today - Questions and answers (Sept. 21, 2023), https://www.pubaffairsbruxelles.eu/eu-institution-news/european-chips-act-enters-into-force-today-questions-and-answers/; Regulation (EU) 2023/1781 of the European Parliament and of the Council of 13 September 2023 establishing a framework of measures for strengthening Europe's semiconductor ecosystem (EU Chips Act), 2023 O.J. (L 229) 1.; and Regulation (EU) 2023/1781 of the European Parliament and of the Council of 13 September 2023 establishing a framework of measures for strengthening Europe's semiconductor ecosystem (EU Chips Act), 2023 O.J. (L 229) 1.

Additionally, insights from the global innovation systems of the US, EU, and China can inform best practices for fostering dual-use technology and securing funding.[27] This is not simply a matter of emulating successful models but of building a robust European capacity that makes the EU a more capable and indispensable partner for the United States. A strong, technologically sovereign EU is the cornerstone of a strategic transatlantic alliance aimed at securing democratic leadership in the global technology race against rising techno-authoritarianism. This transatlantic partnership extends beyond mere technological and economic cooperation; it is fundamentally a political alliance rooted in shared democratic values. As argued by scholars like Robert Person and Michael McFaul, the enduring strength of an alliance like NATO lies not only in its collective defense commitments but in its role as a political community where democracies converge to set standards, build consensus, and reinforce liberal norms.[28] In the same vein, a transatlantic tech alliance for quantum must function as a forum for setting the normative and ethical standards for this new technological era. By jointly developing regulatory frameworks, promoting responsible innovation, and establishing common principles for data governance and security, the EU and the US can ensure that the quantum future is shaped by democratic principles, not authoritarian designs. This political dimension is what transforms a simple technology partnership into a powerful bulwark for the liberal international order.

From a U.S. foreign policy perspective, a technologically sovereign EU is not a competitor but a prerequisite for a resilient democratic bloc capable of setting global norms. The true strength of such an alliance lies not merely in harmonized export controls but in a common, democratically legitimated approach to the technology's use, an argument developed further in the U.S. grand strategy for quantum technology.[29] This view reinforces the imperative for the EU to build a robust, values-based quantum ecosystem, positioning it as a more capable and indispensable partner in shaping the global technological order.

---

[25] Regulation (EU) 2024/1689 of the European Parliament and of the Council of 13 June 2024 laying down harmonised rules on artificial intelligence and amending certain Union legislative acts (EU AI Act), 2024 O.J. (L 1689) 1, https://eur-lex.europa.eu/eli/reg/2024/1689/oj/eng.

[26] *See* Dep't for Sci., Innovation & Tech., A Pro-Innovation Approach to AI Regulation: Government Response (2025), https://www.local.gov.uk/our-support/cyber-digital-and-technology/cyber-digital-and-technology-policy-team/pro-innovation; and Dep't for Sci., Innovation & Tech., A Pro-Innovation Approach to AI Regulation, Policy Paper (Mar. 29, 2023), https://www.gov.uk/government/publications/ai-regulation-a-pro-innovation-approach/white-paper.

[27] *See* e.g., Dep't of Def., Scaling Nontraditional Defense Innovation (2025), https://innovation.defense.gov/Portals/63/DIB%20Scaling%20Nontraditional%20Defense%20Innovation%20250113%20PUBLISHED.pdf; and Varieties of Capitalism: The Institutional Foundations of Comparative Advantage (Peter A. Hall & David Soskice eds., 2001).

[28] Robert Person & Michael McFaul, *Why NATO Is More Than Democracy's Best Defense*, J. of Democracy, Apr. 2024, at https://www.journalofdemocracy.org/online-exclusive/why-nato-is-more-than-democracys-best-defense/.

[29] *See* Kop, M., *Quantum Grand Strategy*, forthcoming (2025).

Crucial lessons in safety, security, non-proliferation verification, export controls, and international collaboration can be gleaned from the long-standing governance of nuclear energy through bodies like the IAEA and treaties such as the NPT.[30] Lastly, the ethical guidelines developed for genetics, AI, and nanotechnology can provide essential moral guidelines for this new technological frontier.[31]

A core pillar of the EU Quantum Act should be the concept of Responsible Quantum Technology (RQT), a framework ensuring that research and innovation align with societal demands and enhance planetary welfare.[32] This involves operationalizing a set of guiding principles—such as the 10 Principles for Responsible Quantum Innovation—to proactively manage the ethical, legal, socio-economic, and policy implications (Quantum-ELSPI) of the technology.[33]

---

[30] *See* Treaty on the Non-Proliferation of Nuclear Weapons art. IV, July 1, 1968, 21 U.S.T. 483, 729 U.N.T.S. 161, https://disarmament.unoda.org/wmd/nuclear/npt/text/; William Walker, A Perpetual Menace: Nuclear Weapons and International Order (2012); and Ionut Suseanu, *The NPT and IAEA Safeguards*, IAEA Bulletin, https://www.iaea.org/bulletin/the-npt-and-iaea-safeguards.
[31] *See* Chris Jay Hoofnagle & Simson Garfinkel, Law and Policy for the Quantum Age (2022); Mark A. Lemley, IP in a World Without Scarcity, 90 N.Y.U. L. Rev. 460 (2015); and Frank Pasquale, The Black Box Society: The Secret Algorithms That Control Money and Information (2015).
[32] Kop, M. *et al. supra* note 20 (Harvard RQT). *See also* Owen, R., von Schomberg, R., & Macnaghten, P. (2021). An unfinished journey? Reflections on a decade of responsible research and innovation. Journal of Responsible Innovation, 1–17. https://doi.org/10.1080/23299460.2021.1948789; and Roberson, T., Leach, J., & Raman, S. (2021). Talking about public good for the second quantum revolution: Analysing quantum technology narratives in the context of national strategies. Quantum Science and Technology, 6(2), 25001. https://doi.org/10.1088/2058-9565/abc5ab.
[33] Kop, M. *et al., 10 Principles for Responsible Quantum Innovation*, 9 Quantum Sci. & Tech. 035013 (2024), https://doi.org/10.1088/2058-9565/ad3776. *See* also Coenen, C., Grinbaum, A., Grunwald, A. et al. Quantum Technologies and Society: Towards a Different Spin. Nanoethics 16, 1–6 (2022). https://doi.org/10.1007/s11569-021-00409-4; Kiesow Cortez E, Yakowitz Bambauer J and Guha S 2023 A quantum policy and ethics roadmap (available at: https://ssrn.com/abstract=4507090); Perrier E 2021 Ethical quantum computing: a roadmap (arXiv:2102.00759); Mauritz Kop, Ethics in the Quantum Age, 2021 Phys. World 34 (12) 31, https://iopscience.iop.org/article/10.1088/2058-7058/34/12/34; Coenen C and Grunwald A 2017 Responsible research and innovation (RRI) in quantum technology Ethics Inf. Technol. 19 277–94 https://link.springer.com/article/10.1007/s10676-017-9432-6; Hoffmann CH, Flöther FF. Why business adoption of quantum and AI technology must be ethical. Research Directions: Quantum Technologies. 2024;2:e4. doi:10.1017/qut.2024.5, https://www.cambridge.org/core/journals/research-directions-quantum-technologies/article/why-business-adoption-of-quantum-and-ai-technology-must-be-ethical/04B38DA86F9FB14C73C4CA132724C565; Bano, M., Ali, S. & Zowghi, D. Envisioning responsible quantum software engineering and quantum artificial intelligence. Autom Softw Eng 32, 69 (2025). https://doi.org/10.1007/s10515-025-00541-5; Umbrello, S.: Ethics of quantum technologies: A scoping review. Int. J. Appl. Phil. (2024), https://www.pdcnet.org/ijap/content/ijap_2024_0999_4_8_201; and Possati, Luca M. (2023). Ethics of Quantum Computing: an Outline. Philosophy and Technology 36 (3):1-21, https://link.springer.com/article/10.1007/s13347-023-00651-6

The Act must also address the fundamental need for standardization, certification, and performance benchmarking.[34] Adopting a 'standards-first' philosophy, as advocated by Aboy *et al.*, prioritizes the establishment of technical standards during the current R&D phase, offering a strategic advantage for early-stage governance before widespread application.[35] The framework should incorporate targeted export controls to manage dual-use risks[36] while upholding the principles of proportionality and subsidiarity to ensure compatibility with existing EU sectoral legislation, such as the AI Act and the Medical Device Regulation.[37] To provide central expertise and coordination, a proposed EU Office of Quantum Technology Assessment (OQTA), inspired by bodies like the former US Office of Technology Assessment (OTA), CERN, and the IAEA, could be established.[38] The framework must aim to bridge the gap between high-level regulatory oversight and practical engineering implementation.

### 1.3. Scope and Objectives of the Proposed Legislation

The EU Quantum Act should have a broad scope, encompassing quantum computing, sensing, networking, quantum-AI hybrids, and the vital supply chain of critical minerals that underpins them.[39] Its primary objectives should be multifaceted. The legislation must first and foremost promote innovation to foster a thriving, globally competitive EU quantum ecosystem that includes hardware, software, and networking capabilities.[40] Concurrently, it must ensure societal benefit and manage risk by carefully balancing the technology's promise against its perils. This involves embedding ethical considerations from the outset, promoting equitable access to prevent a 'quantum divide',[41] and safeguarding against dual-use misuse through guardrails learned from AI, nano, and nuclear governance.[42] To provide a positive, universally understood and globally resonant mission, a key objective should be to direct and incentivize the development of quantum technologies towards addressing global challenges as articulated in the United Nations Sustainable Development Goals (SDGs).[43] It must also pioneer new legal tools to protect fundamental rights in the face of novel threats. This includes exploring novel privacy-enhancing techniques (PETs), such as a copyright-based framework for personal biometrics -including face, voice, iris, fingerprints- inspired by recent Danish legislation, to provide individuals with ownership and control through enforcement mechanisms over their digital likeness and safeguard against quantum-AI enabled deepfakes and identity theft.[44] Lastly, the Act must establish effective governance structures for standardization, certification, benchmarking, balanced intellectual property frameworks, competition oversight, and targeted export controls.[45] This will require a sophisticated mix of hard and soft law, including codes of conduct and Quantum Impact Assessments (QIAs), alongside support for open innovation and self-regulation where appropriate.[46] Such a balanced system would avoid imposing excessive burdens on SMEs, a lesson from the significant compliance challenges that regulations like the General Data Protection Regulation (GDPR) and the EU AI Act have placed on startups and scale-ups.[47]

---

[34] *See* e.g., Majidy, S., Wilson, C. and Laflamme, R. (2024) *Building Quantum Computers: A Practical Introduction*. Cambridge: Cambridge University Press, https://doi.org/10.1017/9781009417020

[35] Aboy, M., Gasser, U., I. Cohen, G., and Kop, M., *Quantum technology governance: a standards-first approach*, Science 389, 575-578 (2025), https://www.science.org/doi/10.1126/science.adw0018

[36] *See* e.g., European Comm'n, Directorate-Gen. for Trade, *Exporting dual-use items*, https://policy.trade.ec.europa.eu/help-exporters-and-importers/exporting-dual-use-items_en (last visited July 21, 2025).

[37] European Commission, AI Act, Shaping Europe's Digital Future, https://digital-strategy.ec.europa.eu/en/policies/regulatory-framework-ai; and Regulation (EU) 2017/745 of the European Parliament and of the Council of 5 April 2017 on medical devices, amending Directive 2001/83/EC, Regulation (EC) No 178/2002 and Regulation (EC) No 1223/2009 and repealing Council Directives 90/385/EEC and 93/42/EEC[1], 2017 O.J. (L 117) 1, https://eur-lex.europa.eu/legal-content/EN/TXT/?uri=CELEX%3A02017R0745-20250110

[38] See Emily G. Blevins, Cong. Rsch. Serv., R46327, The Office of Technology Assessment; History, Authorities, Issues, and Options (2020), https://www.congress.gov/crs-product/R46327; and U.S. Gov't Accountability Off., *The Office of Technology Assessment*, Oct 13, 1977, https://www.gao.gov/products/103962

[39] *See* Regulation (EU) 2024/1252 of the European Parliament and of the Council of 11 April 2024 establishing a framework for ensuring a secure and sustainable supply of critical raw materials (European Critical Raw Materials Act), 2024 O.J. (L 1252) 1; and Regulation (EU) 2024/1252 of the European Parliament and of the Council of 11 April 2024 establishing a framework for ensuring a secure and sustainable supply of critical raw materials (European Critical Raw Materials Act), 2024 O.J. (L 1252) 1.

[40] *See* e.g., European Quantum Industry Consortium, Strategic Industry Roadmap (2025), https://www.euroquic.org/strategic-industry-roadmap-2025/; and Jonathan Ruane et al., MIT Initiative on the Digital Econ., The Quantum Index Report 2025 (2025). https://doi.org/10.48550/arXiv.2506.04259.

### 1.4. A Quantum Event Horizon in Quantum Policy Making

The challenge of governing quantum technologies is compounded by what can be termed the "quantum event horizon"—a metaphor highlighting the profound uncertainty surrounding their future development, applications, and societal impacts.[48] Much like an astrophysical event horizon marks a boundary of predictability, this technological event horizon signifies a point beyond which forecasting becomes exceptionally difficult.[49] Many possible quantum use cases remain speculative, and unforeseen breakthroughs—a potential "ChatGPT moment" for quantum—could rapidly alter the landscape. This inherent unpredictability strongly resonates with the Collingridge dilemma, which posits that attempting to control a technology is easiest in its early stages when knowledge is limited, but becomes increasingly difficult once the technology is widespread, its impacts are better understood, and its trajectory is largely fixed. This dynamic creates a "governance tipping point"—a quantum governance point of no return—and an urgent need to establish robust governance before quantum-AI systems become uncontrollable or cause widespread disruption, such as the cryptographic breach of "Q-Day."[50]

---

[41] *See* Ten Holter C, Inglesant P, Srivastava R and Jirotka M 2022 Bridging the quantum divides: a chance to repair classic(al) mistakes? Quantum Sci. Technol. 7 044006, https://iopscience.iop.org/article/10.1088/2058-9565/ac8db6; and K. Sabeel Rahman, The New Utilities: Private Power, Democratic Control, and the Future of the American State, 49 Colum. Hum. Rts. L. Rev. 1 (2017). *See generally* John Rawls, A Theory of Justice (1971).

[42] *See* also Gary Marchant et al., Learning From Emerging Technology Governance for Guiding Quantum Technology (August 09, 2024). Arizona State University Sandra Day O'Connor College of Law Paper No. 4923230, UNSW Law Research Paper No. 24-33, Available at SSRN: https://ssrn.com/abstract=4923230

[43] *See* GESDA OQI Intelligence Report on Quantum Diplomacy for the Sustainable Development Goals (SDGs) 2 (2d ed. 2024), https://open-quantum-institute.cern/wp-content/uploads/2025/03/GESDA_OQI_Intelligence-Report-2024_Final.pdf

[44] Amelia Nierenberg, *Denmark, Testing E.U.'s Limits, Moves to Copyright A.I.-Generated Deepfakes*, N.Y. Times (July 10, 2025), https://www.nytimes.com/2025/07/10/world/europe/denmark-deepfake-copyright-ai-law.html.

[45] *See* e.g., Kop, M*., Regulating Transformative Technology in The Quantum Age: Intellectual Property, Standardization & Sustainable Innovation* (October 7, 2020). Stanford - Vienna Transatlantic Technology Law Forum, Transatlantic Antitrust and IPR Developments, Stanford University, Issue No. 2/2020, https://purl.stanford.edu/ng658sy7588

[46] *See* Kop, M., *Quantum Technology Impact Assessment,* Futurium, European Commission, (Apr. 20, 2024), https://futurium.ec.europa.eu/en/european-ai-alliance/best-practices/quantum-technology-impact-assessment.

[47] *See* e.g., Martin, N., Matt, C., Niebel, C. *et al. How Data Protection Regulation Affects Startup Innovation.* Inf Syst Front 21, 1307–1324 (2019). https://doi.org/10.1007/s10796-019-09974-2; and Regulation (EU) 2016/679 of the European Parliament and of the Council of 27 April 2016 on the protection of natural persons with regard to the processing of personal data and on the free movement of such data (General Data Protection Regulation), 2016 O.J. (L 119) 1.

This metaphorical event horizon reveals several imperatives for the EU Quantum Act. It highlights the necessity of an agile, adaptive, and anticipatory governance architecture capable of responding to unexpected developments.[51] It also serves as a stark warning against the dangers of technological lock-in and path dependency, as delaying ethical, legal, and societal considerations risks allowing harmful or inequitable applications to become entrenched. This underscores the importance of early intervention, as the current nascent stage of quantum technology offers the most significant window of opportunity to embed shared values, responsible innovation principles, and appropriate standards into its very fabric.

Furthermore, the metaphor implicitly warns that the first actor to achieve breakthrough quantum capabilities might gain a dominant position, creating a point of no return in global technological leadership. This highlights the strategic need for the EU Quantum Act to foster a vibrant pan-European quantum ecosystem, create first-mover advantages, and actively shape global technical standards to avoid a future dominated by misaligned technological norms, such as techno-authoritarianism.[52] Finally, drawing on the Copenhagen interpretation, the metaphor subtly reminds us that the quantum realm's counterintuitive nature may limit the applicability of governance models extrapolated from previous technologies, demanding novel and tailored approaches.[53] The qualitative and quantitative exceptionality of quantum suggests that simply extrapolating from past technological developments will not be sufficient.[54]

---

[48] *See* Kop, M., *Quantum Event Horizon: Addressing the Quantum-AI Control Problem Through Quantum-Resistant Constitutional AI*, Futurium, European Commission (June 11, 2025), https://futurium.ec.europa.eu/en/european-ai-alliance/blog/quantum-event-horizon-addressing-quantum-ai-control-problem-through-quantum-resistant-constitutional
[49] *See* De Wolf, R. (2017). The potential impact of quantum computers on society. Ethics and Information Technology, 19, 271–276.
[50] For further reading about Q-Day, *see e.g.*, Ian Monroe, *"Q Day" Is Coming: Is the World Prepared?*, Ctr. for Int'l Governance Innovation (Nov. 7, 2024), https://www.cigionline.org/articles/q-day-is-coming-is-the-world-prepared/.
[51] *See* Perrier, E., The Quantum Governance Stack: Models of Governance for Quantum Information Technologies, Digital Society, Quantum-ELSPI TC, 1, 22, Springer Nature, (2022), https://link.springer.com/article/10.1007/s44206-022-00019-x
[52] *See* e.g., Paul Scharre, Ctr. for a New Am. Sec., *The Dangers of the Global Spread of China's Digital Authoritarianism* (2023) (testimony before the U.S.-China Economic and Security Review Commission), https://www.cnas.org/publications/congressional-testimony/the-dangers-of-the-global-spread-of-chinas-digital-authoritarianism.

As such, the EU Quantum Act represents a vital effort to navigate this uncertainty proactively and steer quantum development towards beneficial outcomes before the event horizon is crossed. This challenge is particularly acute at the intersection of quantum and AI, where the prospect of bad actors repurposing these powerful agents—or the agents themselves posing an unmanageable control problem—creates a governance challenge that traditional oversight mechanisms cannot solve. The solution may lie in novel paradigms designed to address the risk of technology progressing too fast for classical guardrails, like Quantum-Resistant Constitutional AI, where ethical principles are hardwired into the system's architecture in a way that is secure against both classical and quantum circumvention.[55] This includes 'algorithmic regulation' mechanisms. This concept could be legally operationalized through a novel fiduciary framework, where advanced AI systems are designated as 'quantum-agentic stewards' – as metaphorical 'quantum guardians' with a legally enforceable duty to ensure quantum operations align with constitutional principles and fundamental rights. These systems would be governed by a formal constitution (Constitutional AI) and secured with quantum-resistant cryptography, creating a new framework for technological stewardship. Such a fiduciary arrangement would shift the paradigm from external human oversight, which is unfeasible at quantum speeds, to embedded, automated, and legally accountable supervision. As quantum computing presents a level of technological innovation threatening to outpace human know-how, fiduciary law offers a body of theory and practice directed towards acting on behalf of others to relieve the techno-social burden on citizens, a burden which engenders leaps of faith.[56] This aligns with the broader need for any quantum governance model to systematically identify stakeholder rights, interests, and duties and the appropriate instruments by which such impacts are managed.[57]

---

[53] However, while the 'event horizon' metaphor effectively captures the profound uncertainty and the risk of path dependency, it must be tempered by the historical reality of technological diffusion. As some commentators contend, the belief in a single, imminent 'tipping point'—a moment of sudden, irreversible transformation—largely flies in the face of technological history. With rare exceptions, such as the atomic bomb, progress has been incremental, with technologies like the internet and smartphones taking decades to reshape society through gradual, widespread adoption. Therefore, the EU Quantum Act should be designed to govern not a single, cataclysmic event, but a continuous and uneven process of diffusion. This perspective reinforces the need for an adaptive, modular framework that can manage generations of less powerful, yet still impactful, quantum technologies as they integrate into society. *See also* Schmidt & Xu, *supra* note 22.
[54] *See* also Der Derian, J., & Wendt, A. (2020). 'Quantizing international relations': The case for quantum approaches to international theory and security practice. Security Dialogue, 51(5), 399–413. https://doi.org/10.1177/0967010620901905.
[55] Kop, *supra* note 47 (event horizon)
[56] Robert Herian, *As Much as Faith: A Speculation on Quantum Computing with Fiduciary Law in Public Governance*, in Public Governance and Emerging Technologies: Values, Trust, and Regulatory Compliance 217, 218 (Jurgen Goossens, Esther Keymolen & Antonia Stanojević eds., 2025).
[57] *See generally* Perrier, *supra* note 50 (Stack) (developing an actor-instrument model for quantum governance based on stakeholder rights, interests, and obligations).

This section has laid the groundwork for the proposed EU Quantum Act by outlining the transformative yet complex nature of quantum technologies and establishing the Act's fundamental rationale, scope, and objectives. The framework proposed is dual-pronged, designed to simultaneously foster innovation through strategic investment while imposing clear regulatory guardrails based on risk. It is also modular, designed to be adaptive across the technology's lifecycle with components—such as risk-based tiers, guiding principles, and regulatory sandboxes—that can be adjusted as quantum systems evolve. The following sections will build upon this foundation by examining relevant precedents and detailing specific components of the proposed framework.

## 2. BENCHMARKING SEMICONDUCTOR STRATEGIES: LESSONS FROM THE EU AND US CHIPS ACTS

This section undertakes a comparative analysis of the European Union's and the United States' respective Chips Acts, focusing on their strategic objectives, funding structures, innovation promotion mechanisms, and plans to ensure supply chain security within the semiconductor sector. The primary aim is to distill relevant insights and models that can inform the development of effective strategies for fostering innovation and managing dependencies within the EU Quantum Act. By examining these large-scale industrial policies, we can identify both successful models and potential challenges applicable to the nascent quantum technology landscape.

### 2.1. Comparative Analysis of Objectives, Funding Mechanisms, and Innovation Strategies

### 2.1.1. EU Chips Act

The European Union's strategy is multifaceted, aiming to bolster EU competitiveness and resilience in semiconductors, support the digital and green transitions, and strengthen overall technological leadership.[58] A central goal is to reduce global dependency by aiming for a 20% global market share by 2030, thereby achieving a form of technological sovereignty built on international interdependency.[59] This particular focus on "tech sovereignty" is highly relevant for shaping the parallel goal of quantum strategic autonomy.

To achieve these objectives, the EU plans to mobilize over €43 billion in public and private investment by 2030 through a three-pillar structure. The first pillar, the "Chips for Europe Initiative," leverages €11.15 billion in public funds to attract private investment for R&I

---

[58] *See* European Commission, European Chips Act - Questions and Answers (Sept. 21, 2023), https://ec.europa.eu/commission/presscorner/detail/en/qanda_23_4519; PubAffairs Bruxelles, European Chips Act enters into force today - Questions and answers (Sept. 21, 2023), https://www.pubaffairsbruxelles.eu/eu-institution-news/european-chips-act-enters-into-force-today-questions-and-answers/; and Regulation (EU) 2023/1781 of the European Parliament and of the Council of 13 September 2023 establishing a framework of measures for strengthening Europe's semiconductor ecosystem (EU Chips Act), 2023 O.J. (L 229) 1.
[59] *See* e.g., International Cooperation On Semiconductors (ICOS), Strategy EU CHIPS ACT, https://icos-semiconductors.eu/strategy-eu-chips-act/.

leadership.⁶⁰ The second pillar establishes a security of supply framework to attract investment in production capacity, while the third creates a coordination mechanism through the European Semiconductor Board to respond to disruptions.⁶¹ This structure also includes a €2 billion "Chips Fund" specifically for startups and SMEs. Notably, the Chips Act allows state aid for "first-of-a-kind" facilities, though these approval processes can be slow, and much of the funding represents a redirection of existing funds rather than new capital.⁶² The EU's innovation strategy adopts a comprehensive ecosystem approach, seeking to strengthen R&T leadership in smaller, faster chips; build capacity in design, manufacturing, and packaging; provide access to design tools and pilot lines; and support startups and SMEs with equity finance, all while addressing the quantum skills shortage and seeking to better understand global supply chains.

### 2.1.2. US CHIPS Act

In contrast, the US CHIPS Act is driven by more explicit geopolitical motivations. Its primary objectives are to strengthen US supply chain resilience, counter China's technological ascent, boost domestic research and manufacturing, and fortify national security by reducing reliance on foreign semiconductor production.⁶³

The funding mechanism reflects this urgency, authorizing approximately $280 billion, with $52.7 billion directly appropriated. This "new money" includes $39 billion in manufacturing subsidies, a 25% investment tax credit, and $13 billion for R&D and workforce training.⁶⁴ Further allocations include $2 billion for Department of Defense microelectronics R&D and fabrication, and $500 million for the State Department to coordinate international supply chain security.⁶⁵ The Act also offers loan guarantees, demonstrating a robust use of direct funding and tax credits. The American innovation strategy emphasizes a direct path from "lab-to-fab," incorporating the Endless Frontier Act to boost domestic high-tech research.⁶⁶ This is operationalized through the CHIPS R&D Office, which oversees programs like the National Semiconductor Technology Center (NSTC) and the National Advanced Packaging

---

⁶⁰ European Comm'n, *European Chips Act: The Chips for Europe Initiative* (Nov. 4, 2024), https://digital-strategy.ec.europa.eu/en/factpages/european-chips-act-chips-europe-initiative.
⁶¹ See: European Comm'n, *European Chips Act: Security of supply and resilience* (Nov. 4, 2024), https://digital-strategy.ec.europa.eu/en/factpages/european-chips-act-security-supply-and-resilience (Pillar II); and European Comm'n, *European Chips Act: Monitoring and crisis response* (Mar. 28, 2025), https://digital-strategy.ec.europa.eu/en/factpages/european-chips-act-monitoring-and-crisis-response (Pillar III).
⁶² *See* e.g., U.S. Semiconductor Legislation and Policies, European Chips Act, https://www.european-chips-act.com/USA_Semiconductor_Legislation.html; and
⁶³ *See* CHIPS and Science Act of 2022, Pub. L. No. 117-167, 136 Stat. 1366, *supra* note 23.
⁶⁴ *See* e.g., Nazak Nikakhtar et al., A World of Chips Acts: The Future of U.S.-EU Semiconductor Collaboration, CSIS (2025), https://www.csis.org/analysis/world-chips-acts-future-us-eu-semiconductor-collaboration.
⁶⁵ Nat'l Inst. of Standards & Tech., CHIPS for America Fact Sheet: Federal Incentives (2025), https://www.nist.gov/document/chips-america-fact-sheet-federal-incentives.
⁶⁶ *See* Endless Frontier Act, H.R. 2731, 117th Cong. (2021), https://www.congress.gov/bill/117th-congress/house-bill/2731.

Manufacturing Program (NAPMP) to facilitate technology transfer, support workforce training, and deepen the understanding of the ecosystem.[67]

### 2.1.3. Approaches to Supply Chain Security in the Semiconductor Industry

The two blocks also reveal different philosophies in their approaches to supply chain security. The EU Chips Act aims to ensure security of supply and resilience primarily by reducing dependency and building international partnerships with like-minded countries, a strategy described as "international interdependency". Coordination through the European Semiconductor Board is a central component of this partnership-centric strategy.[68]

The US, on the other hand, has adopted a more assertive and geopolitical posture. It seeks to bolster the domestic supply chain by reducing foreign reliance and implementing restrictions on investment in "foreign countries of concern."[69] A key tool is the "Expansion Clawback," which allows the government to reclaim funds if recipient firms expand advanced manufacturing capacity in such countries.[70] In addition, the Act establishes the International Technology Security and Innovation (ITSI) Fund, a $500 million vehicle for coordinating with international partners to secure and diversify semiconductor supply chains, with a focus on the Indo-Pacific and the Americas.[71] This combination of domestic incentives and foreign restrictions marks a more confrontational stance.

### 2.1.4. U.S. Quantum-Specific Legislative Efforts

Beyond the broad industrial strategy of the CHIPS Act, the U.S. has advanced several quantum-specific bills that reveal a distinct innovation philosophy centered on speed, market-driven solutions, and national security.[72] These efforts, often characterized by a "permissionless innovation" model, contrast with the EU's more precautionary and values-based orientation. One legislative effort, the Quantum Sandbox for Near-Term Applications Act of 2025, aims to establish public-private partnerships to fast-track the development of quantum and quantum-hybrid applications.[73] By creating a "Quantum Sandbox," the bill seeks to break down barriers to accessing quantum hardware and foster innovative solutions for public and private sector challenges, reflecting a strong emphasis on rapid prototyping and deployment.

---

[67] *See* e.g., Nat'l Inst. of Standards & Tech., *National Advanced Packaging Manufacturing Program*, NIST, https://www.nist.gov/chips/research-development-programs/national-advanced-packaging-manufacturing-program.

[68] European Commission, *European Semiconductor Board (ESB)*, E.C., https://ec.europa.eu/transparency/expert-groups-register/screen/expert-groups/consult?lang=en&groupID=3932.

[69] S*ee* Exec. Order No. 14,105, Addressing United States Investments in Certain National Security Technologies and Products in Countries of Concern, 88 Fed. Reg. 54,867 (Aug. 14, 2023); and Provisions Regarding Access to Americans' Bulk Sensitive Personal Data and Government-Related Data by Countries of Concern, 89 Fed. Reg. 15,575 (proposed Jan. 8, 2025) (to be codified at 28 C.F.R. pt. 202.

[70] *See* e.g., Aidan Arasasingham & Gregory C. Allen, Sourcing Requirements and U.S. Technological Competitiveness, CSIS (2025), https://www.csis.org/analysis/sourcing-requirements-and-us-technological-competitiveness.

[71] U.S. Dep't of State, The U.S. Department of State International Technology Security and Innovation Fund, https://www.state.gov/the-u-s-department-of-state-international-technology-security-and-innovation-fund/.

Other bills focus on strengthening domestic manufacturing and supply chains. The Advancing Quantum Manufacturing Act of 2025 and the Support for Quantum Supply Chains Act exemplify the U.S. focus on building a resilient domestic ecosystem.[74] The former mandates the establishment of a "Quantum Manufacturing USA institute" to create an end-to-end manufacturing capability for quantum components and systems, while the latter directs the National Institute of Standards and Technology (NIST) to expand partnerships to accelerate the domestic quantum supply chain.[75] These initiatives are squarely aimed at enhancing U.S. competitiveness and reducing reliance on foreign, particularly Chinese, technology and materials. To ensure strategic oversight, the Advancing Quantum Manufacturing Act also seeks to improve coordination between funding agencies like the Department of Energy (DOE) and the National Science Foundation (NSF) to prevent duplication and ensure research activities cover all essential quantum-enabling technologies, highlighting a whole-of-government strategy to maintaining a U.S. lead in the field.[76]

Complementing these initiatives are bills aimed at reinforcing the foundational research and strategic leadership that underpin the entire U.S. quantum enterprise. Prime examples include the National Quantum Initiative Reauthorization Act and The Department of Energy Quantum Leadership Act of 2025.[77] The former seeks to continue and expand the original National Quantum Initiative, which coordinates quantum R&D across federal agencies, ensuring sustained, long-term investment in fundamental science and workforce development.[78] The latter specifically empowers the Department of Energy to accelerate the translation of basic research into practical applications, directing it to establish and upgrade user facilities and research centers that are indispensable for the entire ecosystem. Together, these acts demonstrate a comprehensive strategy that not only targets near-term applications and supply chain resilience but also fortifies the long-term scientific leadership elemental for sustained progress.

---

[72] *See* National Quantum Initiative Act, Pub. L. No. 115-368, 132 Stat. 5092 (2018); Exec. Order No. 13,885, Establishing the National Quantum Initiative Advisory Committee, 84 Fed. Reg. 45,619 (Aug. 30, 2019); and Exec. Order No. 14,073, Enhancing the National Quantum Initiative Advisory Committee, 87 Fed. Reg. 27,679 (May 9, 2022).
[73] Quantum Sandbox for Near-Term Applications Act of 2025, H.R. 3220, 119th Cong. (2025), https://www.congress.gov/bill/119th-congress/house-bill/3220.
[74] Advancing Quantum Manufacturing Act of 2025, S. 1343, 119th Cong. (2025), https://www.congress.gov/bill/119th-congress/senate-bill/1343.
[75] Support for Quantum Supply Chains Act, H.R. 3788, 119th Cong. (2025), https://www.congress.gov/bill/119th-congress/house-bill/3788.
[76] S. 1343, 119th Cong. § 2 (2025), *supra* note 73.
[77] Committee on Science, Space & Technology, H.R. 6213 - The National Quantum Initiative Reauthorization Act (November 23, 2023), https://science.house.gov/2023/11/the-national-quantum-initiative-reauthorization-act; S.579 - Department of Energy Quantum Leadership Act of 2025 118th Cong. (2025) (proposing over $2.5 billion in funding for quantum research and development under the Department of Energy), https://www.congress.gov/bill/119th-congress/senate-bill/579/text.
[78] H.R.6227 - National Quantum Initiative Act, 115th Congress (2017-2018), June 26, 2018, https://www.congress.gov/bill/115th-congress/house-bill/6227 or National Quantum Initiative Act, Pub. L. No. 115-368, 132 Stat. 5092 (2018).

### 2.1.5. EU Quantum Cybersecurity Preparedness Initiatives

While the EU does not have a single, consolidated piece of legislation with a title equivalent to the U.S. Quantum Computing Cybersecurity Preparedness Act,[79] it is actively pursuing "Q-Day" preparedness through a variety of strategic initiatives and agency mandates. The European approach is more distributed and integrated into broader cybersecurity and technology strategies. Key components include the European Quantum Communication Infrastructure (EuroQCI) Initiative,[80] a major project to build a secure, pan-European quantum communication network to safeguard vital infrastructure.[81] Alongside this, the European Union Agency for Cybersecurity (ENISA)[82] is actively working on the transition to Post-Quantum Cryptography (PQC), providing guidance and analysis to EU institutions and Member States on migrating their cryptographic systems to be secure against quantum computers. These efforts are situated within the overarching policy framework of the Quantum Europe Strategy, which includes pillars dedicated to infrastructure and dual-use security. While the US has a specific Act mandating a plan for migration, the EU is pursuing the same goal through a combination of building new secure infrastructure (EuroQCI) and guiding the migration of existing systems (via ENISA). A dedicated EU Quantum Act would be a logical next step to harmonize and codify these distributed efforts into a single, coherent legislative instrument.

### 2.2. Applicability to the Quantum Technology Landscape

These precedents offer lessons for the EU Quantum Act. For funding, the EU QA could adopt a multi-layered strategy that combines existing mechanisms like Horizon Europe with a dedicated "Quantum Fund". However, it should also consider incorporating significant direct public investment, tax incentives, and loan guarantees inspired by the US model to support capital-intensive quantum infrastructure, alongside streamlined state aid processes. The framework should explicitly permit and encourage funding for dual-use quantum R&D.[83]

---

[79] Quantum Computing Cybersecurity Preparedness Act, Pub. L. No. 117-260, 136 Stat. 2389 (2022), https://www.govinfo.gov/app/details/PLAW-117publ260.
[80] European Comm'n, *The European Quantum Communication Infrastructure (EuroQCI) Initiative*, https://digital-strategy.ec.europa.eu/en/policies/european-quantum-communication-infrastructure-euroqci.
[81] *ibid*.
[82] European Union Agency for Cybersecurity, https://www.enisa.europa.eu/.
[83] See also Thomas Brent, EIC Will Invest in Dual-Use Start-ups, Commission Says, Sci.Business (2025), https://sciencebusiness.net/news/european-innovation-council/eic-will-invest-dual-use-start-ups-commission-says.

In terms of innovation, the EU should aim for a comprehensive ecosystem that supports research, provides infrastructure access through testbeds and networks, and nurtures startups, with a strong emphasis on "lab-to-market" translation. The U.S. model of a "Quantum Sandbox" offers a compelling mechanism for accelerating this transition by providing a controlled environment for near-term application development.[84] Addressing the quantum skills gap will also be paramount.

Regarding supply chain security, the EU should combine its partnership and diversification plan with a more strategic management of critical mineral dependencies. This must include a sophisticated, data-driven methodology for assessing geopolitical risks to the supply of Critical Raw Materials (CRMs).[85] For instance, a Quantum Criticality Index (QCI) could be developed to identify and prioritize specific chokepoints—such as the supply of niobium, indium, or rare earths like ytterbium, which are often processed in China—enabling targeted strategies for stockpiling, diversification, or domestic development.[86] The Act should also implement targeted export controls inspired by US national security guardrails and the EU's own economic security focus, carefully balancing collaboration with the need to safeguard strategic advantages.[87] The targeted U.S. legislative efforts to bolster domestic quantum manufacturing and supply chains underscore the urgency for the EU to develop its own robust industrial capacity to ensure strategic autonomy.[88] As detailed in Section 11, this focus on industrial capacity is a core component of the broader U.S. strategy to win the global technology race, providing a clear benchmark for the EU's own ambitions for "full-stack quantum sovereignty."

Furthermore, the EU should explore innovative public financing mechanisms inspired by emerging U.S. industrial policy, where the government is considering taking non-voting equity stakes in semiconductor firms in exchange for US CHIPS Act grants.[89] An analogous 'equity-for-funding' model within the EU Quantum Act could serve a dual purpose: it would ensure taxpayers share in the financial upside of successful ventures, and more strategically, it would create a powerful incentive for recipient companies to adhere to the Act's long-term principles of responsible innovation, ethical conduct, and societal benefit.

---

[84] Quantum Sandbox for Near-Term Applications Act of 2025, H.R. 3220, 119th Cong. (2025), *supra* note 71.
[85] *See* e.g., Mans, U., Rabbie, J. & Hopman, B. Critical Raw Materials for Quantum Technologies, Quantum Delta NL White Paper (2023); Regulation (EU) 2024/1252 of the European Parliament and of the Council of 11 April 2024 establishing a framework for ensuring a secure and sustainable supply of critical raw materials (European Critical Raw Materials Act), 2024 O.J. (L 1252) 1; and U.S. Dep't of Energy, Critical Materials Strategy (2021).
[86] *See* Min-Ha Lee, Andrew J. Grotto & Mauritz Kop, Methodology for Assessing Geopolitical Risk to Critical Raw Materials Supply Chains and Its Application to Quantum Computing (2025) (working paper, Stanford Center for International Security and Cooperation), https://cisac.fsi.stanford.edu/publications; and Das, S., Chatterjee, A., & Ghosh, S. (2023). A First Order Survey of Quantum Supply Dynamics and Threat Landscapes. arXiv. https://arxiv.org/abs/2308.09772.
[87] *See* e.g., Philip Luck, *The Hidden Risk in Rising U.S.-PRC Tensions: Export Control Symbiosis*, CSIS (Apr. 2025), https://www.csis.org/analysis/hidden-risk-rising-us-prc-tensions-export-control-symbiosis.
[88] S. 1343, 119th Cong. § 2 (2025), *supra* note 73.
[89] *See* e.g., *See* Andrea Shalal, David Shepardson, Nandita Bose and Max A. Cherney, *US examines equity stake in chip makers for CHIPS Act cash grants, sources say*, Reuters (Aug. 19, 2025), https://www.reuters.com/business/media-telecom/us-examines-equity-stake-chip-makers-chips-act-cash-grants-sources-say-2025-08-19/.

This section's comparative analysis of the EU and US Chips Acts and recent quantum-specific legislation reveals valuable precedents for the EU Quantum Act regarding funding structures, innovation ecosystem support, and supply chain security. Key lessons include the prospective benefits of multi-layered funding incorporating direct investment and incentives, the importance of fostering the entire innovation cycle from lab to market, and the need for strategic methodologies to supply chain resilience that balance international partnerships with targeted controls. These insights provide a foundation for designing effective industrial policy components within the EU QA.

## 3. NAVIGATING THE AI GOVERNANCE LANDSCAPE: INSIGHTS FROM THE EU AND THE UK

This chapter explores existing and proposed models for governing Artificial Intelligence (AI) in the European Union and the United Kingdom, recognizing AI as a relevant precedent for regulating powerful, general-purpose technologies. It examines the EU AI Act's comprehensive risk-based framework, the principles of the proposed AI Liability Directive that complemented it, and the UK's contrasting principles-based, pro-innovation strategy. The objective is to synthesize best practices and identify applicable regulatory mechanisms and philosophies that can inform the development of a balanced and effective governance framework for quantum technologies and quantum-AI hybrids within the EU Quantum Act.

### 3.1. The EU AI Act's Risk-Based Product Safety Regime and Reinforcement Mechanisms

The cornerstone of the EU AI Act is its risk-based arrangement, which classifies AI systems into unacceptable, high, limited, and minimal risk levels.[90] This tiered logic strategically focuses regulatory burdens on high-risk systems, defined as those posing serious risks to health, safety, or fundamental rights. This model is highly adaptable for the quantum domain, where applications could be similarly categorized based on societal impact, risk profile, and dual-use potential, using the AI Act's criteria for sectors like critical infrastructure and healthcare as a direct inspiration.[91] As comparative studies by scholars like Gasser & Halim affirm, this horizontal, risk-based model is useful for emerging technologies, though it requires specific adaptations to address quantum's unique characteristics, such as its profound dual-use nature and unpredictable development trajectory.[92]

---

[90] *See* e.g., Autoriteit Persoonsgegevens, EU AI Act risk groups, https://www.autoriteitpersoonsgegevens.nl/en/themes/algorithms-ai/eu-ai-act/eu-ai-act-risk-groups.
[91] European Commission, AI Act, Shaping Europe's Digital Future, https://digital-strategy.ec.europa.eu/en/policies/regulatory-framework-ai, *supra* note 36.

Following this logic, the EU AI Act establishes a list of prohibited practices, banning AI applications that violate fundamental rights and European values, such as social scoring, cognitive manipulation, the exploitation of vulnerabilities, and certain uses of biometric and emotion recognition. The principle of prohibiting the exploitation of vulnerabilities is particularly vital.[93] An EU Quantum Act could establish a similar list of unacceptable quantum risks, such as the capability to break vital public-key encryption or enable rights-infringing surveillance.

For systems identified as high-risk, the EU AI Act imposes stringent requirements that must be met before market placement. These include comprehensive risk assessment and mitigation, high-quality datasets, activity logging, detailed technical documentation, clear user information, robust human oversight, and high standards of cybersecurity, accuracy, and robustness. Furthermore, such arrangements require conformity assessments and registration in a central EU database, with mandated post-market monitoring to ensure ongoing compliance.[94] These obligations serve as a direct template for governing high-risk quantum systems, ensuring responsible development through rigorous standards and transparent oversight. The concepts of a central database, along with requirements for accuracy, robustness, and cybersecurity, are especially relevant for the quantum context.[95]

To ensure user awareness, the Act also includes transparency obligations, which mandate that humans are informed when interacting with AI systems like chatbots and require the clear labeling of AI-generated content. This emphasis on transparency, encompassing traceability and explainability, is essential for building trust in quantum-AI hybrids.[96] Where feasible, explainability is a principal component to ensuring accountability.[97]

Enforcement is managed through a dual structure that combines national market surveillance authorities with a centralized EU AI Office, which holds specific responsibility for General-

---

[92] *See* Halim, Noha Lea and Gasser, Urs, Vectors of AI Governance - Juxtaposing the U.S. Algorithmic Accountability Act of 2022 with The EU Artificial Intelligence Act (May 9, 2023), https://ssrn.com/abstract=4476167; and Gasser, U., and Almeida V., A layered model for AI governance. In IEEE Internet Computing, vol. 21, no. 6, pp. 58-62, (November/December 2017), doi: 10.1109/MIC.2017.4180835.
[93] Regulation (EU) 2024/1689 of the European Parliament and of the Council of 13 June 2024 (EU AI Act), 2024 O.J. (L 1689), *supra* note 24.
[94] *ibid*
[95] *See* Waldherr, A., Cohen, I. G., & Kop, M., *Quantum Trials: An FDA for Quantum Technology*, 1 Stanford Center for RQT Research Series, Stanford Law School (2024). https://purl.stanford.edu/ky321vv7240.
[96] *See* e.g., European Parliament, *EU AI Act: first regulation on artificial intelligence* (June 8, 2023), https://www.europarl.europa.eu/topics/en/article/20230601STO93804/eu-ai-act-first-regulation-on-artificial-intelligence.
[97] *See* Frank Pasquale, The Black Box Society: The Secret Algorithms That Control Money and Information (2015).

Purpose AI (GPAI) models.[98] Penalties are significant and tiered according to the nature of the breach, ranging up to €35 million or 7% of global annual turnover for prohibited practices.[99] An EU Quantum Act will similarly need clear enforcement bodies, such as the proposed OQTA, and a schedule of deterrent penalties. To avoid disproportionate regulatory burdens, the Act should also consider compliance exceptions for SMEs. The EU AI Act and GDPR have demonstrated that complex legal standards can impose significant costs and administrative burdens on smaller firms, stifling innovation.[100] The use of AI-driven tools to automate the creation of a Quantum Technology Quality Management System (QT-QMS) offers one possible model for streamlining compliance.[101]

Finally, it is important to learn from the application of the New Legislative Framework (NLF) model in the AI Act.[102] While the NLF's emphasis on harmonized standards and conformity assessment is a key strength, critiques indicate that its reliance on maximum harmonization could unduly restrict Member States' ability to legislate on specific societal impacts and that its enforcement mechanisms may lack robust avenues for public complaint and redress. Therefore, the EU Quantum Act must be carefully drafted to provide strong protections while allowing for necessary national flexibility.[103]

### 3.2. The Proposed EU AI Liability Directive and its Implications for Emerging Technologies

While the legislative proposal for a dedicated EU AI Liability Directive was not adopted in its original form,[104] its core principles offer important lessons for establishing liability regimes for emerging technologies, and the principles from the proposal were instead partially integrated into the main EU AI Act. Originally intended to complement the AI Act, the directive aimed to harmonize non-contractual liability rules for damage caused by AI. It introduced two mechanisms: a disclosure obligation, allowing claimants to request information from providers, and a rebuttable presumption of causality, which would have presumed a breach of duty caused the damage if a provider failed in its duty of care.[105]

---

[98] *See* European Comm'n, *European AI Office*, Shaping Europe's Digital Future, https://digital-strategy.ec.europa.eu/en/policies/ai-office; and European Comm'n, *Commission Decision of 24.1.2024 establishing the European AI Office*, C(2024) 501 final (Jan. 24, 2024), https://digital-strategy.ec.europa.eu/en/library/commission-decision-establishing-european-ai-office.
[99] *See* e.g., Mauritz Kop, *The Daiki EU AI Act Compliance Solution: Navigating Mandatory AI Governance for Global Enterprises*, Daiki (June 23, 2025), https://dai.ki/blog/the-daiki-eu-ai-act-compliance-solution-navigating-mandatory-ai-governance-for-global-enterprises/.
[100] *ibid*
[101] Aboy *et al., supra* note 34. *See also*: ISO 9001 Quality management systems, https://www.iso.org/standards/popular/iso-9000-family and https://www.iso.org/standard/62085.html
[102] European Commission, The 'New Legislative Framework', https://single-market-economy.ec.europa.eu/single-market/goods/new-legislative-framework_en.
[103] See, e.g., Michael Veale & Frederik Zuiderveen Borgesius, Demystifying the Draft EU Artificial Intelligence Act, 22 Computer L. Rev. Int'l 1 (2021), https://doi.org/10.9785/cri-2021-220402.
[104] *See* e.g., Caitlin Andrews, "European Commission withdraws AI Liability Directive from consideration," IAPP News, 16 July 2024, https://iapp.org/news/a/european-commission-withdraws-ai-liability-directive-from-consideration.

The proposed legislation also sought to expand the definition of a "product" to include software and AI, broaden the scope of liability to include manufacturers and suppliers, and update the concept of a "defect" to account for self-learning systems and cybersecurity vulnerabilities.[106] These principles—particularly the shifts in the burden of proof, access to information, and expanded definitions of liability and defect—remain directly relevant for establishing a quantum liability regime, especially for high-risk applications and complex, unpredictable systems. The emerging question of legal personhood or agenthood for advanced quantum-AI systems is also relevant here, as it will impact how liability is allocated based on societal consensus and frameworks for responsible technology.[107]

### 3.3. The Principles-Based UK Pro-Innovation AI Regulation

In contrast to the EU's detailed legislative style, the UK has adopted a non-statutory, cross-sectoral framework based on five core principles: safety, security, and robustness; transparency and explainability; fairness; accountability and governance; and contestability and redress.[108] This model empowers existing regulators to apply these principles within their specific domains, thereby leveraging deep sectoral expertise.[109] The UK's strategy emphasizes flexibility, adaptability, and an initial reticence for bespoke rules, though it anticipates the future need for binding requirements for the most advanced systems. A central function within the Department for Science, Innovation and Technology (DSIT) supports coordination across sectors.[110] This flexible, principles-based model offers a valuable counterpoint. The EU Quantum Act could incorporate a similar set of high-level principles alongside specific rules, allowing for adaptation and leveraging sectoral regulators for implementation. A phased plan, establishing principles first before codifying specific rules, could also be considered.

### 3.4. Synthesizing Best Practices for Regulating Quantum-AI Hybrids and other Quantum Applications

In the final analysis, a hybrid, modular regulatory structure seems most appropriate for the EU Quantum Act. This strategy would attempt to navigate the trade-offs between different regulatory styles, as highlighted by Mökander & Floridi, by seeking sufficient specificity for legal certainty while retaining the flexibility needed for rapid technological change.[111] It would also guard against the risk of vagueness leading to ineffective implementation. The challenge for regulators is to navigate the four primary models of innovation governance—the

---

[105] European Comm'n, *Liability rules for artificial intelligence*, https://commission.europa.eu/business-economy-euro/doing-business-eu/contract-rules/digital-contracts/liability-rules-artificial-intelligence_en.
[106] *ibid.*
[107] Kop, *supra* note 47. (event horizon)
[108] Dep't for Sci., Innovation & Tech. (2025), *supra* note 25.
[109] Dep't for Sci., Innovation & Tech. (2023), *supra* note 26.
[110] *ibid.*
[111] Mökander & Floridi et al., *supra* note 12

Precautionary Principle, Responsible Innovation (RI), Permissionless Innovation (PI), and the Innovation Principle—noting the common divergence where national strategies often default to a PI model focused on economic growth, while experts increasingly call for an RI approach that proactively integrates ethical and security concerns.[112] Achieving the right balance is paramount, especially when contrasted with China's explicit "accelerationist" policy for AI, which operates on the principle that "a lack of development would be the biggest safety risk," thereby prioritizing speed over precautionary governance. Navigating this trade-off between fostering innovation and implementing necessary safeguards—a central challenge in AI governance—is critical for the EU QA's design, especially given the profound dual-use risks of quantum technologies.[113]

To achieve this, the EU Quantum Act should:

1. Adopt a risk-based classification system, similar to the AI Act, but tailored specifically to quantum risks, including dual-use prospects.
2. Establish a clear set of prohibitions for applications that pose unacceptable risks.
3. Incorporate a set of overarching principles, inspired by the UK model and frameworks like Stanford's RQT, to provide flexibility and normative guidance. These could include Safety, Transparency, Fairness, Accountability, Societal Benefit, Dual-Use Mitigation, and Long-term thinking/ Intergenerational equity. Applying such principles early is central to shaping the technology's trajectory before the 'quantum event horizon' limits intervention options.
4. Address the unique risks of quantum-AI hybrids specifically, ensuring coherence with the AI Act. This includes evolving the concept of human oversight from direct "human-in-the-loop" control, which is unfeasible for high-speed QAI systems, to a strategic "human-on-the-loop" model focused on meta-level reviews and goal-setting. Additionally, the regulatory agility needed for fast-evolving quantum-AI hybrids can be achieved by linking compliance to harmonized, certifiable standards, such as a QT-QMS, which can be updated more rapidly than the legislative text itself.[114] This system-level certification, leading to a CE mark for high-risk products, provides a predictable yet adaptable pathway for innovators, a model proven effective in the medical technology field.
5. Emphasize rigorous risk assessment, transparency, and accountability, drawing from both the EU and UK models. This could involve specific instruments like Quantum Impact Assessments (QIAs), inspired by precedents in genetics and AI.
6. Consider a phased implementation, beginning with principles and risk classification before introducing more specific rules as the technology matures.
7. Establish a dedicated EU body, like the proposed OQTA, to provide coordination, expertise, and guidance.

---

[112] *See* Drífa Atladottir et al., *A Quantum of Responsibility? A Comparison of National Quantum Governance Frameworks and Expert Views*, 4 Digit. Soc. 54 (2025), https://doi.org/10.1007/s44206-025-00205-7.
[113] Halim & Gasser, *supra* note 86.
[114] Aboy *et al., supra* note 34.

8. Draw ethical inspiration from genetics governance, adopting concepts such as independent ethics committees (or "Quantum Ethics Boards"), a strong application of the precautionary principle, and robust pillars for equitable access, consent, and accountability.
9. Reject a purely 'permissionless innovation' model. The failures of this approach in the internet era—leading to market monopolization, systemic erosion of privacy, and a persistent 'pacing problem' where law lags behind technology—provide a powerful doctrinal justification for the necessity of proactive, *ex-ante* regulation for quantum technologies.

To summarize, the examination of EU and UK AI governance approaches provides valuable insights for the EU Quantum Act. It highlights the utility of a risk-based framework combined with overarching principles, the importance of addressing liability, and the need for adaptive mechanisms. Synthesizing these elements suggests a path towards a balanced regulatory strategy for quantum technologies that fosters innovation while upholding safety and fundamental rights.

## 4. GLOBAL INNOVATION ECOSYSTEMS: A COMPARATIVE PERSPECTIVE

This section shifts focus to the global landscape, comparing the innovation systems related to advanced technologies in the United States, the European Union, and China. It analyzes their respective strengths and weaknesses, particularly concerning the development of dual-use technologies and the associated funding mechanisms. The goal is to identify successful models and pitfalls that can inform the EU Quantum Act's strategy for fostering a competitive and strategically sound European quantum ecosystem.

### 4.1. Analysis of Innovation Systems in the US, EU, and China concerning Advanced Technologies

The global innovation landscape is dominated by three distinct models. The United States currently leads in AI, boasting superior talent, research, development, and hardware capabilities.[115] Its ecosystem is characterized by a strong focus on research-to-commercialization pathways, fueled by robust private sector and venture capital funding, which results in strong high-tech exports and patent generation.[116] This market-driven, "permissionless innovation" model is further reinforced by recent legislative efforts like the Quantum Sandbox for Near-Term Applications Act, which prioritizes speed and public-private partnerships to accelerate development.[117] However, this leadership position faces a significant challenge from the rapid catch-up of China.[118]

---

[115] *See* e.g., Exec. Off. of the President, Council of Econ. Advisers, *AI Talent Report* (Jan. 14, 2025), https://bidenwhitehouse.archives.gov/cea/written-materials/2025/01/14/ai-talent-report/.
[116] See e.g., Daniel Castro, Michael McLaughlin & Eline Chivot, Who Is Winning the AI Race: China, the EU or the United States?, Ctr. for Data Innovation (Aug. 2019), https://datainnovation.org/2019/08/who-is-winning-the-ai-race-china-the-eu-or-the-united-states/.

In contrast, the European Union, while outperforming China in some innovation metrics, lags the US, particularly in the task of translating its world-class research into industrial and commercial success.[119] The EU champions a values-based, ethical approach, embedding principles like human rights and the rule of law into its regulatory frameworks. This can create structural tension with the sheer velocity of the US and Chinese systems. The recently announced "Quantum Europe Strategy" aims to address this gap by creating a more unified and competitive ecosystem based on five strategic pillars: R&I, Infrastructure, Ecosystem Strengthening, Dual-Use, and Skills.[120]

Meanwhile, China is rapidly closing the AI gap with the US, establishing a lead in technology adoption and data availability.[121] Its system is defined by powerful government support and investment, especially in strategic and dual-use technologies, and a foundational emphasis on civil-military integration.[122] As a leading provider of scientific publications in future technologies, China employs a state-led, top-down strategy, leveraging immense national resources to accelerate progress, often unencumbered by the ethical debates prevalent in democracies.[123] This approach is exemplified by its recent "AI+ Plan," which reflects a deeply techno-optimistic, even "accelerationist," philosophy. The plan is predicated on the idea that "a lack of development would be the biggest safety risk," thereby prioritizing the rapid, society-wide application and integration of AI over precautionary governance. It treats technological diffusion as a whole-of-society campaign, aiming to reshape the entire economy and social fabric.[124] China's commitment is evidenced by an estimated $15 billion in public funding for quantum R&D and a clear lead in quantum communications, where it boasts the world's largest quantum communication network.[125]

---

[117] Quantum Sandbox for Near-Term Applications Act of 2025, H.R. 3220, 119th Cong. (2025), *supra* note 68.
[118] *See* Graham Allison, Destined for War: Can America and China Escape Thucydides's Trap? (2017); and William Walker, A Perpetual Menace: Nuclear Weapons and International Order (2012).
[119] *See* European Quantum Flagship, Strategic Research and Industry Agenda 2030 (2024), https://qt.eu/about-quantum-flagship/strategic-research-and-industry-agenda-2030; and How is the EU performing in innovation?, European Commission - EU Sci. Hub (June 27, 2024), https://joint-research-centre.ec.europa.eu/jrc-news-and-updates/how-eu-performing-innovation-2024-06-27_en.
[120] European Commission, Quantum Europe Strategy, *supra* note 10.
[121] *See e.g.* Robert D. Atkinson et al., *China Is Rapidly Becoming a Leading Innovator in Advanced Industries*, ITIF (Sept. 16, 2024), https://itif.org/publications/2024/09/16/china-is-rapidly-becoming-a-leading-innovator-in-advanced-industries/
[122] *See* e.g., Meia Nouwens and Helena Legarda, Int'l Inst. for Strategic Stud., China's Pursuit of Dual-Use Technologies (Dec. 2018), https://www.iiss.org/research-paper/2018/12/emerging-technology-dominance/; Opinions on Deepening the Reform of the Scientific and Technological System and Speeding up the Building of a National Innovation System, Ministry of Sci. and Tech. (Nov. 19, 2012), https://en.most.gov.cn/pressroom/201211/t20121119_98014.htm; State Council, 13th Five-Year Plan for National Economic and Social Development (2016-2020) (China); and State Council, 14th Five-Year Plan for National Economic and Social Development and the Long-Range Objectives Through the Year 2035 (2021).
[123] *See* e.g., Margot Schüller and Yun Schüler-Zhou, United States–China Decoupling: Time for European Tech Sovereignty, GIGA Focus (2025), https://www.giga-hamburg.de/en/publications/giga-focus/united-states-china-decoupling-time-for-european-tech-sovereignty

### 4.2. Strengths and Weaknesses in Fostering Dual-Use Technologies and Funding Mechanisms

These differing philosophies are clearly reflected in their handling of dual-use technologies. The United States has a well-established focus on dual-use innovation, supported by government initiatives and significant venture capital investment in defense-oriented startups, reflecting the strategic importance placed on these technologies for both economic and national security.[126]

The European Union has historically focused its R&D efforts on the civilian sector, but there is a growing recognition of the importance of dual-use capabilities.[127] A significant policy shift is marked by the European Innovation Council (EIC) now being permitted to invest in dual-use technologies, signaling a move towards a more integrated approach.[128] The Quantum Europe Strategy explicitly includes a pillar for "Space and Dual-Use Potential Quantum Technologies" to integrate secure, sovereign capabilities into Europe's security and defense strategies. This acknowledges that investing in dual-use capability can be a responsible act of deterrence and is necessary for national security.[129]

For China, the emphasis on civil-military fusion is a core national strategy.[130] This integrated system ensures that technological advancements simultaneously benefit the civilian economy and national defense, with government funding allowing for the rapid translation of research into application.

### 4.2.1. Dual-Use as a Deterrence Strategy

Navigating the current geopolitical reality requires a strategic reframing of dual-use technology. Rather than viewing it solely as a risk to be contained, the EU must recognize that developing and investing in dual-use quantum capabilities is a responsible and necessary act of deterrence.

---

[124] *See* Irene Zhang, *China's New AI Plan*, CHINATALK (Sept. 9, 2023), https://www.chinatalk.media/p/chinas-new-ai-plan.
[125] *See* e.g., Antonia Hmaidi & Jeroen Groenewegen-Lau, *China's Long View: Quantum Tech Has US and EU Playing Catch-Up*, MERICS (2023), https://merics.org/en/report/chinas-long-view-quantum-tech-has-us-and-eu-playing-catch
[126] *See* Def. Advanced Rsch. Projects Agency, https://www.darpa.mil/.
[127] *See* Pau Alvarez-Aragones, *supra* note 7; and LERU, Options for enhancing support for research and development involving technologies with dual use potential (2025), https://www.leru.org/files/Publications/Dual-use-Funding_LERU-Statement.pdf.
[128] *See* Brent, supra note 78; and Dual-use research, European Def. Agency, https://eda.europa.eu/what-we-do/all-activities/activities-search/dual-use-research.
[129] *See* also, European Commission, Quantum Europe Strategy, *supra* note10.
[130] *See* e.g., Elsa Kania, The Dual-Use Dilemma in China's New AI Plan: Leveraging Foreign Innovation Resources and Military-Civil Fusion, Lawfare (2025), https://www.lawfaremedia.org/article/dual-use-dilemma-chinas-new-ai-plan-leveraging-foreign-innovation-resources-and-military-civil.

In an environment of escalating technological competition, possessing advanced capabilities in areas like quantum sensing, secure communication, and computation is essential for maintaining national security, protecting critical infrastructure, and ensuring technological sovereignty. This strategic posture does not contradict the EU's values-based approach; rather, it provides the strength and autonomy required to uphold those values in a contested global landscape. Therefore, the EU Quantum Act's industrial policy must not only permit but actively encourage strategic investment in dual-use applications as a cornerstone of the Union's long-term security and resilience.

### 4.3. Identifying Successful Models and Potential Pitfalls for the EU Quantum Act

From this comparative analysis, several lessons emerge for the EU Quantum Act. The EU should learn from the US model by placing a greater emphasis on the research-to-commercialization link and actively encouraging private and venture capital investment, perhaps through a dedicated Quantum Fund and other incentives. Adopting models for public-private partnerships, such as the proposed "Quantum Sandbox," could accelerate near-term applications and bridge the gap between research and industry.[131] Furthermore, the EU could establish a European agency for high-risk, high-reward research with clear commercialization pathways, modeled on the US Defense Advanced Research Projects Agency (DARPA).[132] Such an agency could run innovation competitions and prize challenges[133] -in addition to direct government funding mechanisms- to accelerate breakthroughs in quantum and dual-use technologies, strengthening the EU's defense capabilities and strategic autonomy. This model acknowledges that many transformative civilian technologies, from the internet to GPS, have spurred from breakthrough military and space innovation. To ensure strategic success, a European quantum strategy must extend beyond fostering innovation and instead pursue holistic integration of the technology across the industrial and security sectors, develop a resilient "full-stack" European quantum infrastructure, and proactively address workforce transitions and skills gaps.[134]

From China, the EU can learn the value of a clear strategic vision, strong EU-level commitment, and dedicated funding for quantum areas that serve both economic competitiveness and national security, including infrastructure and secure networks.[135] This includes learning from China's long-term, five-year planning cycles, and Germany's traditional longer-term views on industrial planning, to ensure the EU's quantum strategy is sustainable and forward-looking.[136]

---

[131] Quantum Sandbox for Near-Term Applications Act of 2025, H.R. 3220, 119th Cong. (2025), *supra* note 68. *See* on quantum sandboxes also Kop, M. (2020), *supra* note 44 (TTLF 2020). *Compare* to Silicon Valley style innovation models such as the Stanford Quantum Incubator, *see*: https://events.stanford.edu/event/stanford-quantum-incubator-workshop-10
[132] *See* DARPA, *supra* note 119.
[133] Daniel J. Hemel & Lisa Larrimore Ouellette, *Innovation Policy Pluralism*, 128 Yale L.J. 544 (2019), https://www.yalelawjournal.org/article/innovation-policy-pluralism
[134] *See* Colin H. Kahl & Jim Mitre, *The Real AI Race: America Needs More Than Innovation to Compete With China*, Foreign Affs. (July 9, 2025), https://www.foreignaffairs.com/united-states/real-ai-race.
[135] *See* e.g., Nouwens & Legarda, *supra* note 116.

Crucially, the EU must also avoid its own historical pitfalls. This means moving away from overly restricting funding to civilian-only applications and instead actively facilitating dual-use R&D. It must also address persistent commercialization challenges by strengthening academia-industry collaboration and providing robust support for deep tech startups. Finally, it is essential to streamline state aid and support mechanisms to avoid the kinds of delays seen with the EU Chips Act.[137]

Moreover, the EU must be attuned to the fragmented and often opaque nature of the emerging commercial quantum market. Recent analysis highlights divergent commercialization strategies: some vendors focus on high-value, bundled deployments, while others prioritize shipping a higher volume of physical systems.[138] This simple dichotomy can be misleading, however, as broad cloud access also serves as a primary vehicle for scaling user ecosystems, not just for securing high-value contracts. This complex landscape underscores the need for policymakers to seek greater transparency on deal structures and market dynamics to accurately assess leadership and support a diverse and competitive European ecosystem.

In addition, the EU must chart a course that avoids the central pitfall of the current U.S. AI strategy. The American fixation on the uncertain goal of AGI risks distracting from the immense value of applying currently available machine intelligence.[139] China, in contrast, is less concerned with surpassing human intelligence and far more focused on deploying existing technology across its economy, from agriculture to healthcare, thereby building public enthusiasm and a formidable real-world advantage. The EU Quantum Act must therefore embrace a dual approach: it must continue to fund and pursue frontier research, while simultaneously making it a core strategic priority to foster a vibrant ecosystem of practical, everyday applications. Applying the powerful quantum models that already exist will start a flywheel of public support and tangible progress, ensuring the EU does not cede leadership by focusing too narrowly on a distant and uncertain finish line. The EU must also address the fragmentation of its research and funding efforts across Member States, which has led to duplication and inefficient use of resources. Understanding the differing regulatory philosophies underpinning these systems, as highlighted by comparative governance studies, is imperative for the strategic positioning of the EU Quantum Act.[140]

---

[136] *See* Kop, M., *Abundance & Equality*, in Scarcity, Regulation, and the Abundance Society, Mark A. Lemley ed., Lausanne, Switzerland: Frontiers in Science, 2022, https://www.frontiersin.org/articles/10.3389/frma.2022.977684/full
[137] *See* European Commission, European Chips Act, *supra* note 22.
[138] *See* e.g., Matt Swayne, *In Initial Stages of Quantum Computing Commercialization, Sales Stats Show IBM Leads in Quantum Deal Value, IQM in Units Sold*, The Quantum Insider (Aug. 19, 2025), https://thequantuminsider.com/2025/08/19/in-initial-stages-of-quantum-computing-commercialization-sales-stats-show-ibm-leads-in-quantum-deal-value-iqm-in-units-sold/.
[139] *See* Schmidt & Xu, *supra* note 22.

A more detailed breakdown of the quantum-specific innovation ecosystems reveals distinct priorities and designs. In *quantum computing and simulation*, the US benefits from strong fundamental research through its National Quantum Initiative (NQI), major private sector players like Google and IBM, a growing startup ecosystem, and a focus on fault-tolerance and applications. The EU displays strong academic research via its Quantum Flagship, fosters diverse hardware architectures, and is expanding national and EU-level initiatives like EuroHPC to provide hardware access and build a broad ecosystem.[141] China, by contrast, directs major state investment toward achieving specific computational benchmarks and strategic applications, resulting in a less diverse public ecosystem.

In *quantum sensing and metrology,* the US has strong, defense- and industry-driven R&D, with applications in navigation, medicine, and environmental monitoring seeing growing commercialization. The EU has a historically strong research base in this area, with applications in the medical and automotive sectors, and a focus on translating research into market-ready products. China's significant investment is geared toward national priorities like resource exploration and defense, accompanied by strong patenting activity.

For *quantum networking and communication*, the US has growing government initiatives, such as Department of Energy testbeds, with a focus on developing core components and protocols for a future quantum internet, where universities play a strong role.[142] The EU has a pronounced focus on secure communication through its EuroQCI initiative, deploying Quantum Key Distribution (QKD) networks and researching enabling technologies like quantum repeaters. Here, governance must learn from the national security challenges encountered with 5G infrastructure, emphasizing the need for trusted vendors and responsible networking standards. China is the global leader in long-distance QKD, notably via satellite, and is undertaking major state-driven infrastructure projects for secure government and military communications.

Regarding *quantum-AI hybrids*, the US leverages its strong AI base and significant interest from both the private sector and agencies like DARPA to focus on performance breakthroughs. The EU has a strong research foundation in both AI and quantum through programs like the Quantum Flagship and Horizon Europe, with a particular focus on foundational Quantum Machine Learning (QML) and its responsible and ethical development. China utilizes its strong AI capabilities and is likely making significant state investments to gain a strategic advantage in areas like pattern recognition and optimization, driven by its civil-military fusion strategy.

---

[140] *See* e.g., Kop, M., *Towards an Atomic Agency for Quantum-AI: A Comparative Analysis of AI & Quantum Technology Regulation and Innovation Models in the U.S., EU and China* (May 5, 2025), https://doi.org/10.48550/arXiv.2505.11515; and Varieties of Capitalism: The Institutional Foundations of Comparative Advantage (Peter A. Hall & David Soskice eds., 2001).
[141] Council Regulation (EU) 2021/1173 of 13 July 2021 on establishing the European High Performance Computing Joint Undertaking, 2021 O.J. (L 256) 3.
[142] *See* e.g., Nat'l Quantum Initiative Advisory Comm., Quantum Networking: Findings and Recommendations for Growing American Leadership (2024).

In the domain of *quantum materials and devices*, the US has a robust materials science base in its national labs and universities, focusing on new qubit designs, error correction hardware, and advanced fabrication techniques linked to the CHIPS Act. The EU also has a strong materials research base through its Flagship and national programs, focusing on novel materials and platforms while developing its own supply chains. China is making massive state investments aimed at achieving self-sufficiency in critical materials and fabrication, reflected in its high volume of patenting in this area.[143]

Finally, in *fundamental physics research*, the US is home to world-leading universities and labs with strong funding for basic science from the NSF and DOE, supporting diverse theoretical and experimental approaches. The EU has excellent academic centers with a long tradition of collaborative foundational science, backed by significant funding from the ERC, the Quantum Flagship, and national bodies. China is rapidly increasing its investment and output, focused on building world-class institutions and talent, with the state prioritizing research areas deemed to be of strategic importance.

By comparing the US, EU, and Chinese innovation systems, this section has highlighted key strategic considerations for the EU Quantum Act. The analysis indicates the EU should emulate the US focus on commercialization and private investment, adopt China's strategic commitment (particularly regarding dual-use), and avoid its own historical pitfalls related to funding restrictions and commercialization gaps. These comparative insights, including the domain-specific overview, are elemental for designing policies within the EU QA that effectively boost Europe's global competitiveness in quantum technologies.

## 5. ESTABLISHING STANDARDS, BENCHMARKS, VERIFICATION, AND CERTIFICATION

This section addresses the area of standardization, verification, certification, and benchmarking within the rapidly evolving field of quantum technology. It highlights the current nascent state of standardization efforts compared to mature technology sectors and discusses the inherent challenges posed by rapid innovation and platform diversity. The section explores the significant opportunities that unified standards offer for interoperability, trust-building, and safety, outlining the essential role the EU Quantum Act must play in proactively supporting and coordinating these vital initiatives.

### 5.1. The Nascent State of Quantum Standardization

Unlike mature technology sectors, the quantum field currently lacks comprehensive, widely adopted standardization frameworks.[144] While initial efforts are underway to establish common terminology, preliminary metrics, and testing methods, these face significant challenges.[145] An analysis of the current state reveals a fragmented landscape. There are no universal standards for quantum computing, where metrics like qubit count, fidelity, and coherence time exist but are not standardized for reliable comparison across platforms. Similarly, the domains of quantum sensing and networking lack comprehensive certification, although some security protocols are beginning to emerge. Benchmarking, for instance through metrics like quantum volume, remains an active area of research, but a firm consensus has yet to be reached. In this context, a "standards-first" approach could provide a valuable mechanism for governing R&D before broader regulation is imposed.[146]

---

[143] *See* e.g., State Council, Next Generation Artificial Intelligence Development Plan (July 8, 2017) (China).
[144] *See* e.g., Kop, supra note 44 (TTLF 2020)

### 5.2. Challenges and Opportunities

The path to standardization is impeded by several challenges. The rapid pace of innovation means that standards risk becoming obsolete quickly or, conversely, stifling novelty. This is compounded by the diversity of quantum platforms, the lack of mature systems and empirical data for validation, and the sheer complexity of the underlying technology, which demands highly specialized expertise.[147] Technical standards are not merely technical; they are vessels for values. Beyond their technical function, standards are a primary mechanism for embedding universal ethical principles, normative values, and democratic norms into the technological substrate.[148] This can be achieved by explicitly requiring that the design, development, and risk management processes codified within standards—such as a QT-QMS—include auditable criteria for fairness, transparency, accountability, and respect for fundamental rights.[149] By making these values an integral part of the certification process, the EU can ensure that technical interoperability is accompanied by normative alignment.

Despite these hurdles, the opportunities presented by standardization are immense. Unified standards would lead to improved communication and understanding across the field. They would also foster interoperability, which is required for market growth and reducing vendor lock-in. Standards are necessary for building user confidence and trust, and for ensuring safety and security by defining clear protocols, testing procedures, and benchmarks, particularly for dual-use applications with national security implications.[150]

---

[145] *See* e.g., Lubinski, T., Czarnik, P., Cincio, L., Sornborger, A., Coffrin, C., & Coles, P. J. (2023). *Application-Oriented Performance Benchmarks for Quantum Computing.* IEEE Transactions on Quantum Engineering, 4, 1-22, https://doi.org/10.1109/TQE.2023.3253761; Seskir, Z. et al, Quantum Sci. Technol. 8 024005 (2023) DOI 10.1088/2058-9565/acb6ae, https://iopscience.iop.org/article/10.1088/2058-9565/acb6ae; and OECD, Recommendation of the Council on Artificial Intelligence (2019, rev. 2024).
[146] Aboy *et al., supra* note 34.
[147] *See*.e.g., ISO/IEC JTC 3 Quantum technologies, https://www.iso.org/committee/10138914.html; and ISO/IEC FDIS 4879 Information technology Quantum computing Vocabulary, https://www.iso.org/standard/80432.html.
[148] Kop, *supra* note 139; Matt Perault & J. D. Williams, AI and Product Safety Standards Under the EU AI Act, Carnegie Endowment for Int'l Peace (Mar. 2024), https://carnegieendowment.org/research/2024/03/ai-and-product-safety-standards-under-the-eu-ai-act; and Corina Pintea, The European approach to regulating AI through technical standards, Pol'y Rev. (2025), https://policyreview.info/articles/analysis/regulating-ai-through-technical-standards.
[149] Aboy *et al., supra* note 34.

### 5.3. The Role of the EU Quantum Act in Standardization

Therefore, the EU Quantum Act has a determining role to play. It must prioritize and support standardization, verification, certification, and benchmarking initiatives across all quantum domains, including computing, sensing, networking, and QAI hybrids. It should recognize the distinct advantages of a 'standards-first' philosophy, as advocated by Aboy *et al*., for governing the technology during its early R&D stages, thereby laying the foundations for interoperability, market development, and future regulatory compliance.[151] This method leverages voluntary, consensus-driven standards to create a flexible and globally harmonized governance ecosystem, which can serve as the groundwork for more targeted regulation as the technology matures.[152] The Act should also fund dedicated metrology and standardization research and ensure engagement in international standards bodies like ISO.[153] The strategic importance of establishing responsible standards early—before path dependencies solidify as the 'quantum event horizon' approaches—cannot be overstated.

To translate these principles into practice, the Act should propose the development of a harmonized, certifiable Quantum Technology Quality Management System (QT-QMS) standard under ISO/IEC.[154] It is built on the principle of certifying an organization's entire management system, rather than individual products—a model successfully pioneered in the medical technology sector. As envisioned, this procedure would entail certification of a company's QT-QMS by an independent, accredited body (akin to a notified body in the MedTech field), establishing a verifiable framework for quality, transparency, and accountability.

This standard would serve as the primary mechanism for providers to demonstrate compliance with the Act's regulatory requirements. The QT-QMS provides an auditable framework for operationalizing the principles for Responsible Quantum Technology (RQT), embedding ethical, legal, societal, policy, and interoperability (ELSPI) considerations directly into the development lifecycle. For high-risk systems, this system-level certification against the harmonized QT-QMS standard would then become the foundation for demonstrating conformity and achieving a CE mark to gain market access, mirroring the successful model of ISO 13485 for medical devices.[155] This arrangement provides regulatory agility, as the standard can be updated periodically by technical experts to reflect technological progress without requiring a lengthy amendment of the core legislative text.

---

[150] *See* e.g., *Quantum Benchmarking Working Group P7131*, IEEE, https://sagroups.ieee.org/7131/.
[151] Aboy *et al., supra* note 34.
[152] *ibid*.
[153] *See* e.g., Provisions on the Management of Deep Synthesis in Internet Information Services (promulgated by the Cyberspace Admin. of China et al., Nov. 25, 2022, effective Jan. 10, 2023) (China).
[154] Aboy *et al., supra* note 34.
[155] *ibid*.

To implement this, the EU should leverage its existing bodies, such as CEN and CENELEC, with potential coordination from the proposed OQTA.[156] It should also consider the role of ENISA and ETSI in quantum cryptography certification, ensuring alignment with international efforts like the NIST Post-Quantum Cryptography (PQC) standardization process.[157] This includes considering algorithms such as CRYSTALS-Kyber, CRYSTALS-Dilithium, FALCON, and SPHINCS+ as key benchmarks for ensuring cybersecurity against quantum threats and addressing 'store now, decrypt later' vulnerabilities.[158] The Act could also explore regulated registries or databases for quantum algorithms to enhance transparency and aid in export control. The operationalization of these standards can be achieved through a dedicated Quantum Technology Quality Management System (QT-QMS), which provides a structured, auditable framework for managing the entire lifecycle of quantum technologies, ensuring compliance with emerging regulations, and translating high-level principles into verifiable RQT metrics.[159]

The foregoing analysis underscores the foundational importance of standardization, verification, certification, and benchmarking for the responsible development and adoption of quantum technologies. Despite the challenges, the prospective benefits for interoperability, trust, safety, and market growth are immense. The EU Quantum Act must therefore incorporate strong provisions to actively foster and guide standardization efforts, including alignment on post-quantum cryptography, ensuring the EU plays a leading role in shaping the technical underpinnings of the quantum future.

## 6. LEARNING FROM NUCLEAR GOVERNANCE: REGULATION AND NON-PROLIFERATION

This section examines the decades-long experience of governing nuclear technology, encompassing both fission and fusion, as a source of valuable lessons for regulating quantum technologies. It explores the stringent safety and security regimes, the international non-proliferation frameworks (NPT, IAEA, NSG), and the models for large-scale international collaboration found in the nuclear field. The objective is to identify principles and mechanisms applicable to the EU Quantum Act for managing the risks, particularly dual-use implications, and fostering responsible international cooperation in the quantum domain.

### 6.1. The Nuclear Governance Precedent

---

[156] *See* European Comm. for Standardization (CEN) & European Comm. for Electrotechnical Standardization (CENELEC), Joint Technical Comm. 22, Quantum Technologies, (CEN-CLC/JTC 22), https://www.cencenelec.eu/areas-of-work/cen-cenelec-topics/quantum-technologies/.
[157] *See* European Telecommunications Standards Inst. (ETSI), Industry Specification Grp. on Quantum Key Distribution (ISG QKD), https://www.etsi.org/committee/qkd; and ENISA, supra note 81.
[158] *See* Nat'l Inst. of Standards & Tech., U.S. Dep't of Com., *Post-Quantum Cryptography*, https://csrc.nist.gov/projects/post-quantum-cryptography.
[159] Aboy *et al., supra* note 34.

The governance of nuclear technology, developed over decades in response to its power and potential dangers, offers valuable insights for establishing a robust regulatory framework for quantum technologies, particularly concerning safety, security, international collaboration, and the management of dual-use capability.[160]

Regulatory frameworks governing nuclear fission and fusion are characterized by a paramount emphasis on safety and security.[161] This includes stringent regulations covering the entire lifecycle—from design and construction to operation and decommissioning of facilities—with a strong focus on preventing accidents and mitigating their consequences. Comprehensive licensing and permitting processes are mandatory, involving rigorous technical assessments and ongoing oversight by independent regulatory bodies with significant authority and expertise. Environmental protection is also a consideration, with detailed regulations governing the handling and disposal of radioactive materials.

In parallel to domestic safety regimes, international non-proliferation efforts in the nuclear domain are guided by the fundamental principle of preventing the spread of nuclear weapons. Go-to instruments include treaties like the Nuclear Non-Proliferation Treaty (NPT), which establishes obligations for states based on three pillars: preventing proliferation, pursuing disarmament, and promoting the peaceful uses of nuclear energy.[162] International organizations like the International Atomic Energy Agency (IAEA) play a crucial role in verifying the peaceful use of nuclear materials through safeguards agreements and inspections, while also facilitating international cooperation and providing technical assistance.[163] Furthermore, export controls on nuclear materials, equipment, and technology, often guided by frameworks like the Nuclear Suppliers Group (NSG) guidelines, are an important mechanism for preventing proliferation of weapons of mass destruction by restricting their transfer to unauthorized actors.[164]

Beyond regulation and control, the nuclear field, particularly in fusion research like the ITER project, exemplifies a model of large-scale international collaboration and investment.[165] This model is designed to tackle complex scientific and engineering challenges that are beyond the capacity of single nations. This collaborative approach, which pools global resources and expertise, is reminiscent of the CERN model in particle physics and offers a compelling precedent for accelerating progress in capital-intensive fields like quantum computing.[166]

---

[160] *See* e.g. William Walker, A Perpetual Menace: Nuclear Weapons and International Order (2012).
[161] Debs A, Monteiro NP. *Nuclear Politics: The Strategic Causes of Proliferation*. Cambridge University Press; 2016, https://doi.org/10.1017/9781316257692.
[162] Treaty on the Non-Proliferation of Nuclear Weapons art. IV, July 1, 1968, 21 U.S.T. 483, 729 U.N.T.S. 161, https://disarmament.unoda.org/wmd/nuclear/npt/text/.
[163] Int'l Atomic Energy Agency, *Safeguards and Verification*, https://www.iaea.org/topics/safeguards-and-verification.
[164] Nuclear Suppliers Grp., *NSG Guidelines*, https://www.nuclearsuppliersgroup.org/index.php/en/guidelines/nsg-guidelines.
[165] ITER, *ITER - The Way to New Energy*, https://www.iter.org/.
[166] *See* e.g., CERN, *A Quantum Leap for Antimatter Measurements*, (July 23, 2025), https://home.cern/news/news/physics/quantum-leap-antimatter-measurements.

## 6.2. Applicability to Quantum Technology Governance

The principles and mechanisms underpinning nuclear governance have significant relevance for the regulation of quantum technologies. Specifically, the EU Quantum Act can draw inspiration from the nuclear model's unwavering focus on safety and security, especially for high-risk quantum applications with possible military uses. This could involve implementing rigorous risk assessment protocols, establishing security standards for quantum facilities and data, and creating independent oversight bodies to ensure compliance.

Moreover, the principles of international cooperation and information sharing, which are central to nuclear governance, should be applied to quantum technologies to address global security concerns, share best practices, and prevent an uncontrolled "quantum arms race"[167]. A central goal of international quantum governance must be to mitigate the "quantum security dilemma," in which the competitive race for a strategic advantage incentivizes nations to prioritize speed and secrecy over safety, thereby increasing global instability and the risk of conflict.[168] The EU should actively engage in international dialogues on quantum security, and the collaborative investment model seen in fusion provides a useful template for large-scale quantum infrastructure projects, linking to the concept of a "CERN for Quantum/AI."

To manage proliferation risks, targeted dual-use export controls mirroring those for nuclear materials and technologies will be essential.[169] These could be informed by the NSG guidelines and adapted from existing dual-use frameworks like the Wassenaar Arrangement, which could be strengthened for this purpose. The EU Quantum Act should define clear criteria for such controls based on prospective military or other harmful applications. Lastly, the IAEA's safeguards system offers insights for developing verification regimes for certain high dual-use application quantum technologies, should future international agreements necessitate them. This is a key tenet of the proposal to establish an 'Atomic Agency for Quantum-AI' dubbed 'International Quantum Agency (IQA)' to oversee a global treaty or accord, leveraging the IAEA's expertise in safeguards implementation to manage dual-use risks.[170]

Established practices within nuclear governance offer significant insights for the EU Quantum Act. The paramount focus on safety and security, the multi-layered non-proliferation regime involving treaties (NPT), verification (IAEA), and export controls (NSG), and the successful models of large-scale international scientific collaboration (ITER/CERN) provide valuable templates. Applying these lessons can help the EU QA effectively manage risks, ensure security, and foster responsible international cooperation in the quantum era.

---

[167] *See* e.g., Martin Giles, *The U.S. and China Are in a Quantum Arms Race That Will Transform Warfare*, MIT Tech. Rev. (Jan. 3, 2019), https://www.technologyreview.com/2019/01/03/137969/us-china-quantum-arms-race/.

[168] *See* Henry A. Kissinger, Eric Schmidt & Craig Mundie, *War and Peace in the Age of Artificial Intelligence*, Foreign Affs. (Nov. 18, 2024), https://www.foreignaffairs.com/united-states/war-and-peace-age-artificial-intelligence.

[169] *See* e.g., European Union, *Dual-Use Export Controls*, https://eur-lex.europa.eu/EN/legal-content/summary/dual-use-export-controls.html.

[170] Kop, *supra* note 139.

## 7. QUBITS FOR PEACE: A GLOBAL GOVERNANCE FRAMEWORK

To effectively govern the dual-use nature of quantum technologies and steer their development toward peaceful, beneficial outcomes, this paper proposes a "Qubits for Peace" initiative. This framework extrapolates from the historical "Atoms for Peace" proposal and the contemporary "Chips for Peace" concept for AI, advocating for a global governance structure built on three pillars: safety regulation, benefit-sharing, and non-proliferation.

### 7.1. Historical Precedent: "Atoms for Peace"

In his 1953 address to the United Nations, U.S. President Dwight D. Eisenhower outlined his "Atoms for Peace" vision.[171] Facing a world overshadowed by the "dark chamber of horrors" of atomic warfare, Eisenhower proposed a new path. He suggested that the governments principally involved in nuclear development should "begin to make joint contributions of fissionable material to an Atomic Power Authority of the United Nations."[172] This international body would be responsible for impounding, storing, and protecting these materials, and devising methods to apply them to the "peaceful pursuits of mankind," such as providing electrical energy to power-starved areas of the world.[173] The core idea was to transform the "greatest of destructive forces" into a great constructive force, dedicating strength to serve the needs, rather than the fears, of the world.[174]

### 7.2. Contemporary Analogy: "Chips for Peace"

Drawing inspiration from this historical precedent, the "Chips for Peace" framework has been proposed as a multilateral initiative for the safe and beneficial governance of frontier artificial intelligence.[175] Leveraging influence over the global semiconductor ecosystem, this framework rests on three interrelated commitments. First, member states commit to safety regulation by governing domestic frontier AI development to reduce risks to public safety and global security, ensuring AI systems meet consistent safety standards. Second, they agree to benefit-sharing by ensuring the advantages of safe AI are distributed broadly to compensate for the costs of regulation and non-proliferation. Third, they pursue non-proliferation by coordinating policies, including export controls on AI hardware and cloud computing capacity, to prevent non-member states from developing high-risk AI systems that could undermine the agreement.[176] This framework aims to create a cooperative structure that balances innovation with safety, loosely inspired by the NPT and the IAEA model.

---

[171] Dwight D. Eisenhower, U.S. President, Address Before the General Assembly of the United Nations on Peaceful Uses of Atomic Energy (Dec. 8, 1953), https://www.eisenhowerlibrary.gov/research/online-documents/atoms-peace.
[172] *ibid.*
[173] *ibid.*
[174] *ibid.*
[175] Cullen O'Keefe, *Chips for Peace: How the U.S. and Its Allies Can Lead on Safe and Beneficial AI*, Lawfare (July 10, 2024), https://www.lawfaremedia.org/article/chips-for-peace--how-the-u.s.-and-its-allies-can-lead-on-safe-and-beneficial-ai. *See* also: Natasza Gadowska, *Seizing Global AI Regulation: Chips for Peace as America's Last Call to Lead*, Harv. J.L. & Tech. (May 21, 2025), https://jolt.law.harvard.edu/digest/seizing-global-ai-regulation-chips-for-peace-as-americas-last-call-to-lead.

### 7.3. A Framework for "Qubits for Peace"

Extrapolating from these models, a "Qubits for Peace" initiative would establish a global governance framework for quantum technologies, structured around the same three pillars. The first pillar, Safety Regulation for Quantum Technologies, would involve proactive ethical research and the establishment of robust safety standards for quantum systems. Under this pillar, the EU Quantum Act can lead this effort by mandating Quantum Technology Impact Assessments (QIAs) to evaluate ethical, legal, and societal impacts throughout a system's lifecycle, particularly for high-risk and dual-use applications. Furthermore, agile regulatory sandboxes could be used to test quantum-AI hybrids, and a body like the proposed OQTA would coordinate the development of technical standards to translate high-level principles into practice.[177] The second pillar, Benefit-Sharing and Equitable Access, would focus on preventing a "quantum divide" through the equitable redistribution of the benefits and risks of quantum technologies. This would be achieved by promoting collaborative research platforms for quantum and AI, modeled on CERN or ITER, to foster coordinated and responsible innovation on a global scale, and by democratizing access to quantum resources and supporting less-resourced regions to ensure the fruits of the quantum revolution are shared by all of humanity. Finally, the third pillar, Non-Proliferation of High-Risk Quantum Capabilities, would concentrate on creating a global quantum weapons of mass destruction non-proliferation framework, for instance through a UN Quantum Treaty modeled on the 2024 UN AI Resolution and the NPT, for instance titled the 'Global Quantum Accord'.[178] An 'Atomic Agency for Quantum-AI' labelled 'International Quantum Agency (IQA)', inspired by the IAEA, could be established to oversee treaty compliance, manage non-proliferation risks, and implement international safeguards. This would necessarily involve a classification system for dual-use quantum technologies and internationally coordinated export controls on high-risk capabilities, such as those for cryptanalysis, calibrated to avoid stifling legitimate innovation.[179]

### 7.4. Relevance for the EU Quantum Act

The "Qubits for Peace" framework provides a vital strategic and normative compass for the EU Quantum Act. It aligns with the EU's values-based regulatory model, as seen in the GDPR and the AI Act, while providing a clear path for global leadership. By embedding the principles of "Qubits for Peace" into its own legislation, the EU can champion a model of quantum governance that is not only competitive but also responsible, safe, and equitable. This stands in contrast to more assertive models of technology diplomacy, such as the U.S. AI Action Plan's promotion of "full-stack export packages" to build a geopolitical alliance, thereby offering the world a distinct, values-driven alternative (see Section 11 for a full comparison).[180] This course of action reinforces the goals of the Quantum Europe Strategy[181]—to ensure strategic autonomy while fostering international cooperation—and provides a concrete structure for achieving a "Quantum Acquis Planétaire" that prevents fragmentation and promotes a stable, secure, and beneficial global quantum ecosystem.[182] Establishing such a global quantum acquis is not just an ideal, but a pragmatic imperative to prevent a fractured and unstable geopolitical landscape. This requires a holistic global quantum governance framework that skillfully integrates international treaties, coordinated export controls, harmonized standards, and coherent national regulations. As Perrier notes, while no formal international treaty for adversarial cyber technologies currently exists, their use is already framed within existing jurisprudence, making the proactive development of quantum-specific norms a critical strategic task.[183]

---

[176] *ibid.*
[177] Kop, *supra* note 139 (Atomic).
[178] *ibid. See* also United Nations G.A. Res. 78/241, Seizing the opportunities of safe, secure and trustworthy artificial intelligence systems for sustainable development (Mar. 21, 2024).
[179] Kop, *supra* note 139.

## 8. ENSURING COHERENCE WITH THE EXISTING EU REGULATORY LANDSCAPE

This section addresses the need for the proposed EU Quantum Act to integrate seamlessly and coherently within the complex web of existing European Union legislation. It examines the potential interactions and overlaps with regulations such as the EU AI Act, the Medical Device Regulation (MDR)[184], the Machinery Directive/Regulation[185], and rules governing the financial sector. The objective is to highlight areas requiring careful alignment and coordination to avoid regulatory conflicts, duplication, or gaps, thereby ensuring legal certainty and facilitating the smooth adoption of quantum technologies across various sectors.

### 8.1. The Imperative of Regulatory Coherence

The proposed EU Quantum Act will need to operate within the existing body of EU legislation, ensuring coherence and avoiding unnecessary duplication or conflict. This requires a careful analysis of the interplay between the EU QA and other relevant cross-sectoral and sectoral regulations.

### 8.2. Interaction with Key Legislative Frameworks

A primary area of concern is the relationship with the EU AI Act. The EU QA should be designed to be fully compatible with the AI Act, particularly regarding aspects such as risk assessment methodologies, transparency obligations for quantum-AI hybrid systems, and conformity assessment procedures.[186] The EU QA must clearly articulate how its regulations interact with those of the AI Act, establishing specific guidelines or clarifications. This is especially salient for governing quantum-AI hybrids, which may require novel legal-technical taxonomies to delineate regulatory boundaries and ensure that unique synergistic effects and emergent properties are addressed.[187]

---

[180] *See* e.g., https://www.csis.org/analysis/experts-react-unpacking-trump-administrations-plan-win
[181] European Commission, Quantum Europe Strategy, *supra* note 10.
[182] Kop, *supra* note 139.
[183] Perrier, *supra* note 13.
[184] Regulation (EU) 2017/745, 2017 O.J. (L 117) 1, https://eur-lex.europa.eu/eli/reg/2017/745/oj/eng
[185] *Regulation (EU) 2023/1230 of the European Parliament and of the Council of 14 June 2023 on machinery* replaces Directive 2006/42/EC on machinery, *see* Regulation (EU) 2023/1230, 2023 O.J. (L 165) 1, https://osha.europa.eu/en/legislation/directive/regulation-20231230eu-m

Beyond AI, compatibility with sectoral legislation, such as the Medical Device Regulation (MDR) and the Machinery Directive/Regulation, will be vital for quantum technologies intended for use in specific applications like healthcare or industrial control systems. The EU QA should ensure that its requirements do not conflict with established safety and performance standards for medical devices or machinery. Clarity will be needed regarding which regulatory regime takes precedence or how they complement each other. To avoid issues seen previously, legislators should learn from the legal uncertainty and confusion that arose in the context of the EU AI Act and the MDR.[188]

In the finance sector, the EU QA will need to consider the profound implications of quantum computing, particularly its capacity to compromise existing cryptographic tactics.[189] The Act should promote the development and adoption of quantum-safe cryptography standards[190], for example aligned with NIST PQC standards, to safeguard financial systems and data. This would ensure alignment with relevant financial regulations concerning data security and resilience, such as the GDPR, the Network and Information Systems Security (NIS) Directive, and the Digital Operational Resilience Act (DORA).[191] Synergies could also be explored in leveraging quantum-enhanced AI for applications like fraud detection and risk management.

Across other major market verticals, including energy, logistics, space, and defense, the EU QA should aim to foster synergies and avoid conflicts. In energy, it should consider how to incentivize quantum-enhanced solutions within the EU's energy policy framework. For logistics, it should promote the responsible use of quantum optimization while considering existing transportation regulations. In space, alignment with current space law is needed for quantum sensing and communication applications. In defense, careful coherence with EU defense policies is required, establishing specific guidelines for military applications while ensuring alignment with the Act's overall objectives. NATO has already identified quantum as an Emerging and Disruptive Technology (EDT) and is developing its own strategy to ensure the Alliance is "quantum-ready," highlighting the need for the EU QA to align with broader transatlantic security objectives.[192] Coordination clauses and Commission guidance on the interplay between regulations will be imperative for all these sectors.

---

[186] European Commission, AI Act, Shaping Europe's Digital Future, https://digital-strategy.ec.europa.eu/en/policies/regulatory-framework-ai, *supra* note 36.
[187] *See* Kop, *supra* note 139.
[188] *See* e.g., Aboy, M., Minssen, T. & Vayena, E. Navigating the EU AI Act: implications for regulated digital medical products. *npj Digit. Med.* **7**, 237 (2024). https://doi.org/10.1038/s41746-024-01232-3
[189] *See* e.g., Urs Gasser & Fabienne Marco, *Cracking the Quantum Future of Finance*, Duckbucks (May 9, 2025), https://duckbucks.com/a/cracking-the-quantum-future-of-finance.
[190] *See* e.g., Vidick, T. and Wehner, S. (2023) *Introduction to Quantum Cryptography*. Cambridge: Cambridge University Press, https://doi.org/10.1017/9781009026208
[191] *See* Regulation (EU) 2016/679, 2016 O.J. (L 119) 1., Directive (EU) 2022/2555, 2022 O.J. (L 333) 80, Regulation (EU) 2022/2554, 2022 O.J. (L 333) 1; Regulation (EU) 2016/679 of the European Parliament and of the Council of 27 April 2016 on the protection of natural persons with regard to the processing of personal data and on the free movement of such data (General Data Protection Regulation), 2016 O.J. (L 119) 1; Regulation (EU) 2022/2065 of the European Parliament and of the Council of 19 October 2022 on a Single Market For Digital Services (Digital Services Act), 2022 O.J. (L 277) 1; Regulation (EU) 2022/1925 of the European Parliament and of the Council of 14 September 2022 on contestable and fair markets in the digital sector (Digital Markets Act), 2022 O.J. (L 265) 1; and Directive 2002/58/EC of the European Parliament and of the Council of 12 July 2022 concerning the processing of personal data and the protection of privacy in the electronic communications sector (ePrivacy Directive), 2002 O.J. (L 201) 37.

### 8.3. Principles for Ensuring a Harmonized Environment

To aid coherence across diverse quantum-AI implementations and market verticals, the Quantum Act can be guided by the principle of functional equivalence, which focuses on regulating behaviors and uses rather than the technology itself. Insights from comparative AI governance analyses are foundational for ensuring this coherence, particularly for regulating quantum-AI hybrids and aligning the QA with the principles and structure of the AI Act.

Likewise, ensuring coherence requires fostering institutional plasticity. This means encouraging existing regulatory bodies, such as the European Medicines Agency (EMA)[193] and the notified bodies involved with the MDR and Machinery Directive/Regulation, to develop the necessary expertise and adaptive capacity to evaluate quantum-specific aspects effectively.

Achieving coherence between the EU Quantum Act and the existing body of EU law is essential for its effective implementation and for providing legal certainty to innovators and users. Careful consideration of interactions with the AI Act, sectoral regulations like the MDR and Machinery Directive/Regulation, and financial rules is necessary. Mechanisms such as coordination clauses, Commission guidance, and promoting institutional plasticity will be vital to navigate overlaps and ensure a harmonized regulatory environment for quantum technologies.

### 9. TOWARDS A ROBUST GOVERNANCE STRUCTURE FOR QUANTUM TECHNOLOGY

This section addresses the question of how to structure the governance of quantum technologies within the European Union. It explores candidate institutional models, drawing inspiration from established international organizations like CERN and the IAEA, and introduces the proposal for a dedicated EU Office of Quantum Technology Assessment (OQTA). Additionally, it considers complementary governance mechanisms such as regulatory forums, foresight techniques, and the overarching goal of international coordination, including non-proliferation efforts.

---

[192] *See* NATO, NATO's Quantum Technologies Strategy (2024), https://www.nato.int/cps/en/natohq/official_texts_221777.htm; NATO, NATO's First-Ever Quantum Strategy (2024), https://www.nato.int/cps/en/natohq/news_221724.htm; and NATO, NATO's Digital Transformation Implementation Strategy (2024), https://www.nato.int/cps/en/natohq/official_texts_229801.htm
[193] European Medicines Agency, https://www.ema.europa.eu/en/homepage.

### 9.1. Institutional Models and Precedents

To begin, established international organizations offer valuable models for EU quantum governance. The CERN model, for instance, demonstrates the success of international collaboration in fundamental research, characterized by shared governance through its Council, a culture of open science, and the effective management of large-scale shared infrastructure.[194] This, alongside the collaborative investment seen in nuclear fusion research like ITER, highlights the benefits of collective investment and strongly informs the prospect for a "CERN for Quantum/AI." Similarly, the IAEA model provides a template for an organization that simultaneously promotes peaceful use while inhibiting military applications. It achieves this by establishing safety standards, verifying peaceful use through its safeguards system, and facilitating international cooperation and technical assistance, thus covering both promotion and risk mitigation. Its verification role is particularly relevant for future quantum non-proliferation agreements.[195]

### 9.2. A Proposed EU Office of Quantum Technology Assessment (OQTA)

Drawing on these precedents, we propose the creation of a dedicated EU Office of Quantum Technology Assessment (OQTA), inspired by the IAEA, CERN, the former US OTA, and the EU AI Office. This body would serve a multitude of functions. It would be tasked with facilitating EU-wide collaboration among researchers, industry, and policymakers, and could manage shared infrastructure like quantum testbeds. A core responsibility would be to conduct horizon scanning and comprehensive risk-benefit assessments, with a particular focus on dual-use potential, while also developing ethical guidelines and providing independent policy advice to the Commission, Parliament, and Council. The OQTA would also play a role in coordinating standardization efforts—proactively driving the development of standards as a key early-stage governance tool—and overseeing export controls. Operationally, it would support regulatory enforcement, serve as a central stakeholder forum, oversee codes of conduct, and evaluate the results from regulatory sandboxes. Its mandate would also include overseeing a mandatory Quantum Technology Impact Assessment (QIA) process for high-risk applications, ensuring a systematic evaluation of risks to fundamental rights and security before market entry. It would also be charged with promoting institutional plasticity and expertise within existing EU bodies, monitoring the intellectual property landscape and market competition dynamics, and actively working to avoid regulatory fragmentation across Member States—all while navigating the deep uncertainty of the 'quantum event horizon'.[196]

---

[194] CERN Council, https://council.web.cern.ch/en/content/news.
[195] *See* e.g., Int'l Atomic Energy Agency, Press Release, *Independent Review Assesses IAEA's Internal Safety Regulatory System for First Time, Finds Well-Established Framework* (Oct. 9, 2024), https://www.iaea.org/newscenter/pressreleases/independent-review-assesses-iaeas-internal-safety-regulatory-system-for-first-time-finds-well-established-framework.
[196] *See* Kop, *supra* note 47 (event horizon).

### 9.3. Complementary Governance Mechanisms and Global Ambitions

Beyond a central office, additional governance ideas can complement this structure. A Regulatory Forum, similar to the UK's Digital Regulation Cooperation Forum (DRCF), could be established to ensure regulators across sectors become "quantum-ready," fostering integrated development and preventing fragmentation.[197] This can be supported by foresight techniques, such as systematic horizon scanning, to anticipate future regulatory needs proactively. The governance system itself should be modular and adaptive, designed in adaptable units for maximum flexibility and responsiveness.[198] This could involve using risk-based tiers, sector-specific rules, "plug-and-play" modules, regulatory sandboxes, and regular reviews, drawing on frameworks like PAGIT which adapts governance to Technology Readiness Levels (TRLs).[199] Such a system can be operationalized through tools like a Quantum Technology Quality Management System (QT-QMS), which translates high-level principles into auditable, practical workflows.[200]

A particularly innovative set of complementary mechanisms may be needed to address the core challenge of technology progressing faster than traditional legal and regulatory cycles. To this end, we propose the "Quantum-Agentic Stewardship" framework, a novel model of technological stewardship designed for the quantum-AI era. This framework is composed of three interlocking concepts:

First is the idea of Quantum-Resistant Constitutional AI. This involves embedding core, fundamental principles—derived from sources like the EU Charter of Fundamental Rights[201]— directly into the operational architecture of an advanced AI system. This is not merely a high-level policy goal but a technical implementation where the AI's decision-making processes are constrained by a hard-coded constitution. To be effective against the threats posed by quantum computing, this entire framework must be secured with post-quantum cryptography, making it resistant to both classical and quantum-based attacks.[202]

Second, to give this technical architecture legal force, we propose the imposition of a legally enforceable fiduciary duty upon these " Quantum-Agentic Steward" systems. This represents a significant evolution of agency law. By designating these advanced AI systems as fiduciaries, the law would obligate them to act with undivided loyalty and in the best interests of their human beneficiaries, ensuring their operations remain aligned with the embedded constitutional principles.[203] This creates a powerful legal hook for accountability, moving beyond mere technical compliance to a legally recognized duty of care, grounded in established legal principles.[204]

---

[197] *See* Digital Regulation Cooperation Forum (DRCF), https://www.drcf.org.uk/.
[198] *See* Henry T. Greely, Governing Emerging Technologies—Looking Forward with Horizon Scanning and Looking Back with Technology Audits, Glob. Pub. Pol'y & Governance, (2022) https://doi.org/10.1007/s43508-022-00045-y.
[199] Tait, Joyce & Banda, Geoffrey & Watkins, Andrew. (2017). Proportionate and Adaptive Governance of Innovative Technologies (PAGIT): A Framework to Guide Policy and Regulatory Decision Making, http://www.innogen.ac.uk/sites/default/files/2019-04/PAGIT%20FrameworkReport-Final_170717.pdf
[200] Aboy *et al., supra* note 34.
[201] Charter of Fundamental Rights of the European Union, 2012 O.J. (C 326) 391.
[202] *See* NIST, *supra* note 157.

Third, the operational enforcement of this duty would be carried out through algorithmic regulation.[205] This is the dynamic component of the framework, where the " Quantum-Agentic Steward" system actively monitors and steers quantum processes in near real-time. It would function as an automated, embedded supervisor, ensuring that all operations adhere to the system's constitutional mandate and its fiduciary obligations. This model of internal, automated oversight is a necessary response to the "quantum-AI control problem"—the recognition that external human oversight is unfeasible at the speed and complexity at which these integrated systems will operate.[206] This technological stewardship model, however, complicates the quantum-AI control problem, as endowing agents with such supervisory capabilities and hybrid classical-quantum reasoning could create new, unforeseen risks if not perfectly aligned with human values and goals.

On a more ambitious scale, the EU should explore the feasibility of a "CERN for Quantum/AI," a large-scale international research center dedicated to quantum and AI collaboration that promotes open science and shared infrastructure. This leads to the ultimate goal of a Global Non-Proliferation Treaty for Quantum/AI. The EU should proactively initiate discussions towards a global treaty inspired by NPT/IAEA principles—preventing misuse, promoting peaceful use, and ensuring transparency, accountability, verification, and enforcement—to manage global security risks and prevent a quantum arms race. Establishing such norms proactively is of the utmost importance given the likelihood for irreversible technological trajectories associated with crossing the 'quantum event horizon'. International organizations such as UNESCO and the OECD can play a vital role in facilitating this dialogue, building consensus on ethical principles, and laying the groundwork for such agreements.[207] Consistent with the 'Brussels Effect' observed in other domains[208], a well-defined EU Quantum Act has the capacity to significantly shape global norms for responsible quantum technology. The ultimate institutional embodiment of such a treaty could be an 'Atomic Agency for Quantum-AI,' or 'International Quantum Agency (IQA)' modeled on the IAEA, to oversee compliance and manage global non-proliferation risks.[209]

---

[203] *See* Lynn M. LoPucki, *Algorithmic Entities*, 95 Wash. U. L. Rev. 887 (2018), https://openscholarship.wustl.edu/law_lawreview/vol95/iss4/7/; and Jack M. Balkin, *The Three Laws of Robotics in the Age of Big Data*, 78 Ohio St. L.J. 1217 (2017), https://openyls.law.yale.edu/entities/publication/7e7ed54c-9dfe-42b1-be8f-15ad8c083958.
[204] *See* e.g., Restatement (Third) of Agency § 8.01 (Am. L. Inst. 2006).
[205] *See* Karen Yeung, *Algorithmic Regulation: A Critical Interrogation*, 2 Reg. & Governance 1 (2018).
[206] *See* Nick Bostrom, Superintelligence: Paths, Dangers, Strategies (2014).
[207] *See* OECD (2025), "A quantum technologies policy primer", OECD Digital Economy Papers, No. 371, OECD Publishing, Paris, https://doi.org/10.1787/fd1153c3-en; and OECD, Recommendation of the Council on Artificial Intelligence (2019, rev. 2024).
[208] Anu Bradford, The Brussels Effect: How the European Union Rules the World (2020), https://scholarship.law.columbia.edu/books/232/
[209] Kop, *supra* note 139.

Establishing a robust governance structure is paramount for effectively managing quantum technologies. This section has explored various models and mechanisms, advocating for a dedicated EU body (OQTA) complemented by adaptive regulatory tools, strong international coordination facilitated by organizations like UNESCO and OECD, and ambitious collaborative projects. Such a structure aims to provide the necessary expertise, oversight, and flexibility to navigate the complexities of the quantum era responsibly.

## 10. RECOMMENDATIONS FOR A SOCIETALLY BENEFICIAL QUANTUM ECOSYSTEM

Building upon the analysis of regulatory precedents and governance structures, this section presents specific recommendations aimed at fostering a quantum ecosystem within the EU that is both innovative and societally beneficial. It outlines targeted actions, obligations, and incentives for key stakeholder groups. These recommendations are structured according to the Quadruple Helix innovation model, which emphasizes the critical interplay between academia (university), industry, government (policymakers), and the public (civil society and end-users).[210] The successful implementation of these recommendations is essential for translating the framework of the EU Quantum Act into tangible positive outcomes.

To cultivate a thriving and responsible quantum ecosystem, specific actions, obligations, and incentives should target key stakeholder groups.

### 10.1. For Academia (Research Institutions & Researchers)

The academic community serves as the foundation of the quantum ecosystem and thus bears significant responsibilities. Evident obligations for research institutions and researchers should include conducting fundamental and applied research that adheres to the highest ethical standards and the principles outlined in this Act. They must also implement and comply with rigorous dual-use risk assessment protocols, for example mandated or guided by the OQTA, and ensure transparency in research methodologies and findings where appropriate, balancing openness with security considerations. Moreover, academia has an obligation to actively participate in public engagement initiatives and integrate Ethical, Legal, Societal, and Policy Implications (ELSPI) into both research projects and curricula.[211]

To encourage these behaviors, a system of clear incentives is required. This should include access to increased and sustained public funding through mechanisms like Horizon Europe and a dedicated EU Quantum Fund, with eligibility explicitly linked to adherence to responsible research principles. Compliant research groups should be granted preferential access to shared EU quantum infrastructure, such as testbeds and computing resources. Additionally, the EU should provide targeted grants for interdisciplinary projects focusing on quantum technologies and their ELSPI aspects, and establish awards, recognition, and career incentives for researchers and institutions demonstrating leadership in responsible quantum innovation.

---

[210] *See* Florian Schütz, Marie Lena Heidingsfelder & Martina Schraudner, *Co-shaping the Future in Quadruple Helix Innovation Systems: Uncovering Public Preferences toward Participatory Research and Innovation*, 5 SHE JI: J. DESIGN, ECON. & INNOVATION 128 (2019), https://doi.org/10.1016/j.sheji.2019.04.002.
[211] Kop, M. Quantum-ELSPI: A Novel Field of Research. *DISO* **2**, 20 (2023). https://doi.org/10.1007/s44206-023-00050-6

## 10.2. For Policymakers (EU Institutions & Member States)

Policymakers are responsible for creating and maintaining the regulatory environment. Their core obligations must include ensuring the consistent and effective implementation and enforcement of the EU Quantum Act across the Union, and guaranteeing the independence, resources, and mandate of the OQTA and national competent authorities. They are also obligated to conduct regular reviews and updates of the Act to maintain its relevance, and to actively engage in international cooperation on quantum governance, standards, and non-proliferation, while promoting public awareness and democratic debate. To avoid the regulatory fragmentation and public backlash that plagued biotechnology, the Act must embed mechanisms for mandatory, broad public and multi-stakeholder engagement in its governance structures.

While policymakers are driven by policy goals rather than direct incentives, the successful implementation of the Act offers significant rewards. These include strengthening EU technological leadership, competitiveness, and strategic autonomy; enhancing national and economic security through responsible risk management; increasing public trust in the governance of emerging technologies; and realizing the societal benefits of quantum applications in key sectors.

To achieve these goals, policymakers should undertake several specific actions. Akin to the AI literacy provisions in the EU AI Act, and heeding calls from scholars, the Quantum Act should include measures to promote a general understanding of quantum technologies among the public, policymakers, and civil society.[212] This includes actively promoting quantum literacy within regulatory, judicial, and administrative bodies to ensure informed oversight. Policymakers must also develop innovation-friendly immigration policies to attract and retain top global quantum talent and support EU-level initiatives for building a diverse quantum workforce. Lastly, they must plan for the future by addressing the immense energy demands of quantum-AI systems, which involves strengthening the power grid and investing in giga-watt scale energy facilities, including modern nuclear reactors.

## 10.3. For Industry (Developers, Providers, Deployers)

The industry stakeholders who develop, provide, and deploy quantum technologies have a direct responsibility for their products' impact. Their obligations must include strict compliance with the Act's risk-based requirements, particularly for high-risk systems, which involves rigorous risk and conformity assessments and post-market monitoring. Where required, they must conduct Quantum Impact Assessments (QIAs) to evaluate broader societal impacts[213], and adhere to all established EU and international standards and certification schemes. Industry must also maintain transparency regarding system capabilities, limitations, and data usage, implement robust cybersecurity measures, and cooperate fully with the OQTA and national authorities.

---

[212] *See* e.g., T. Dekker, F. Martin-Bariteau, *Can. J. Law Technol.* **20**, 179 (2023).

In return for meeting these obligations, industry should be offered a compelling set of incentives. These include access to dedicated EU funding mechanisms, such as an EU Quantum Fund and the EIC, and potentially streamlined state aid for strategic projects. Industry should be given opportunities to participate in and shape standardization processes, influencing market development, and gain access to shared EU research and testing infrastructure. An incentive could be the creation of an "EU Quantum Trust Mark" or certification that signifies compliance, thereby enhancing market access, competitiveness, and user trust.[214] These measures thereby provide increased legal certainty and a predictable regulatory environment, and could be complemented by facilitating participation in patent pools or collaborative licensing initiatives under fair, reasonable, and non-discriminatory (FRAND) terms.[215]

## 10.4. For End Users (Citizens, Businesses, Public Sector)

Finally, end-users are not merely passive recipients but active participants in the ecosystem. Their obligations include utilizing quantum technologies responsibly and in accordance with applicable laws, and reporting suspected non-compliance, security vulnerabilities, or harmful incidents through designated channels, which should include robust whistleblower protections. The incentives for end-users are rooted in the benefits of a well-regulated ecosystem. These include access to safe, reliable, and secure quantum applications and services; enhanced data security and privacy through the promotion of quantum-safe cryptography; and meaningful opportunities for participation in public consultations. End-users should also have access to clear information about the quantum systems they interact with and effective redress mechanisms in case of harm. The overarching incentive is the promotion of equitable access to the benefits derived from quantum technologies.

In essence, fostering a thriving quantum ecosystem requires a concerted effort from all stakeholders. This section has outlined responsibilities and incentives for academia, policymakers, industry, and end-users, aiming to align their actions with the overarching goals of responsible innovation, societal benefit, and European leadership in the quantum domain. By structuring these recommendations within the Quadruple Helix model, we emphasize that a truly successful quantum strategy must be co-created through the dynamic interaction of all four societal pillars. The successful engagement of all these actors will be vital for the effective implementation of the EU Quantum Act.

---

[213] *See* e.g., Quantum Delta N.L., *Quantum Delta NL launches Exploratory Quantum Technology Assessment (EQTA)* (Apr. 13, 2023), https://quantumdelta.nl/news/quantum-delta-nl-launches-exploratory-quantum-technology-assessment-eqta.

[214] For recommendations to establish trusted and certified Quantum Technologies, *see* Kop, M., *Establishing a Legal-Ethical Framework for Quantum Technology*, Yale Journal of Law & Technology, The Record, (Mar. 30, 2021), https://yjolt.org/blog/establishing-legal-ethical-framework-quantum-technology.

[215] On FRAND terms, *see* Layne-Farrar, Anne; Padilla, A. Jorge; Schmalensee, Richard (2007), *Pricing Patents for Licensing in Standard-Setting Organizations: Making Sense of FRAND Commitments*, Antitrust Law Journal. **74**: 671, https://heinonline.org/HOL/LandingPage?collection=journals&handle=hein.journals/antil74&div=23&id=&page=

## 11. STRATEGIC INSIGHTS FROM U.S. AND CHINESE AI PLANS: A COMPARATIVE ANALYSIS

While the EU QA rightly draws lessons from semiconductor and AI-specific legislation, it is also instructive to analyze the broader strategic posture of major geopolitical actors. The recently unveiled U.S. "Winning the AI Race: America's AI Action Plan", especially when viewed in conjunction with its parallel suite of Quantum legislative proposals, presents a significant strategic document whose principles, ambitions, and identified shortcomings offer valuable lessons for the proposed European Quantum Act.[216] This combined U.S. strategy can be understood through the analogy of an "American Digital Silk Road"—a coordinated, multi-technology initiative aimed at establishing a U.S.-led techno-economic sphere of influence. While this "road" is fundamentally different from China's state-directed model, its ambition to project power through technology, standards, and alliances, and its underlying philosophy—prioritizing rapid innovation, deregulation, and aggressive international diplomacy to establish technological leadership offers a salient point of comparison.

Integrating insights from this analogy and contrasting it with China's recently published "AI+ Plan" can help the EU QA to preemptively address potential weaknesses, sharpen its unique value proposition, and enhance its overall strategic coherence.[217] This analysis outlines key areas where the U.S. and Chinese AI plans can inform the development of the EU QA, focusing on strengthening its industrial policy, refining its regulatory model, and ensuring its geopolitical resilience.

### 11.1. Adopting and Refining the "Full-Stack" Ecosystem Approach

A lauded feature of the U.S. plan is its comprehensive "full-stack" approach, which views the entire AI ecosystem as a single, integrated strategic asset.[218] This perspective spans from foundational elements like energy and data centers—whose permitting processes are to be accelerated via Executive Order[219]—to semiconductors, the talent pipeline, and cybersecurity.

---

[216] *See* America's AI Action Plan, AI.GOV, https://www.ai.gov/action-plan.
[217] *See* e.g., *State Council released Opinions on deepening the implementation of the Artificial Intelligence Plus Action (Guofa [2025] No. 11)*, DIGITAL POLICY ALERT (Aug. 26, 2025), https://digitalpolicyalert.org/event/33043-state-council-released-opinions-on-deepening-the-implementation-of-the-artificial-intelligence-plus-action-guofa-2025-no-11 and https://www.gov.cn/zhengce/content/202508/content_7037861.htm
[218] *White House Unveils America's AI Action Plan*, THE WHITE HOUSE (July 2025), https://www.whitehouse.gov/articles/2025/07/white-house-unveils-americas-ai-action-plan/.
[219] Exec. Order on Accelerating Federal Permitting of Data Center Infrastructure (July 2025), https://www.whitehouse.gov/presidential-actions/2025/07/accelerating-federal-permitting-of-data-center-infrastructure/.

The U.S. strategy effectively weaponizes this full stack as the primary export of its own "Digital Silk Road." The EU QA's proposed two-pillar structure, combining regulation with a "Chips Act-style" industrial policy, is already aligned with this thinking. However, the U.S. plan's explicit articulation of this concept is a powerful narrative tool. It is therefore recommended that the EU QA formally adopt and even expand upon the "full-stack" terminology. It should explicitly frame its industrial policy as a "Full-Stack Quantum Sovereignty" initiative. This would involve mapping and securing the entire value chain, encompassing upstream elements like critical raw materials and specialized equipment; midstream components such as semiconductor fabrication and data infrastructure; downstream applications including quantum software and algorithm development; and cross-stream necessities like cybersecurity, harmonized standards, and robust intellectual property frameworks.

11.2. Counter-Posing a European Model of International Diplomacy

This "full-stack" concept extends into the plan's novel and ambitious model of international diplomacy. Through a dedicated Executive Order, the strategy promotes the export of "full-stack AI export packages" to U.S. allies and partners, aiming to build an "AI alliance" based on American values and standards to counter Chinese influence.[220] This represents a highly assertive form of technological diplomacy and is the primary mechanism for building the "American Digital Silk Road." It aims to create long-term strategic dependencies where partner nations are reliant on American technology and security guarantees.

The EU, with a potential "Qubits for Peace" initiative and focus on a rules-based international order, can offer a compelling alternative. The EU QA should develop its own version of a "full-stack export" model, framed not just as a commercial or geopolitical tool, but as a "Quantum for Global Good" package. This package would be offered to partners and developing nations, bundling access to European quantum hardware and cloud platforms with the export of the EU's regulatory and ethical frameworks, the promotion of harmonized standards, and explicit commitments to human rights and the rule of law. This approach provides a clear alternative to both the assertive U.S. diplomatic style and China's model, which, while framed around UN-centric global cooperation and support for the Global South, operates within a broader state-led strategic context. This tactic directly leverages the EU's strength as a regulatory superpower and aligns with the goals of the World Bank, addressing a noted criticism of the U.S. plan's lack of ambitious financial solutions for the Global South.[221] The shared goal of the transatlantic tech alliance is to ensure democratic values prevail. A technologically sovereign EU is a more valuable ally to the US than a dependent one. This approach can attract nations that might be wary of the more assertive U.S. diplomatic style, thereby strengthening the overall position of the democratic bloc. Ultimately, a technologically sovereign EU is a more valuable ally to the

---

[220] Exec. Order on Promoting the Export of the American AI Technology Stack (July 2025), https://www.whitehouse.gov/presidential-actions/2025/07/promoting-the-export-of-the-american-ai-technology-stack/; and Nat'l Sec. Presidential Memorandum on United States Government-Supported Research and Development National Security Policy, NSPM-33 (Jan. 14, 2021).
[221] *Experts React: Unpacking the Trump Administration's Plan to Win the AI Race*, CTR. FOR STRATEGIC & INT'L STUD. (July 2025), https://www.csis.org/analysis/experts-react-unpacking-trump-administrations-plan-win-ai-race.

United States than a dependent one, creating a stronger, more resilient transatlantic partnership capable of jointly setting the global rules for the quantum age.[222]

Framing the Act around the SDGs provides a positive, universally understood, and globally resonant mission, shifting the narrative from merely managing risk to proactively steering technology for the common good. This would also strengthen the EU QA's proposed "Quantum for Global Good" diplomatic package, giving it substantive grounding and a clear set of internationally recognized targets. Furthermore, the Open Quantum Institute (OQI) at CERN provides a successful, operational model for a multi-stakeholder initiative focused on quantum for the SDGs, offering a blueprint the EU can learn from and scale.[223]

### 11.3. Learning from the U.S. Plan's Incoherencies and Challenges

The "American Digital Silk Road" analogy is not without its limitations, which can reveal strategic opportunities for the EU. Unlike China's state-directed model, the U.S. approach is fundamentally market-driven and reliant on its private sector, which introduces dynamism but also fragmentation. More importantly, experts have identified significant execution challenges and internal policy contradictions within the U.S. plan.[224] The EU QA can be designed to avoid these pitfalls from the outset. The U.S. plan's pro-innovation goals, driven by a domestic agenda that advocates for cutting regulations, may clash with its restrictive trade and immigration policies. This deregulatory push, which is also accompanied by distinct ideological directives such as an Executive Order aimed at "Preventing Woke AI in the Federal Government,"[225] raises concerns about federal overreach and the erosion of civil liberties protections. To avoid these issues, the EU QA must be explicitly designed for maximum coherence with other cornerstone EU policies, including the Digital Single Market, trade agreements, and talent mobility initiatives. The legislative text should include provisions that mandate alignment to prevent internal contradictions. Additionally, the EU QA should position its "modular and two-pillar framework" as a superior alternative, arguing that a predictable, risk-based regulatory environment—which lets innovation breathe within clear ethical guardrails[226]—is more conducive to sustainable, long-term investment and public trust than a volatile, deregulated market. Lastly, acknowledging the criticism that the U.S. plan may lack sufficient funding, the EU QA's industrial policy component must be backed by a concrete, substantial, and long-term financial commitment, similar in scale and ambition to the EU Chips Act.

### 11.4. Incorporating Robust Benchmarking for Security and Societal Impact

---

[222] *See* Kop, *supra* note 28 (Grand Strategy).
[223] *See* GESDA OQI Report 2024, *supra* note 42.
[224] *ibid*.
[225] Exec. Order on Preventing Woke AI in the Federal Government (July 2025), https://www.whitehouse.gov/presidential-actions/2025/07/preventing-woke-ai-in-the-federal-government/.
[226] *See generally* Urs Gasser & Viktor Mayer-Schönberger, Guardrails: Guiding Human Decisions in the Age of AI (2024), https://press.princeton.edu/books/hardcover/9780691150680/guardrails

A well-documented omission in the U.S. plan is the lack of benchmarks and testbeds for evaluating AI models in high-stakes foreign policy and national security scenarios.[227] This is a major gap that the EU can fill, reinforcing its focus on safety, security, and societal benefit. The EU QA should mandate the creation of European Quantum Testbeds and Benchmarks with a dual focus. The first is assessing technical performance of quantum computers, sensors, and networks against standardized metrics. The second is ensuring societal and security resilience by developing use cases and simulations to test quantum systems in high-stakes scenarios, including modeling their impact on critical infrastructure, financial markets, and national security systems, as well as their capacity for misuse. This directly addresses the U.S. plan's oversight and aligns with the EU QA's risk-based model.

The U.S. AI Action Plan and its associated quantum initiatives, when analyzed through the "American Digital Silk Road" analogy, alongside the contrasting strategic logic of China's "AI+ Plan," provide an invaluable strategic tool for refining the European Quantum Act. By strategically adopting concepts like the "full-stack" ecosystem, offering a values-driven alternative for international technology diplomacy, designing for policy coherence to avoid U.S.-style contradictions, and pioneering robust security and societal benchmarking, the EU can position its Quantum Act as a world-leading model for governing transformative technologies. This plan ensures that the EU does not simply react to U.S. or Chinese initiatives but proactively defines a competitive, resilient, and ethically grounded path to quantum leadership.

## 12. NEW PHILOSOPHIES FOR THE QUANTUM-AI ERA

The emergence of quantum-AI technologies represents more than mere technological progression; it signals a profound philosophical inflection point. The principles of quantum mechanics do not just enable new forms of computation; they fundamentally challenge the classical, deterministic worldview that has shaped Western thought, and by extension its legal and ethical systems, for centuries. Effectively and responsibly navigating this new era requires not only novel regulatory tools but also new philosophical perspectives on reality, knowledge, and our relationship with technology. The sheer productivity and autonomous oversight capabilities of the 'agentic stewards' discussed previously would challenge the foundational assumptions of market capitalism, necessitating new philosophical frameworks.

First, the probabilistic and uncertain nature of quantum phenomena calls for an epistemology of humility. Classical governance often operates on an assumption of predictability and control, a tradition stretching from Plato's ideal forms[228] to the rationalism of Kant.[229] The quantum realm, however, as described by Heisenberg's uncertainty principle, is governed by probabilities, not certainties.[230] This requires a shift in our worldview on regulation, moving away from rigid, deterministic rules towards adaptive systems that acknowledge the inherent

---

[227] *Experts React: Unpacking the Trump Administration's Plan to Win the AI Race*, supra note 220.
[228] Plato, The Republic (c. 375 BCE).
[229] Immanuel Kant, Critique of Pure Reason (1781).
[230] Heisenberg, *supra* note 15.

Page 54 of 80

limits of our predictive capabilities. This philosophical stance underpins the practical need for mechanisms like regular reviews, regulatory sandboxes, and the dynamic, data-driven oversight envisioned for the OQTA. The intricate, self-similar patterns of fractal geometry, recently discovered in quantum materials, can be seen as a metaphor for this new reality—a "secret code of creation" that reveals order within chaos, operating on principles that extend beyond simple linear logic.[231]

Second, the convergence of quantum-AI with the promise of technological abundance may necessitate a transition toward a post-capitalist, post-scarcity economy. As technologies begin to eradicate scarcity in sector after sector, the foundational assumption of economics—and the legal systems like intellectual property built upon it—is eroded. This creates a tension between technological potential and social reality, a 'poverty paradox' where abundance for some coexists with scarcity for others. A new framework of distributive justice is needed, moving beyond classical liberalism. A post-Rawlsian principle such as Equal Relative Abundance (ERA) offers a path forward. ERA would combine Rawls's focus on improving the position of the least advantaged with desert-based principles, allowing for unequal rewards for unequal contributions only to the extent that they raise the baseline for all of society, ensuring that the fruits of technological progress are distributed equitably.[232] In such an economy, the primary challenge shifts from production to equitable distribution and the cultivation of post-materialist values like solidarity, creativity, and self-actualization, echoing the ideas of thinkers from Marx to Maslow.[233]

Third, the phenomenon of entanglement, or non-locality, suggests a deeply interconnected reality that has direct implications for morality.[234] Traditional Western ethics, from Aristotle's focus on individual virtue to Kant's emphasis on autonomous will, often centers on the individual as the primary moral agent. Entanglement, however, provides a powerful metaphor for a more relational or networked ethics, where responsibilities, actions, and consequences are understood as distributed across systems. This new ethical framework, inspired by the non-local, interdependent nature of quantum mechanics, models moral duties not as isolated obligations but as interconnected functions within a broader system. It supports the development of legal doctrines like 'entangled liability', where accountability for a harm caused by a complex, multi-agent system is shared among the operators of its constituent parts, reflecting a physical reality that classical individualism cannot capture.[235] This perspective supports the development of legal doctrines like 'entangled liability', where accountability is shared among the operators of interconnected systems, reflecting a physical reality that classical individualism cannot capture. On a deeper level, it invites us to consider that we may all be entangled in a cosmic sense, a notion that resonates with some interpretations of quantum mechanics that suggest consciousness itself may have quantum properties, and that we are all "walking quantum computers."[236] This aligns with Floridi's concept of the 'infosphere', an informational environment where we exist as interconnected 'inforgs' (informational organisms), and where our moral actions are defined by their contribution to the health of this shared space.[237]

---

[231] *See* also Christopher R. Ast et al., *Sensing the Quantum Limit in Scanning Tunneling Microscopy*, 10 Nature Commc'ns 2882 (2019), https://pmc.ncbi.nlm.nih.gov/articles/PMC5059741/.
[232] Rawls, *supra* note 40.
[233] Karl Marx, Das Kapital (1867); and Abraham H. Maslow, *A Theory of Human Motivation*, 50 Psych. Rev. 370 (1943).
[234] Einstein, Podolski & Rosen, s*upra* note 19.
[235] *See* e.g., David Vladeck, *Machines Without Principals: Liability Rules and Artificial Intelligence*, 89 Wash. L. Rev. 117 (2014), https://digitalcommons.law.uw.edu/wlr/vol89/iss1/6/.

Finally, the prospect of advanced quantum-AI agents necessitates a pragmatic form of post-humanist governance.[238] As we delegate increasingly complex cognitive tasks to these systems, the traditional distinction between the human user and the technological 'tool' begins to erode. The concept of designating advanced AI systems as 'quantum-agentic stewards' with legally binding fiduciary duties is a direct response to this shift. It moves beyond seeing technology as a passive object to be regulated and instead conceptualizes it as an active participant in the governance ecosystem, with legally encoded obligations of care, loyalty, and alignment with fundamental human values.[239] This does not grant such systems personhood, but rather establishes a new category of technological stewardship, essential for maintaining meaningful human control in an era of unprecedented computational power. The EU Quantum Act, by incorporating these forward-looking concepts, can therefore do more than regulate a new technology; it can pioneer a philosophical and governance paradigm fit for the complexities of the 21st century.

It must be emphasized, however, that many of these more profound legal and philosophical challenges will only become fully salient with the advent of universal, fault-tolerant quantum-classical supercomputers. At that future stage, advanced concepts such as the *Quantum Lex Machina*—the expression of law in algorithmic, quantum-native forms—may become necessary. To prepare for such a reality, it may be prudent to begin developing in multidisciplinary settings a set of foundational constitutional principles, a 'Quantum Constitution for the Quantum Age,' to ensure that these powerful future systems remain aligned with fundamental EU values.

## 13. CONCLUSION: CHARTING A COURSE FOR RESPONSIBLE QUANTUM INNOVATION IN THE EU

An adaptive, forward-looking EU Quantum Act, responding to strategic imperatives such as the Quantum Europe Strategy, is central to navigating the complex quantum landscape. The unprecedented capabilities and risks stemming from quantum mechanics' non-classical properties—including superposition, entanglement, and tunneling—defy the intuitive assumptions underpinning existing legal paradigms, demanding a *sui generis* approach, contributing to the emerging *lex specialis* for quantum information technologies. The EU

---

[236] *See* e.g., Der Derian, J., Project Q: War, Peace and Quantum Mechanics, Bullfrog Films (2020), https://www.imdb.com/title/tt8574978/.

[237] Luciano Floridi, The Fourth Revolution: How the Infosphere Is Reshaping Human Reality (2014), https://philpapers.org/rec/FLOTFR-3.

[238] *See* e.g., N. Katherine Hayles, How We Became Posthuman: Virtual Bodies in Cybernetics, Literature, and Informatics (1999).

[239] *See* Lopucki, *supra* note 202; and Balkin, *supra* note 202.

Quantum Act thus represents a strategic opportunity to foster innovation and competitiveness while ensuring societal benefit and safeguarding against inherent dual-use risks. By drawing lessons from semiconductor and AI regulation, learning from the long-standing experience of nuclear governance—including its collaborative models like fusion research—and analyzing global innovation dynamics, the EU can chart a responsible and effective course.

To achieve this, the Act must be constructed as a robust two-pillar instrument. The first pillar, an ambitious industrial and security policy inspired by the EU and US Chips Acts, is a prerequisite for technological sovereignty. Strategic investment in research, supply chains, and a "lab-to-market" pipeline—potentially accelerated by a DARPA-style agency—is a responsible act of deterrence and a necessary foundation for the EU to become an indispensable partner in a transatlantic tech alliance, ensuring democratic values inform the next technological era.

The second pillar is a New Legislative Framework (NLF)-style regulatory structure designed for responsible innovation. This pillar must combine risk-based rules with overarching principles, operationalized through key mechanisms proposed herein: a "standards-first" philosophy embedding fundamental rights through certifiable Quantum Technology Quality Management Systems (QT-QMS); a new legal duty of 'Anticipatory Data Stewardship' to mitigate the 'Q-Day' cybersecurity threat; and novel intellectual property rights to protect foundational algorithms under balanced FRAND terms. This hybrid, modular strategy ensures that governance is adaptive across the technology's lifecycle.

The "quantum event horizon" metaphor serves as a powerful heuristic, emphasizing the deep uncertainty and risk of technological lock-in that necessitates this proactive framework. It underscores the urgency of establishing a robust framework now to navigate the uncharted territory of the quantum age and ensure that its benefits are realized responsibly and equitably across the European Union and the globe. This effort must be guided by a dedicated EU Office of Quantum Technology Assessment (OQTA) providing expert oversight. As the technology matures, this framework must be prepared to incorporate forward-looking paradigms, such as 'algorithmic regulation' and the imposition of a fiduciary duty upon advanced AI systems to act as 'quantum-agentic stewards'.

Such a domestic framework must be a component of a holistic global quantum governance strategy that integrates international treaties, export controls, standards, and regulations. This framework must be supported by sustained investment in research, infrastructure, and education, alongside quantum literacy initiatives, policies that cultivate a skilled workforce, and carefully designed intellectual property and competition rules that encourage open innovation where beneficial. On the international stage, the EU should champion collaboration on standards and security, culminating in a global non-proliferation framework for quantum and AI weapons of mass destruction, inspired by the IAEA/NPT model. This vision, encapsulated by the "Qubits for Peace" initiative and overseen by a proposed 'International Quantum Agency (IQA)', seeks to ensure quantum technologies are developed safely and ethically. By implementing the EU Quantum Act, the European Union can position itself as a global leader, driving prosperity and harnessing the immense promise of quantum technologies for the benefit of all humanity.

## 14. POTENTIAL OUTLINE: EU QUANTUM ACT (EU QA) WITH EXPLANATIONS

This final section synthesizes the preceding analyses into a potential structural outline for the EU Quantum Act. It organizes the key legislative components discussed – including general provisions, the regulatory framework, innovation support, intellectual property strategy, standardization, governance, and international dimensions – into a coherent blueprint. This outline serves as a concrete proposal, integrating the diverse insights gathered and providing a detailed overview of the recommended content and structure for the Act.

### I. Preamble & General Provisions

- **Rationale:** Articulate the transformative promise of quantum technologies (computing, sensing, networking, QAI hybrids) across various sectors (finance, MedTech, energy, logistics, space, defense) and the need for a harmonized EU framework to foster innovation while managing risks, including their dual-use nature. Reference lessons learned from the governance of AI, nanotechnology, and nuclear energy.

    - *Explanation: This section sets the stage, highlighting the significant opportunities presented by quantum technologies, alongside the inherent risks (especially dual-use capabilities and cybersecurity threats) that necessitate proactive, harmonized EU action based on past experiences with transformative technologies. The structured, codified approach inherent in the EU legal tradition, influenced by its Roman Law roots, provides a foundation for the Act's design.*

- **Objectives:**
    - Promote a world-leading, innovative, and competitive quantum ecosystem within the EU.

    - Ensure the development and deployment of quantum technologies are societally beneficial, ethical, and aligned with fundamental EU values. A core objective shall be to direct and incentivize the development of quantum technologies towards addressing global challenges as articulated in the United Nations Sustainable Development Goals (SDGs).

    - Establish legal certainty and a harmonized regulatory environment across all Member States.

    - Safeguard against the risks associated with quantum technologies, particularly their significant dual-use ability.

    - Enhance the EU's strategic autonomy and technological sovereignty in the critical field of quantum technology.

    - Foster international collaboration with like-minded partners while strategically

protecting core EU interests.

- ○ ***Explanation:*** *The Act pursues multiple, interconnected goals: driving innovation to ensure global competitiveness, embedding ethical considerations and societal benefit, managing dual-use risks effectively, achieving strategic autonomy in a designated technological area, fostering beneficial international partnerships, and establishing clear, effective governance structures.*

- **Scope:** Define the Act's scope to cover quantum computing, sensing, networking, quantum-AI hybrids, and the strategically important supply chain of critical minerals needed for quantum device fabrication.

    - ○ ***Explanation:*** *The Act requires a broad scope to encompass the main pillars of quantum technology identified as transformative, while also addressing the dependencies related to critical raw materials needed for hardware development, mirroring concerns in other strategic technology sectors.*

- **Definitions:** Provide clear and precise definitions for key terms related to quantum technologies (e.g., 'quantum system', 'quantum computer', 'quantum sensor', 'quantum network', 'QAI hybrid') and essential regulatory concepts (e.g., 'high-risk application', 'dual-use', 'quantum benchmark', 'conformity assessment').

    - ○ ***Explanation:*** *Precise definitions are fundamental for ensuring legal clarity, consistent interpretation, and uniform application of the Act across diverse quantum fields and throughout all Member States, which is standard practice for effective legislation. Careful consideration must be given to terminology and its underlying physics being subject to constant debate by the scientific community, learning from debates in AI governance (e.g., Mökander & Floridi's analysis of 'AI systems' vs. technology-agnostic 'Automated Decision Systems') to ensure the definitions are both precise, and more future-proof (technology-neutral) against unforeseen technological shifts.*[240]

- **Territorial Scope & Extraterritorial Effect (Inspired by GDPR/AI Act):** Define the geographical application of the Act, explicitly addressing its potential extraterritorial reach.

    - ○ ***Explanation:*** *This provision clarifies that the Act applies primarily to providers and users of quantum technologies operating within the European Union. It also specifies the conditions under which the Act may extend its reach to actors based outside the EU, for instance, if they place quantum systems on the EU market, put them into service within the EU, specifically target EU users with their services, or if the output produced by their quantum system is used within the Union and affects individuals or entities therein. Such a methodology aims to*

---

[240] Mökander & Floridi, *supra* note 12.

*ensure a level playing field for businesses and comprehensive protection for EU citizens and interests, mirroring the established extraterritorial scope of regulations like the GDPR and the EU AI Act.*

- **Relationship with Existing Legislation:** Clarify the interaction between the EU Quantum Act and other relevant EU laws, including the AI Act, GDPR, NIS Directive, Medical Device Regulation (MDR), Machinery Directive/Regulation, financial regulations, and others, ensuring coherence and avoiding conflicts. Mention the role of coordination clauses or Commission guidance and the use of the functional equivalence principle where applicable.

  - ***Explanation:*** *Seamless integration into the existing EU legal landscape is essential. This requires careful articulation of how the Quantum Act interacts with other regulations, particularly concerning quantum-AI hybrids (vis-à-vis the AI Act), quantum applications in specific sectors (e.g., healthcare devices under MDR, industrial systems under Machinery Directive/Regulation), and data protection or cybersecurity aspects (vis-à-vis GDPR, NIS Directive, DORA), thereby preventing regulatory fragmentation and contradictory obligations. This section must mandate policy coherence across domains (e.g., trade, talent mobility, digital single market) to avoid the strategic contradictions and implementation challenges identified in other international technology strategies.*

II. Regulatory Framework: A Hybrid, Modular, and Lifecycle Approach

***Explanation:*** *This section outlines a regulatory approach that combines risk-based rules with guiding principles, designed to be modular and adaptive throughout the technology lifecycle (ex-ante, ex-durante, ex-post), incorporating a 'standards-first' philosophy where appropriate for early-stage governance, necessary for navigating the uncertainty and lock-in associated with a 'quantum event horizon'. To address the risk of technological acceleration outpacing legislative cycles, this Act envisions a dynamic regulatory model. Beyond periodic reviews, the OQTA shall be empowered to develop and oversee 'algorithmic regulation' mechanisms, where certain compliance checks and risk threshold adjustments can be updated in near real-time based on monitoring data, ensuring that the guardrails evolve at a pace commensurate with the technology itself. It also consciously attempts to balance regulatory specificity, needed for clarity and enforcement, with the flexibility required to adapt to quantum's rapid evolution, addressing the inherent trade-offs identified in comparative analyses of legislative styles.[241]*

A. Ex-Ante Regulation (Pre-Market & Development Phase)

***Explanation:*** *Focuses on measures applied before quantum systems are placed on the market or put into service, guiding development and ensuring baseline safety, ethical alignment, and*

---

[241] *See* Kop, *supra* note 139.

*compliance. This phase is instrumental for addressing the Collingridge dilemma by embedding values and standards early, before a 'quantum event horizon' limits intervention possibilities.*

- **Risk-Based Classification (Inspired by EU AI Act):** Establish distinct categories for quantum applications based on their potential risk levels based on the 'pyramid of criticality' (e.g., unacceptable, high, limited, minimal), considering factors such as societal impact, infringement on fundamental rights, safety implications, and dual-use capabilities. Define clear criteria for classifying specific quantum systems as high-risk (comparative analyses of AI governance confirm the relevance of such risk-tiered structures for emerging technologies, adaptable to the specific context).

    - *Explanation: This approach concentrates regulatory scrutiny and burden on the areas with the highest capacity for harm. It involves establishing risk categories tailored specifically to the unique capabilities and possible impacts of quantum technologies, while drawing inspiration from the classification criteria used in the EU AI Act for identifying high-risk applications in sensitive domains.*

- **Prohibited Practices (Inspired by EU AI Act):** Identify and explicitly prohibit specific quantum applications deemed to pose unacceptable risks because they inherently conflict with EU values, fundamental rights, or security interests. Instances could include certain types of quantum-enhanced surveillance or the use of quantum computers to break encryption vital for societal functions without authorization.

    - *Explanation: Similar to the EU AI Act's stance on unacceptable risk AI, this provision establishes clear red lines from the outset, banning quantum applications whose ability for harm fundamentally outweighs any conceivable benefit within the EU's legal and ethical framework.*

- **Obligations for High-Risk Quantum Systems (Pre-Market Aspects):** Mandate stringent requirements focused on the design, development, and pre-market validation phases for systems classified as high-risk. This includes rigorous risk assessment and mitigation design, ensuring data quality where applicable (e.g., for training QAI models), comprehensive technical documentation, and mandatory conformity assessments (leading to a CE marking) before market placement or putting into service.

    - *Explanation: These ex-ante obligations ensure that high-risk quantum systems are designed and validated to meet high standards of safety, ethical alignment, reliability, and performance before they can be introduced to the EU market, drawing on the robust pre-market requirements established by the EU AI Act.*

- **Principles-Based Regulation (Guiding Development):** Incorporate overarching principles for responsible quantum innovation to guide the development phase, allowing flexibility while embedding core values. Candidate principles include: Safety, Security, and Robustness; Transparency and Explainability; Fairness; Accountability and Governance; Contestability and Redress; Sustainability and Societal Benefit; Equitable Access; Privacy Protection; Human Agency and Oversight; Proactive Risk Management

& Dual-Use Mitigation; Fostering Collaboration; and Long-term thinking / Intergenerational equity. These principles can allow flexibility and application by sectoral regulators.

- ○ ***Explanation:*** *Complementing specific rules with high-level principles provides necessary flexibility for a rapidly evolving field like quantum technology. These principles serve as a normative compass during the development process, guiding researchers and developers towards ethically sound and societally beneficial innovation, ensuring key values are considered from the outset. The 10 Principles for Responsible Quantum Innovation (categorized as Safeguarding, Engaging, and Advancing - SEA) offer a concrete framework for operationalizing this approach.[242]*

- **Ethical Frameworks & Impact Assessment (Inspired by Genetics/AI):** Promote the use of specialized Quantum Ethics Boards, analogous to Institutional Review Boards (IRBs) in genetics, for reviewing projects with significant ethical or dual-use implications. Apply the precautionary principle where appropriate for deployments involving high uncertainty and severe risks. Mandate Quantum Impact Assessments (QIAs) for certain applications to evaluate broader societal, ethical, and legal impacts before widespread deployment.

  - ○ ***Explanation:*** *This integrates proactive ethical oversight mechanisms, drawing lessons from the governance of biosciences and AI, to anticipate and mitigate negative consequences before quantum technologies are widely adopted. The design of effective QIAs can benefit from experience gained with the Exploratory Quantum Technology Assessment, and by analyzing the requirements and implementation challenges of impact assessments mandated in other technology regulations, such as those for AI.[243]*

- **Modular Design & Sandboxes:** Utilize regulatory sandboxes as part of a modular design, allowing developers to test innovative quantum technologies in a controlled environment under regulatory supervision prior to seeking full market access.

  - ○ ***Explanation:*** *Sandboxes provide a contained space for experimentation and learning for both innovators and regulators, facilitating the development of appropriate rules for novel technologies without prematurely stifling innovation or exposing the public to undue risk.*

*B. Ex-Durante Regulation (During Operation/Use Phase)*

***Explanation:*** *Covers requirements and oversight applicable while quantum systems are in operation, ensuring continued safety, compliance, and responsible use.*

---

[242] Kop *et al., supra* note 32 (10 principles).
[243] Kop, *supra* note 44 (TTLF).

- **Obligations for High-Risk Quantum Systems (Operational Aspects):** Mandate requirements for the ongoing operation of high-risk systems. This includes ensuring continued accuracy, robustness, and cybersecurity against evolving threats; implementing effective human oversight mechanisms; maintaining comprehensive logging for traceability and auditing; and fulfilling transparency obligations towards users and affected parties during the system's use phase.

    - *Explanation: These obligations ensure that high-risk quantum systems remain safe, secure, reliable, and function as intended throughout their operational lifespan, addressing risks that may emerge or change after deployment.*

- **Principles-Based Regulation (Guiding Deployment & Use):** Reiterate the importance and applicability of the overarching principles (safety, fairness, transparency, accountability, etc.) during the actual deployment and use phases of quantum technologies, guiding operators and users in responsible practices.

    - *Explanation: The ethical and societal principles are not just relevant during development but must continue to guide how quantum technologies are implemented and used in real-world contexts.*

- **Ongoing Oversight:** Define the roles of the OQTA, the Regulatory Forum, and national competent authorities in the continuous monitoring of quantum systems in operation, overseeing compliance with ongoing obligations, and tracking market developments and emerging risks during the use phase.

    - *Explanation: Effective governance requires continuous oversight mechanisms to ensure that quantum technologies operate safely and responsibly after they have been placed on the market.*

C. Ex-Post Regulation (Post-Market Phase)

*Explanation: Addresses measures taken after systems are on the market, focusing on monitoring performance and incidents, establishing liability, and enforcing compliance.*

- **Post-Market Monitoring:** Establish systematic post-market monitoring requirements for providers, particularly for high-risk quantum systems, to collect data on real-world performance, identify unforeseen risks or failures, and inform necessary updates or corrective actions.

    - *Explanation: This ensures that the performance and safety of quantum systems are continuously tracked after deployment, allowing for timely intervention if problems arise.*

- **Centralized Registration Database:** Consider establishing a publicly accessible or authority-managed database for high-risk quantum systems placed on the EU market.

- **Explanation:** *Such a database would facilitate regulatory tracking and oversight, enhance market transparency for users and procurers, and support post-market surveillance activities.*

- **Liability Rules (Inspired by the proposed AI Liability Directive):** Adapt principles from the proposed AI Liability Directive—some of which were integrated into the final AI Act—to address non-contractual liability for damages caused by quantum systems. This could include rules facilitating claimant access to information (disclosure rules) and adjusting the burden of proof regarding causality in cases of non-compliance (presumption of causality). This should also pioneer novel legal doctrines such as 'probabilistic causation' to address harms from non-deterministic systems where the traditional 'but-for' test fails, and establish a 'Quantum Market Share Liability' model for systemic failures where fault cannot be traced to a single actor. It must also consider the implications of the legal status of advanced QAI systems, such as 'Quantum-Agentic Stewards' bound by a fiduciary duty, for liability allocation. To contend with the challenges of non-locality and probabilistic outcomes, this section could pioneer doctrines such as 'entangled liability', where operators of entangled systems share joint and several liability, or establish evidentiary presumptions based on statistical correlations where classical causation is impossible to prove. Consider the implications of the legal status of advanced QAI systems for liability allocation.

    - **Explanation:** *These rules aim to ensure that victims of harm caused by quantum systems have effective means of seeking redress, by addressing the specific challenges posed by the complexity and opacity of these technologies, drawing on tools developed for AI liability and innovating new legal concepts for the quantum era.*

- **A Regulation on the Governance of Autonomous Quantum Systems**

    - **Explanation:** *This Title would legally define the status of 'Quantum-Agentic Stewards' (QAS) and establish their fiduciary duties of care, loyalty, and prudence. It would mandate registration of high-risk quantum agents with the OQTA and set requirements for transparency, auditability, and meaningful human oversight.*

- **Enforcement Mechanisms:** Detail the enforcement actions available to national competent authorities and the OQTA, including the power to demand corrective actions, impose restrictions or withdrawals from the market, and apply significant, tiered penalties for non-compliance, proportionate to the severity of the infringement.

    - **Explanation:** *Robust enforcement mechanisms, including meaningful penalties, are essential to ensure that the rules of the Act are respected and to deter non-compliant behavior.*

III. Innovation, Funding, and Ecosystem Support (Inspired by US&EU Chips Acts)

- **Funding Mechanisms:** Establish significant and sustained public funding streams, combining EU-level programs like Horizon Europe with dedicated national contributions. Create a dedicated "EU Quantum Fund," similar to the Chips Fund, to facilitate access to finance for innovative startups and SMEs, leveraging private investment. Consider the use of tax incentives and loan guarantees, inspired by the US CHIPS Act, to encourage private R&D investment and support the development of capital-intensive quantum infrastructure. Streamline state aid rules and approval processes for strategic, "first-of-a-kind" quantum facilities and projects, learning from delays encountered with the Chips Act.

    - *Explanation: A multi-pronged funding strategy is needed to build a competitive EU quantum ecosystem, combining direct EU and national support, leveraging private capital through dedicated funds and financial instruments, and facilitating necessary state aid for high-cost, high-risk strategic projects while improving administrative efficiency.*

- **Innovation Strategy:** Support the full innovation cycle from fundamental research through applied R&D, prototyping, testing, and commercialization, emphasizing the "lab-to-market" transition. Invest strategically in shared quantum infrastructure, such as computing testbeds, secure networking facilities, specialized foundries, and advanced sensing platforms, ensuring accessibility across the EU. Foster strong collaboration between academia, research institutions, and industry through joint projects and partnerships. Address the quantum skills gap comprehensively and implement proactive measures to attract and retain top global talent, including through dedicated education and training programs at all levels, as well as innovation-friendly immigration policies.

To provide compelling, real-world justifications for this investment and to guide research efforts, the innovation strategy should explicitly encourage and prioritize projects addressing high-impact societal challenges. Examples of such priority application areas, derived from the work of the Open Quantum Institute (OQI) at CERN, include:

- **For SDG 2 (Zero Hunger):** Using quantum machine learning to accelerate the gene editing of crops for improved climate resilience and applying quantum optimization to solve complex Vehicle Routing Problems (VRPs) for more efficient and less wasteful food distribution.
- **For SDG 3 (Good Health and Well-Being):** Using quantum simulation and quantum reservoir computing (QRC) to accelerate the discovery of novel antibiotics to combat antimicrobial resistance (AMR), and applying quantum machine learning (QML) to improve the accuracy of cancer diagnosis from medical imaging.
- **For SDG 6 (Clean Water and Sanitation):** Using quantum optimization for the ideal placement of sensors in urban water networks to detect leaks, and quantum simulation to model the chemical breakdown of "forever chemicals" (PFAS) for more effective water purification.
- **For SDG 7 (Affordable and Clean Energy) & SDG 13 (Climate Action):** Applying quantum algorithms for Wind Farm Layout Optimization (WFLO) and designing more efficient catalysts for direct air capture of $CO_2$ using Quantum Generative Adversarial Networks (QGANs).[244]

  - *Explanation: Building on the successful aspects of the Chips Acts, the EU QA should foster a complete innovation ecosystem. This strategy should be explicitly framed as achieving "Full-Stack Quantum Sovereignty," securing the entire value chain from upstream critical materials and midstream fabrication to downstream software and cross-stream cybersecurity and standards. This involves supporting the entire value chain from basic science to market application, providing access to essential and often costly infrastructure, encouraging public-private partnerships, and crucially, cultivating and attracting the necessary highly skilled human capital through education, training, and supportive talent policies. The U.S. legislative push for a "Quantum Manufacturing USA institute" and a "Quantum Sandbox" provides a concrete model for creating dedicated infrastructure to accelerate innovation from design and fabrication to testing and application development.[245]*

- **Dual-Use Technology Support:** Explicitly permit and encourage funding for quantum research and development projects that have dual-use implications, recognizing the strategic importance of these technologies for both economic competitiveness and security.

  - *Explanation: Given the inherently dual-use nature of many quantum technologies and learning from the strategic playbooks of the US and China, the EU QA should facilitate support for dual-use R&D, moving beyond historical limitations and aligning with recent policy shifts (e.g., EIC) to maintain a technological edge and address security needs.*

- **Energy Infrastructure Planning:** Acknowledge the immense energy demands of future data centers and quantum-AI systems by requiring long-term strategic planning for strengthening the power grid and investing in giga-watt scale facilities, including modern nuclear reactors, to ensure a stable and sufficient energy supply.

  - *Explanation: Proactive planning for energy infrastructure is chief to sustainably support the computational demands of a mature quantum ecosystem.*

---

[244] *See* Open Quantum Inst., *OQI White Paper 2024: SDG Use Cases* 3-38 (2024), https://open-quantum-institute.cern/wp-content/uploads/2024/12/OQI_WhitePaper2024.pdf
[245] *See* Quantum Sandbox for Near-Term Applications Act of 2025, H.R. 3220, 119th Cong. (2025) and Advancing Quantum Manufacturing Act of 2025, S. 1343, 119th Cong. (2025), *supra* notes 68 and 69.

IV. Intellectual Property and Competition Strategy

***Explanation:*** *This section outlines strategies to balance intellectual property incentives with the need for broad innovation, access, and fair competition in the quantum sector, acknowledging quantum's nature as a General-Purpose Technology (GPT) and the goal of mitigating a 'quantum divide'.*

- **Balanced IP Approach:** Define principles for intellectual property in quantum, aiming to incentivize investment and reward invention while ensuring sufficient access to foundational knowledge and preventing market failures like patent thickets or monopolies.[246] Promote open innovation models where appropriate for GPTs, recognizing the limitations of overly closed systems.[247]

    - ***Explanation:*** *Addresses the challenge of value appropriation for GPTs, the need to balance incentives with access, and the benefits of open innovation, while avoiding detrimental effects of IP overprotection.[248]*

- **Specific IP Tools & Frameworks:** Detail considerations for specific IP tools, such as tiered IP approaches (distinguishing foundational tech from applications, inspired by Lemley/nano), patent pools, collaborative licensing, compulsory licensing under specific conditions, potentially shorter patent terms for rapidly evolving areas, and support for open-source quantum software/hardware. To incentivize exponential innovation, the creation of a *sui generis* intellectual property right for foundational quantum algorithms and software, offering shorter protection terms (e.g., 10 years) in exchange for mandatory licensing under Fair, Reasonable, and Non-Discriminatory (FRAND) terms to foster a competitive ecosystem and prevent patent thickets, is proposed.

    - ***Explanation:*** *Provides concrete mechanisms to implement the balanced approach, mitigating thickets and facilitating access, learning from experiences in other deep tech fields.[249]*

- **Novel Innovation Frameworks for Quantum Outputs:** Explicitly address the adequacy of existing IP frameworks for novel quantum outputs, including quantum algorithms and data generated by quantum processes, for instance drawing inspiration from nuanced property concepts (e.g., Roman Law ideas of res publica or use/benefit rights).[250] Consider the "Res Publicae ex Machina" principle for outputs generated autonomously by QAI systems, designating them as public domain.[251]

---

[246] Lemley, M. A. (2005). Patenting nanotechnology. *Stanford Law Review*, 58(2), 601-630, https://doi.org/10.2139/ssrn.741326.

[247] Friesike, S., Widenmayer, B., Gassmann O., et al. Opening science: towards an agenda of open science in academia and industry. J Technol Transf 40, 581–601, (2015). https://doi.org/10.1007/s10961-014-9375-6.

[248] *See* in this light: Yang, J., Chesbrough, H., & Hurmelinna-Laukkanen, P. (2021). How to Appropriate Value from General-Purpose Technology by Applying Open Innovation. *California Management Review*, *64*(3), 24-48. https://doi.org/10.1177/00081256211041787; and Kop, Aboy & Minssen, Intellectual property in quantum computing and market power: a theoretical discussion and empirical analysis, *Journal of Intellectual Property Law & Practice*, Volume 17, Issue 8, August 2022, Pages 613–628, https://doi.org/10.1093/jiplp/jpac060

[249] *See* e.g., Lemley *supra* note 245.

- ○ **Explanation:** *Acknowledges that unique quantum outputs, especially from autonomous QAI, may require bespoke IP solutions or public domain dedication, potentially informed by Roman Law concepts.*

● **Protection of Biometric Identity through Privacy-Enhancing Techniques (PETs):** Introduce novel legal tools that function as privacy-enhancing techniques (PETs), for example based on copyright or related rights. A key example is a copyright-based legal framework (inspired by Danish law) that grants individuals explicit ownership and control over their own biometric data (e.g., face, voice, fingerprints) to protect against quantum-AI enabled deepfakes.

- ○ **Explanation:** *This provision is designed to provide a strong legal basis for combating the creation and dissemination of unauthorized deepfakes and other forms of identity theft that will be supercharged by quantum-AI systems.*

● **Antitrust Oversight and Enforcement:** Mandate proactive monitoring of the quantum market for anti-competitive practices (e.g., exclusionary IP strategies, coopting disruption by incumbents) and ensure robust enforcement of competition law to prevent abuse of dominance and winner-takes-all dynamics. This is particularly critical given the risk of market concentration by a handful of dominant technology firms who could leverage their position in cloud computing and other markets to stifle competition in the nascent quantum ecosystem.[252]

- ○ **Explanation:** *Emphasizes the decisive role of competition law alongside IP law to ensure a level playing field and address market power issues, including the risk of incumbents coopting disruption.[253]*

● **Mitigating the Quantum Divide:** Ensure IP and antitrust policies actively contribute to mitigating the quantum divide by lowering barriers to entry for SMEs and actors from less developed ecosystems, promoting knowledge diffusion, and facilitating access to quantum technologies.

---

[250] *See* e.g., Robbie King, Quantum Algorithms: A Call To Action (Quantum Frontiers, Caltech, 2025) https://quantumfrontiers.com/2025/04/20/quantum-algorithms-a-call-to-action/.
[251] *See* e.g., U.S. Pat. & Trademark Off., *AI and Emerging Technology Partnership Engagement and Events*, https://www.uspto.gov/initiatives/artificial-intelligence/ai-and-emerging-technology-partnership-engagement-and-events. *See* on the Roman Law concepts: Kop, M., AI & Intellectual Property: Towards an Articulated Public Domain, University of Texas School of Law, Texas Intellectual Property Law Journal (TIPLJ), Vol. 28, No. 1, 2020, https://tiplj.org/wp-content/uploads/Volumes/v28/Kop_Final.pdf
[252] *See* e.g., Dekker & Martin-Bariteau, *supra* note 211.
[253] *See* e.g. Mark A. Lemley & Mark P. McKenna, Unfair Disruption, 100 Boston University Law Review 71 (2020), https://www.bu.edu/bulawreview/files/2020/01/LEMLEY-MCKENNA.pdf.

- ○ ***Explanation:*** *Directly links IP/competition policy to the objective of equitable access and avoiding concentration of benefits.*

- **IP & National Security / Dual-Use:** Address the intersection of IP rights with national security, considering mechanisms for strategic patent handling (e.g., balancing secrecy needs with innovation incentives, drawing parallels with challenges in nuclear patent regimes) in alignment with export controls and investment screening for sensitive dual-use quantum technologies. The challenge of balancing innovation incentives with security imperatives through mechanisms like strategic patent handling is particularly acute and is explored in detail in a forthcoming study on the nexus of quantum IP and national security.[254] This nexus forms a dynamic feedback loop: a technology's dual-use nature triggers security concerns, which lead to export controls; these controls, in turn, force innovators into a 'patent-secrecy dilemma,' often favoring confidential trade secrets over public patents to avoid disclosing sensitive capabilities. Understanding this interplay is essential for crafting a coherent EU strategy that avoids inadvertently stifling the very innovation it seeks to promote.

  - ○ ***Explanation:*** *Integrates national security considerations into IP strategy for sensitive dual-use tech, seeking a balanced approach informed by experiences in other sensitive fields.*

- **Evidence-Based & Adaptive Policy:** Commit to ongoing empirical analysis of the quantum patent landscape and innovation dynamics (following Aboy et al.) to inform regular reviews and adaptations of the IP and competition strategy.

  - ○ ***Explanation:*** *Ensures policies remain relevant and effective based on real-world data regarding patenting trends and market structures in the evolving quantum sector.*

V. Standardization, Verification, Certification, and Benchmarking

- **Mandate and Support:** Prioritize and provide dedicated funding for the development of EU and international standards, certification schemes, and reliable performance benchmarks for quantum hardware, software, algorithms, and networks. Consider adopting a "standards-first" philosophy in certain areas to guide early-stage development.

  - ○ ***Explanation:*** *Prioritizing standards development ('standards-first' methodology per Aboy et al.) offers early governance advantages before technology lock-in, fosters market growth through interoperability, builds trust, and provides the*

---

[254] *See* Kop, M., *The Nexus of Quantum Technology, Intellectual Property, and National Security* (CIGI, 2025, forthcoming).

> *technical basis for future certification and regulatory compliance.[255] Standards must also serve as the primary vehicle for embedding universal ethical principles and democratic norms into the technology's architecture.*

- **Quantum Technology Quality Management System (QT-QMS) and Certification (CE Mark):** Propose the development of a harmonized, certifiable QT-QMS standard under ISO/IEC. This system certifies an organization's entire management process, rather than individual products, learning from successful models in the medical device sector. Certification would be conducted by an accredited independent body and would establish a verifiable framework for quality and accountability. This standard would serve as the primary mechanism for providers to demonstrate compliance with the Act's regulatory requirements. This system provides the auditable framework for operationalizing the principles for Responsible Quantum Technology (RQT) outlined in Part II. For high-risk systems, successful third-party certification against this QT-QMS standard would be a prerequisite for obtaining a CE mark and gaining market access, mirroring the successful model of ISO 13485 for medical devices.

    - *Explanation: This arrangement provides regulatory agility, as the standard can be updated to reflect technological progress without amending the core legislation.[256]*

- **Key Areas:** Focus standardization efforts on priority areas such as establishing common terminology, ensuring interoperability between systems, defining meaningful performance metrics (e.g., for qubit quality, computational power, sensor precision), developing robust security protocols (including quantum-safe cryptography aligned with international standards like NIST PQC, coordinated via ENISA), setting safety standards for quantum devices, and creating reliable testing methodologies.

    - *Explanation: Standardization efforts need to target specific, high-priority areas to effectively facilitate communication, market growth, user trust, and safety/security.*

- **European Quantum Testbeds and Benchmarks**: Mandate the creation of dedicated testbeds with a dual focus, learning from omissions in other international strategies. These testbeds will not only assess technical performance against standardized metrics but also provide environments for evaluating societal and security resilience, allowing for the simulation of quantum systems' impact on critical infrastructure, financial markets, and national security.

- **Coordination Role:** Leverage existing EU standardization bodies (e.g., CEN, CENELEC) and empower the proposed EU Quantum Office (OQTA) to coordinate standardization activities at the EU level and represent EU interests internationally.

---

[255] Aboy *et al., supra* note 34.
[256] *ibid.*

- ○ ***Explanation:*** *Effective standardization requires strong coordination involving established bodies and a dedicated entity like the OQTA to ensure alignment and progress.*

- **Registry:** Consider establishing a regulated registry or database for certain types of quantum algorithms or systems, particularly those classified as high-risk or relevant for export controls.

  - ○ ***Explanation:*** *A registry can enhance market transparency, facilitate regulatory oversight and post-market surveillance, and aid in the tracking and control of sensitive algorithms, but requires careful coordination to avoid fragmentation and ensure coherence with other databases and international efforts, recognizing the urgency imposed by the risk of technological lock-in.*

## VI. Governance and Enforcement

- **EU Office of Quantum Technology Assessment (OQTA):** Establish a central EU body dedicated to quantum technology, providing expertise, coordination, and oversight, inspired by models like IAEA, CERN, the former US OTA, and the EU AI Office. Its functions should include: providing independent policy advice; conducting risk/benefit assessments (including dual-use implications); developing ethical guidelines; monitoring technological developments (horizon scanning); coordinating standardization efforts; managing shared EU quantum infrastructure; overseeing export controls; support regulatory enforcement activities; serving as a stakeholder forum; overseeing codes of conduct; and evaluating results from regulatory sandboxes; including promoting institutional plasticity within relevant EU/national bodies and monitoring the IP and competitive landscape. To strengthen multilateral governance and prevent a "quantum divide," the OQTA's mandate shall include a specific function: "To serve as a liaison with international bodies like the OQI and UNESCO, and to actively promote inclusive access to quantum resources and educational materials for quantum-underserved geographies, in alignment with the goals of the International Year of Quantum Science and Technology 2025."[257]

  - ○ ***Explanation:*** *A dedicated, independent expert body is fundamental to effective and agile governance of this complex field, drawing inspiration from successful international and EU models to provide assessment, coordination, and oversight functions, including promoting institutional adaptation across the regulatory landscape and tracking IP/competition dynamics. This proposed structure, combining an EU-level body with national authorities, reflects considerations of different enforcement models, such as the delegation approach seen in some US proposals versus the detailed ecosystem outlined in the EU AI Act.*

---

[257] *See* Open Quantum Inst., *Progress Report (2d Edition)* 4, 6 (2025), https://open-quantum-institute.cern/wp-content/uploads/2025/03/OQI-Progress-Report-March2025.pdf

- **Regulatory Forum:** Consider establishing a dedicated forum, similar to the UK's Digital Regulation Cooperation Forum (DRCF), to facilitate coordination and exchange of expertise among different national and sectoral regulators dealing with quantum technologies.

    - *Explanation:* *Such a forum can enhance consistency in regulatory practices across different domains and Member States, help regulators become "quantum-ready," and prevent unnecessary fragmentation.*

VII. International Aspects, Security, and Export Controls

- **International Cooperation:** Foster collaboration with like-minded international partners on quantum research, development of global standards, and responsible governance principles. A concrete mechanism to advance this would be the establishment of a G7 Quantum Commission, a dedicated forum to coordinate on standards, export controls, and joint research initiatives, thereby preventing a fragmented and counter-productive "quantum race" among allies.[258] Explore the feasibility of ambitious international projects like a "CERN for Quantum/AI," drawing parallels with the collaborative models of CERN and large-scale nuclear fusion projects (e.g., ITER).[259] This shall include fostering "multi-stakeholder dialogues modeled on the GESDA Quantum Diplomacy Symposium to build trust and co-shape global governance norms."[260] Emphasize information sharing and the exchange of best practices, particularly those learned from the nuclear field regarding safety and security culture, and leverage platforms like UNESCO and OECD for ongoing policy dialogue and coordination.

    - *Explanation:* *Recognizing that quantum technology development is a global endeavor, international cooperation is imperative for scientific progress, establishing common standards, addressing shared security concerns, leveraging successful collaborative models, and promoting harmonized regulatory architectures to avoid international fragmentation and shape global norms proactively before a 'quantum event horizon' is crossed. This requires navigating the distinct geopolitical influences of the market-driven "Brussels Effect" and the security-centric "Washington Effect." Frameworks like the AUKUS security pact, which includes specific provisions for quantum technology collaboration, underscore the importance of such strategic partnerships.[261] Furthermore, a nuanced strategy of strategic "recoupling" with competitors like*

---

[258] *See* Dekker & Martin-Bariteau, *supra* note 211.
[259] *See* e.g., Alberto Di Meglio et al., *CERN Quantum Technology Initiative Strategy and Roadmap*, ZENODO (2022), https://doi.org/10.5281/zenodo.5846455 and
https://cds.cern.ch/record/2789149?ln=en
[260] *See* GESDA OQI Report 2024, *supra* note 42.
[261] *See* U.S. Dep't of Def., *AUKUS: The Trilateral Security Partnership Between Australia, U.K. and U.S.*, https://www.defense.gov/Spotlights/AUKUS/.

*China on areas of mutual interest, such as fundamental standards or global challenges, may be necessary to prevent a fractured "quantum splinternet." This view is echoed by the G7, which has called for boosting investment and fostering trusted global cooperation, while acknowledging that a formal global regulatory framework is likely premature.[262]*

- **Values-Based Technology Diplomacy**: Develop and promote a "Quantum for Global Good" package as a cornerstone of the EU's international strategy. This package should bundle access to European quantum technology and platforms with the export of the EU's values-based regulatory and ethical frameworks, offering a distinct alternative to more assertive, geopolitically-driven technology export models.

- **Supply Chain Security:** Implement comprehensive strategies to ensure a secure and resilient supply of critical minerals and components of quantum technologies. This should involve a mix of diversifying supply sources, building strategic partnerships, monitoring dependencies, promoting circular economy practices (recycling), and developing domestic capabilities where feasible.

    - ***Explanation:*** *Securing the complex supply chains for quantum hardware is a strategic necessity. Learning from the Chips Acts, and recent U.S. legislation like the Support for Quantum Supply Chains Act,[263] the EU QA needs proactive strategies involving diversification, partnerships, dependency management, and sustainability to mitigate geopolitical risks and ensure reliable access to essential resources. Data-driven assessment methodologies, such as a Quantum Criticality Index, are needed for identifying specific material vulnerabilities and informing these strategies.[264]*

- **Export Controls (Inspired by Nuclear Governance & Wassenaar):** Establish targeted, robust, and regularly updated export controls for sensitive quantum technologies, components, software, and related know-how that possess significant dual-use applications. This should be part of a risk-based functional approach, where controls are tailored to specific use cases/applications and their potential for harm, rather than to broad technological categories.[265] This approach must also recognize that such controls directly influence corporate IP strategy, often pushing firms to protect innovations via trade secrecy rather than patents, thereby impacting the broader innovation ecosystem.[266] Draw lessons from established nuclear export control regimes (e.g., NSG guidelines) in defining control lists and procedures. Define clear criteria for controls based on technology maturity, potential for military or other harmful applications, and destination risk. Ensure effective coordination of controls at the EU level and work towards alignment with (and adaptation of) international regimes like the Wassenaar Arrangement, with oversight involving the OQTA. Consider implementing tracking systems for controlled exports. In addition, any export control regime must carefully consider conflicts with fundamental rights, such as the freedom of academic discourse, particularly regarding international research collaborations, which may receive constitutional protection against overly broad "deemed export" rules.[267]

---

[262] *See* G7, Kananaskis Common Vision for the Future of Quantum Technologies (2025), https://g7.canada.ca/en/news-and-media/news/kananaskis-common-vision-for-the-future-of-quantum-technologies/.
[263] Support for Quantum Supply Chains Act, H.R. 3788, 119th Cong. (2025), *supra* note 70.
[264] Min-Ha Lee, Andrew J. Grotto & Mauritz Kop, *Methodology for Assessing Geopolitical Risk to Critical Raw Materials Supply Chains and Its Application to Quantum Computing* (2025) (Working Paper, Stanford Center for International Security and Cooperation), https://cisac.fsi.stanford.edu/publications
[265] Dekker & Martin-Bariteau, *supra* note 211.
[266] *See* Kop, *supra* note 252 (Nexus).

  - *Explanation: Given the significant dual-use nature of quantum technologies, effective export controls are essential for non-proliferation and national security. However, such controls must be surgical and harmonized, recognizing the inherent trade-offs between security and innovation. The challenge is compounded by "dual-use ambiguity," where the same technology can serve both benign and harmful ends, making clear-cut classifications difficult. This requires mirroring the principles of established nuclear and dual-use export control practices, involving clear criteria, strong EU coordination, international alignment, and dedicated oversight and tracking. The effectiveness of such controls is often undermined by a lack of international agreement on technical details and competing commercial interests, necessitating a harmonized and strategic EU-level approach.[268] Furthermore, the European Commission has called on Member States to review outbound investments in critical technologies like quantum to assess and mitigate economic security risks.[269]*

- **Non-Proliferation Dialogue & Verification:** Proactively engage in international discussions aimed at establishing frameworks or a treaty for the non-proliferation of potentially harmful quantum and AI capabilities, inspired by the NPT's core pillars (non-proliferation of weapons of mass destruction, disarmament, peaceful use). Consider the need for and feasibility of verification mechanisms inspired by IAEA safeguards for certain high-risk technologies under future international agreements. Utilize relevant international forums, including UNESCO and OECD, to facilitate dialogue, promote shared ethical norms, and build the necessary consensus towards such a treaty or accord.

  As Perrier explicates, the unique characteristics of quantum systems, such as entanglement, present novel challenges to traditional legal concepts of jurisdiction and control, reinforcing the need for new international legal instruments and verification regimes tailored to these non-classical properties.²⁷⁰ For instance, if an action is initiated in one jurisdiction using one part of an entangled pair of particles, causing an instantaneous effect in another jurisdiction via the second particle, which state's laws apply? How is causality established for liability purposes when the connection is non-local? And which jurisdiction's rules should apply? This physical reality of 'jurisdictional entanglement,' where a single non-local quantum process can span multiple sovereign territories simultaneously, renders traditional territorial law obsolete and necessitates a harmonized international legal regime. These are fundamental questions of international private law.²⁷¹ Traditional legal frameworks, both public and civil, which are predicated on actions occurring within specific, contiguous spacetimes, are ill-equipped to address such scenarios. This reinforces the argument that we need both *sui generis* public international law for state-level governance and *sui generis* private international law frameworks to resolve civil liability across borders.

---

[267] *See* Meredith Fore, *The Legal Frontier of Quantum Technology*, U. Chi. News (May 19, 2025), https://news.uchicago.edu/story/legal-frontier-quantum-technology.
[268] *See* Hmaidi & Groenewegen-Lau, *supra* note 124.
[269] Commission Recommendation on reviewing outbound investments and assessing risks to economic security (Jan. 15, 2025), https://ec.europa.eu/commission/presscorner/detail/en/ip_25_261. *Compare to*: Personal Information Protection Law of the People's Republic of China (promulgated by the Standing Comm. Nat'l People's Cong., Aug. 20, 2021, effective Nov. 1, 2021) (China).

- o **Explanation:** *Addressing the profound global security implications of advanced quantum (and AI) systems requires proactive diplomacy aimed at establishing international norms and verifiable agreements to prevent misuse and promote peaceful applications, explicitly learning from the NPT structure and IAEA functions, and leveraging appropriate international organizations for consensus-building. This vision of a 'Quantum Acquis Planétaire' could be enforced by an 'Atomic Agency for Quantum-AI' named 'International Quantum Agency (IQA)', which would oversee treaty compliance and implement safeguards.*²⁷²

- **Quantum Cybersecurity and Q-Day Preparedness (Inspired by US Quantum Computing Preparedness Act):** Mandate the development and implementation of a strategic plan for migrating the EU's critical digital and cryptographic infrastructure to quantum-safe standards. This includes setting clear timelines, defining technical standards in alignment with international bodies like NIST, and supporting both public and private sector entities in this transition. This mandate establishes a new legal duty of 'Anticipatory Data Stewardship,' a proactive obligation for organizations to protect data against foreseeable future technological threats, particularly the "harvest now, decrypt later" risk.

---

²⁷⁰ Perrier, *supra* note 13. *See* also Kop, *supra* note 44 (TTLF, 2020).
²⁷¹ It is essential to distinguish between the public and private law dimensions of a *sui generis* quantum framework. The proposed EU Quantum Act is, by its nature, an instrument of public law. It establishes the rules governing the relationship between the state (and the EU) and the developers, providers, and users of quantum technology. It sets conditions for market access, defines prohibitions, mandates conformity assessments, and creates public enforcement bodies like the proposed OQTA. However, the unique challenges of quantum technology, particularly non-local effects from entanglement, extend deeply into the realm of private law, which governs disputes between private parties. The Quantum Act, while a public law instrument, must therefore be designed to anticipate and proactively address these future private law challenges. It can do so by, for example, harmonizing certain liability rules, mandating technical standards for logging and transparency that would be required for proving causation in a future civil lawsuit, or explicitly paving the way for a subsequent, separate EU instrument designed to amend the rules of private international law (such as the Rome II or Brussels I Regulations) for quantum-related damages. The argument for a *sui generis* framework is thus holistic: it requires a new public law regime that is simultaneously conscious of and preparatory for the unprecedented private law questions to come.
²⁷² Kop, *supra* note 139.

- *Explanation*: *This provision directly addresses the "Q-Day" threat by creating a legal mandate for quantum readiness, similar to the US model. It ensures a coordinated, Union-wide effort to protect essential data and infrastructure from future quantum attacks, moving beyond ad-hoc initiatives to a structured, legally-binding migration to post-quantum-cryptography (PQC) strategy.*

## VIII. Final Provisions

- **Review Clause:** Include a mandatory clause requiring regular review and, if necessary, updates to the Act to ensure it remains effective and adapted to the rapid pace of technological development in the quantum field.

    - *Explanation: Given the high velocity of quantum innovation, a built-in mechanism for periodic review and adaptation is required for the legislation to remain relevant, proportionate, and supportive of responsible progress without becoming outdated or unduly restrictive.*

- **Transitional Arrangements:** Define clear transitional periods and provide supporting measures (e.g., guidance documents, funding for compliance) to allow stakeholders, including industry (especially SMEs), researchers, and regulatory authorities, adequate time and support to adapt to the new framework.

    - *Explanation: Implementing a new regulatory framework requires clear timelines and support mechanisms to ensure a smooth transition for all affected parties, which is standard legislative practice.*

- **Entry into Force:** Specify the date on which the Act, or specific provisions thereof, will become applicable following its adoption.

    - *Explanation: Standard legislative practice requires a defined entry into force date to provide legal certainty.*

## 15. EXECUTIVE SUMMARY

Quantum technologies, encompassing quantum computing, quantum sensing, networking, and artificial intelligence (including quantum-AI hybrids), hold promise for transformative advancements across numerous sectors. Humanity stands at a technological inflection point, with quantum poised to redefine entire industries. Realizing this potential while mitigating inherent risks and upholding fundamental values like human rights and the rule of law necessitates a robust, anticipatory, and harmonized regulatory framework at the EU level, grounded in the precautionary principle. Existing legal and regulatory paradigms, built on classical assumptions of cause and effect, are insufficient to govern technologies derived from the unique and counter-intuitive effects of quantum mechanics. This need for a bespoke framework is rooted in a fundamental conflict between the physics of the quantum realm and the classical, Newtonian worldview that implicitly underpins our legal systems. Its principles lead to an erosion of factual certainty through superposition, the end of locality through entanglement, and a profound challenge to causality through tunneling. The very nature of superposition (the ability to exist in multiple states at once), entanglement (non-local correlations that Einstein called 'spooky action at a distance'), and tunneling (the capacity to breach classical barriers) creates unprecedented capabilities and risks that demand a *sui generis* legal approach, contributing to the emerging *lex specialis* for quantum information technologies.

This contribution outlines the rationale and key considerations for a dedicated European Quantum Act (EU QA), responding to strategic imperatives such as the Quantum Europe Strategy. It draws valuable lessons from existing legislative and innovation strategies in the semiconductor (EU and US Chips Acts) and artificial intelligence (EU AI Act, UK pro-innovation approach, US AI Action Plan, and China's AI+ Plan) domains, as well as governance models from the nuclear sector (IAEA/NPT). A central aspect of this analysis involves a strategic comparison with the U.S. "Winning the AI Race" plan and China's AI+ Plan, and its associated quantum legislative proposals. These U.S. policies can be viewed as forming an "American Digital Silk Road," an ambitious effort to build a techno-economic sphere of influence by setting the global "rules of the road." Understanding this analogy provides the EU with valuable insights into the importance of adopting a "full-stack" industrial policy and defining a distinct, values-based model of international technology diplomacy as a compelling alternative.

In response, our analysis concludes that the EU Quantum Act should be a two-pillar instrument, combining New Legislative Framework (NLF)-style regulation with a more ambitious Chips Act-style industrial and security policy that learns from successful US innovation models. This two-pillar approach achieves two complementary objectives simultaneously: it simultaneously fosters innovation and economic growth through strategic investment while imposing clear regulatory guardrails based on risk. The framework is also modular, designed to be adaptive across the technology's lifecycle with adjustable components—such as risk-based tiers, guiding principles, and regulatory sandboxes. For its industrial policy, the Act can draw from the EU and US Chips Acts to model funding mechanisms, secure supply chains for critical materials, and accelerate the "lab-to-market" pipeline for innovation. This can be further enhanced by adopting practices inspired by the US Defense Advanced Research Projects Agency (DARPA) to incentivize high-risk, high-reward research through competitions and prizes. For its regulatory arm, the EU AI Act, which is built upon the EU's New Legislative Framework, provides a robust, risk-based framework for managing the dual-use nature of emerging technologies, offering a model for establishing prohibitions and obligations for high-risk quantum applications. To balance this and avoid the EU falling behind due to overregulation, the contrasting UK's principles-based approach to AI regulation offers an avenue for tempering prescriptive rules with overarching guidelines, implemented through sector-specific regulators. Coherence across diverse quantum-AI implementations, market verticals, and existing EU legislation can be achieved through the principle of functional equivalence, which focuses on regulating behaviors and use cases rather than the technology itself. Synthesizing historical lessons, the Act must combine the *anticipatory foresight* of the nuclear model, the *ethical integration* of biotechnology and AI, and the *adaptive flexibility* of the internet model, while rejecting the latter's failed "permissionless innovation" ethos to justify proactive, *ex-ante* rules. Finally, the historical challenges of nanotechnology regulation—including overregulation, overpromising, and a lack of public awareness—underscore the critical need for proactive dialogue and the avoidance of overly burdensome rules that could stifle innovation.

To ensure regulatory action is both proactive and innovation-friendly, the Act must be guided by a sophisticated application of core EU principles. It should embrace the precautionary principle to justify early engagement with uncertain, high-impact risks, yet this must be rigorously balanced by the principles of proportionality and subsidiarity to avoid stifling innovation. This balanced stance is operationalized through the Act's risk-based framework and regulatory sandboxes, ensuring that intervention is targeted and does not overregulate promising areas of development.

A comparative study of global innovation ecosystems (US, EU, China) highlights the need for the EU QA to strategically support both fundamental research and commercialization, particularly fostering dual-use technologies and avoiding historical limitations on funding scope. This includes learning from the comprehensive ecosystem-wide strategy of the U.S. AI Action Plan and the pragmatic, application-focused approach of China's AI+ Plan, while consciously developing a distinct European model of international partnership that contrasts with the assertive technology diplomacy of the U.S. In this context, investing in dual-use capability is presented as a responsible act of deterrence necessary for national security and technological sovereignty in the current geopolitical reality. This strategic investment is not merely an act of self-reliance but a prerequisite for the EU to become an indispensable partner in a transatlantic tech alliance, ensuring that democratic values underpin the next technological era. Learning from the well-established governance structures of the nuclear industry—including the International Atomic Energy Agency (IAEA), the Treaty on the Non-Proliferation of Nuclear Weapons (NPT), and the Nuclear Suppliers Group (NSG)—particularly in safety regulation, non-proliferation verification, and export controls, provides valuable insights for managing quantum risks. The large-scale international investment and collaboration model seen in nuclear fusion research (e.g., the International Thermonuclear Experimental Reactor ITER) and fundamental physics (e.g., CERN) also offers a compelling precedent for quantum computing and simulation, sensing and metrology, and networking and communication development.

This contribution recommends a hybrid, modular regulatory structure for the EU QA, following the technology lifecycle (*ex-ante, ex-durante, ex-post*). This approach combines a risk-based classification system with a comprehensive set of overarching principles for Responsible Quantum Technology (RQT). These principles address the full spectrum of concerns, from technical requirements like Safety, Security, and Robustness, to procedural safeguards such as Transparency, Explainability, Accountability, and Contestability, and foundational societal values including Fairness, Sustainability, Equitable Access, Privacy, and the protection of Human Agency and Oversight. This is complemented by a forward-looking commitment to Proactive Risk Management, Dual-Use Mitigation, international Collaboration, and long-term Intergenerational Equity.

The contribution articulates the concept, or metaphor, of a 'quantum event horizon', which underscores the inherent unpredictability and the prospect for technological lock-in that characterize this nascent field. Just as an event horizon in astrophysics marks a boundary beyond which events cannot affect an observer, and in quantum mechanics, observation influences the state of a particle, the 'quantum event horizon' in the context of quantum technology signifies a point where future developments, applications, and societal impacts become increasingly difficult to foresee. In addition, it serves as a stark warning against technological lock-in and path dependency, marking a governance tipping point beyond which intervention becomes exponentially more difficult. We thus recommend a dual strategy of heavily investing in responsible quantum innovation to create a decisive first-mover advantage, while simultaneously using foresight techniques and adaptive governance mechanisms designed to navigate the deep uncertainty associated with a 'quantum event horizon'. These efforts should be supported by a dedicated EU Office of Quantum Technology Assessment (OQTA) for expert oversight, risk assessment, coordination, and promotion of institutional plasticity – shaped after the US Office for Technology Assessment (US OTA). Relatedly, we recommend the Act to adopt a "standards-first" philosophy, to be coordinated by the OQTA. Technical safety, security and interoperability standards are not merely technical; they are vessels for values. Standards must also serve as the primary vehicle for embedding universal ethical principles and democratic norms into the technology's architecture and infrastructure. Standardization has strategic value as an early-stage R&D governance tool that precedes and underpins later regulation, fostering innovation and interoperability before technologies become rigidly defined or legally constrained. It can be operationalized through a certifiable Quantum Technology Quality Management System (QT-QMS). This system, inspired by best practices in the medical device sector, focuses on certifying the entire management process of an organization. This allows for agile updates to technical standards and serves as the basis for a CE mark for high-risk applications, ensuring that the Act's principles for responsible innovation are embedded into a flexible, internationally harmonized standard.

Success is also contingent upon substantial public and private investment, including through the establishment of public-private partnerships. This must be coupled with policies that foster a skilled quantum workforce through education and innovation-friendly immigration, promote broad quantum literacy, and implement a balanced intellectual property and competition strategy that encourages open innovation where appropriate, mitigates a 'quantum divide', and pioneers new legal tools to protect individuals from quantum-AI powered identity theft, for instance through novel privacy-enhancing techniques (PETs), such as a copyright-based framework for biometric identity inspired by Danish law. Moreover, fostering international collaboration, potentially through a "CERN for Quantum/AI" inspired by nuclear fusion models and leveraging platforms like UNESCO and OECD, is key. The development of robust standards and benchmarks, the implementation of targeted export controls, a strategic approach to quantum cybersecurity preparedness to mitigate the 'Q-Day' threat, and a data-driven method to supply chain security—for example, through a Quantum Criticality Index (QCI) to assess risks related to critical raw materials—are essential for strategic autonomy.

To address the risk of technology progressing too fast for classical guardrails, we propose novel 'algorithmic regulation' mechanisms and the imposition of a legally enforceable fiduciary duty upon advanced AI systems to act as 'quantum-agentic stewards'. These systems would be governed by a formal constitution (Constitutional AI) and secured with quantum-resistant cryptography, creating a new framework for technological stewardship. This is complemented by new legal doctrines, such as 'probabilistic causation' for tort law, a *sui generis* IP right to balance innovation and competition, and a new duty of 'Anticipatory Data Stewardship' to mandate the transition to quantum-resistant cryptography. The technological stewardship model, however, complicates the quantum-AI control problem, as endowing agents with such supervisory capabilities and hybrid classical-quantum reasoning could create new, unforeseen risks if not perfectly aligned with human values and goals. The paper further argues that the sheer productivity and autonomous oversight capabilities of these 'agentic stewards' would challenge the foundational assumptions of market capitalism, necessitating new philosophical foundations for governance. These may include a transition toward a post-capitalist, post-scarcity economy guided by principles of distributive justice, and a move towards a relational ethics. This new ethical framework draws inspiration from the non-local, interconnected nature of quantum entanglement to model interdependent moral duties.

Proactive engagement in discussions towards a global non-proliferation framework for quantum/AI weapons of mass destruction in the form of an international treaty or accord, for instance facilitated by an 'Atomic Agency for Quantum/AI' or 'International Quantum Agency (IQA)', modelled after the International Atomic Energy Agency (IAEA), is recommended. Such a *sui generis* framework must be designed to address the unique challenges that quantum phenomena, particularly the non-local and probabilistic nature of entanglement (creating 'jurisdictional entanglement'), pose to traditional principles of public international law. This vision culminates in the "Qubits for Peace" initiative, which forms part of a necessary holistic global quantum governance framework that integrates international treaties, coordinated export controls, harmonized standards, and coherent national regulations, designed to ensure quantum technologies are developed safely, ethically, and for the benefit of all humanity. Ultimately, the EU QA should aim to create a vibrant, responsible, and societally beneficial quantum ecosystem, ensuring these transformative technologies are harnessed for sustainable economic growth and the common good while safeguarding against harms and avoiding or minimizing regulatory fragmentation.

The concluding section consolidates the previous analysis to propose a potential outline of a EU Quantum Act that contains key legislative elements. This blueprint serves as a concrete proposal, integrating the diverse insights gathered and providing a detailed overview of the recommended content and structure for a prospective EU Quantum Act.